\documentclass{ragtime} % NO optional arguments are required
\usepackage{graphicx}

\title[Variation of the primary and reprocessed radiation in the flare--spot
model]%
{Variation of the primary and reprocessed radiation in the flare--spot model}

\author[M. Dov\v{c}iak et al.]%   Now let's start the paper title authors:
       {Michal Dov\v{c}iak\at[]{1,a}  % Makes referencing superscript `1'
        Vladim\'{\i}r Karas\at[]{1}   % ref. superscr. `1,a',
        Giorgio Matt\at[]{2} and \splitauthors          %
        Ren\'{e} W. Goosmann\at[]{1}\\% Termination of authors' block; if
        \ins{1}Astronomical Institute, Academy of Sciences of the Czech
        Republic,\splitins[1]
        Bo\v{c}n\'{\i}~II, CZ-140\,31~Prague, Czech Republic\\%
        \ins{2}Dipartimento di Fisica, Universit\`a degli Studi
        ``Roma Tre'',\splitins[1]
        Via della Vasca Navale 84, I-00146~Roma, Italy\\%
        \ins{a}\Email{dovciak@astro.cas.cz}} % This is how to present E-mail.

\begin{document}

\begin{abstract}
We study light curves and spectra (equivalent widths of the iron line and
some other spectral characteristics) which arise by reprocessing on the surface
of an accretion disc, following its illumination by a primary off-axis source
--- an X-ray `flare', assumed to be a point-like source just above the
accretion disc. We consider all general relativity effects (energy shifts,
light bending, time delays, delay amplification due to the spot
motion) near a rotating black hole.
For some sets of parameters the
reflected flux exceeds the flux from the primary component.
We show that the orbit-induced variations of the equivalent width with
respect to its mean value can be as high as 30\% for an observer's inclination
of 30$^{\circ}$, and much more at higher inclinations. We calculate the
ratio of the reflected flux to the primary flux and the hardness ratio which
we find to vary significantly with the spot phase mainly for small orbital
radii. This offers the chance to estimate the lower limit of the black hole
spin if the flare arises close to the black hole. We show the results for
different values of the flare orbital radius.
\end{abstract}


\section{Introduction}
X-ray spectral measurements of the iron line and the underlying continuum
provide a powerful tool to study accretion discs in active galactic
nuclei (AGN) and Galactic black holes, for a review see \citet{fab00,rey03}.
If a line originates by
reflection of the primary continuum, then its observed characteristics
may reveal rapid orbital motion and light bending near the central black hole.
Spectral characteristics can be employed to constrain the black hole mass
and angular momentum. A particularly important role is played by the
equivalent width (EW), which reflects the intensity of the line versus the
continuum flux as well as the role of general relativity effects in the
source. In order to reduce the ambiguity of the results one needs to perform
spectral fitting with self-consistent models of both the line and
continuum.

Some AGN are known to exhibit iron lines with an EW greater than expected for
a ``classical''
accretion disc. Enhanced values for the EW can be obtained by assuming
an anisotropical distribution of the primary X-rays \citep{ghi91},
significant ionization of the disc matter \citep{mat93}
or iron overabundance \citep{geo91}. \citet{mar96} and \citet{mar00} found,
using an axisymmetric lamp-post
scheme, an anticorrelation between the intensity of the reflection features and
the primary flux. When the primary source is at a low height on the disc axis,
the EW can be increased by up to an order of magnitude with respect
to calculations neglecting general relativity effects.
When allowing the source to be located off the axis of rotation, an
even stronger enhancement is expected \citep{dab01}.
\citet{min03} and \citet{min04} have realised that this so-called light
bending model can naturally explain the puzzling behaviour of the iron line of
MCG--6-30-15, when the line saturates at a certain flux level and then
its EW starts decreasing as the continuum flux increases further.
\citet{nie07} point out that the illumination by radiation which returns to the
disc (following the previous reflection of the primary emission) also
contributes significantly to the formation of the line profile in some
cases. This results into the line profile with a pronounced blue peak
unless the reflecting material is absent within the innermost 2--3
gravitational radii.

In our previous paper \citep{dov04a}, we have proposed that
the orbiting spot model could explain the origin of transient narrow
lines, which have been reported in some AGN X-ray spectra
\citep{tur02,gua03,yaq03} and widely discussed since then. The main
purpose of the current paper is to present accurate
computations of time-dependent EWs and other spectral characteristics within
the framework of the flare-spot model, taking into account a consistent scheme
for the local spectrum reprocessing. The main difference from previous
papers is that the current one combines the primary source power-law
continuum with the reprocessed spectral features. Both components are
further modified by relativistic effects as the signal propagates
towards an observer.

In a parallel paper, \citet{dov07}, we study general relativistic
effects and spectral characteristics (EWs, hardness ratio, etc.) for the
flare-spot model in two model setups --- the Schwarzschild black hole with
a flare arising at radius $7\,r_{\rm g}$ and an extremally rotating black hole
with a flare at $3\,r_{\rm g}$. In the current paper we would like to present
the results of our computations for more values of the flare orbital radius.
We also show that for a given flare radius the resultant spectra do not differ
much for different spins of the black hole.

In Section \ref{model} we describe the model and the approximation used and
in Section \ref{results} we present the results of our calculations. For a more
detailed description of the model and for the equations used we refer the
reader to the paper \citet{dov07}.

\section{Model approximations and limitations}
\label{model}

\begin{figure}
\begin{center}
\includegraphics[width=7cm]{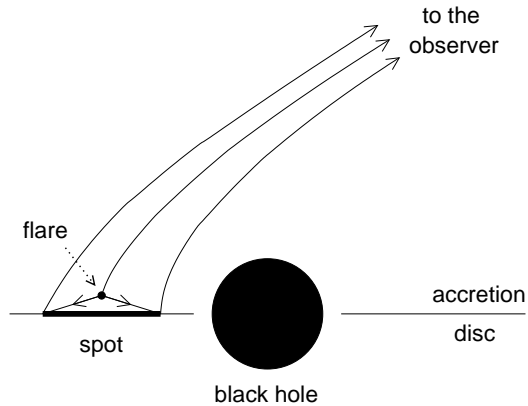}
\end{center}
\caption{A sketch of the model geometry (not to scale).
A localized flare occurs above the
disc, possibly due to magnetic reconnection, and creates a spot
by illuminating the disc surface. The resulting `hot spot' co-rotates with the
disc and contributes to the final observed signal by reprocessing
the primary X-rays.}
\label{fig-model}
\end{figure}

We examine a system composed by a black hole, an accretion disc and a
co-rotating flare with the spot underneath \citep{col03},
see Fig.~\ref{fig-model}. The gravitational field is described in terms of
the Kerr metric \citep{mis73}. Both static
Schwarzschild and rotating Kerr black holes are considered. The
co-rotating Keplerian accretion disc is geometrically thin and optically
thick, therefore we take into account only photons coming from the
equatorial plane directly to the observer. We further assume that the
matter in the accretion disc is cold and neutral.

A flare is supposed to arise in the disc corona due to a magnetic
reconnection event \citep{gal79,pou99,mer01}. Details
of the formation of the flare and its
structure are not the subject of the present paper, instead we assume
that the flare is an isotropic stationary point source with a power-law
spectrum, located very near above the disc. It co-rotates with the accretion
disc. We also assume that a single flare dominates the
intrinsic emission for a certain period of time.

The spot represents the flare-illuminated part of the disc surface. We
consider the spot to be a rigid two dimensional circular feature, with its
centre directly below the flare. Thus the spot does not share the
differential rotation with the disc material. However, the matter in the disc
lit by the flare is in Keplerian motion at the corresponding radii, and so
it has different velocities at different parts of the spot (which is
important when calculating the transfer function for the observer in the
infinity). Because the flare is very close to the disc, the spot does not
extend far from below the flare. We approximate the photon trajectories between
the flare and
the spot by straight lines and we do not consider the energy shift and
abberation due to the different motion of the flare and matter illuminated by
it. Furthermore we neglect the time delay between the photon's emission from
the flare and its later re-emission from the spot.

\begin{figure}
\begin{center}
  \includegraphics[width=4.cm]{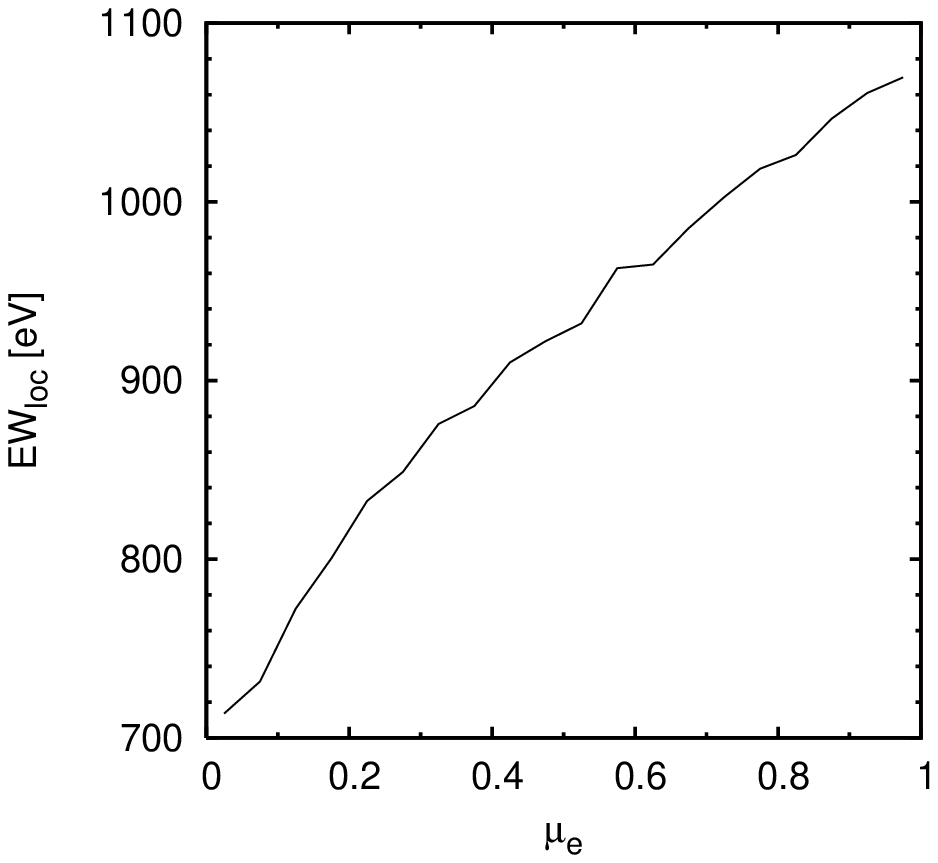}
  \includegraphics[width=4.cm]{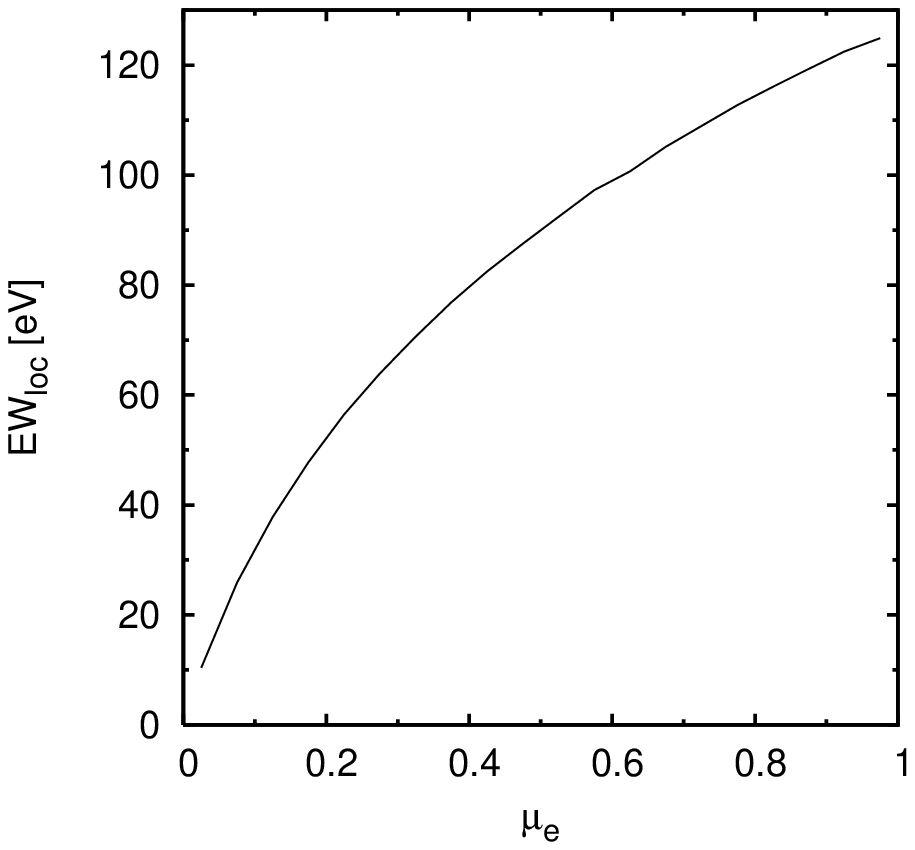}
  \caption{{\bf Left:} The local equivalent width without taking the primary
  flux   into account as a function of the direction of emission.
  {\bf Right:} The same as in the left panel but with the flux from the primary
  source included.}
  \label{fig-ew_loc}
\end{center}
\end{figure}

\begin{figure}
\begin{center}
  \includegraphics[width=4.cm]{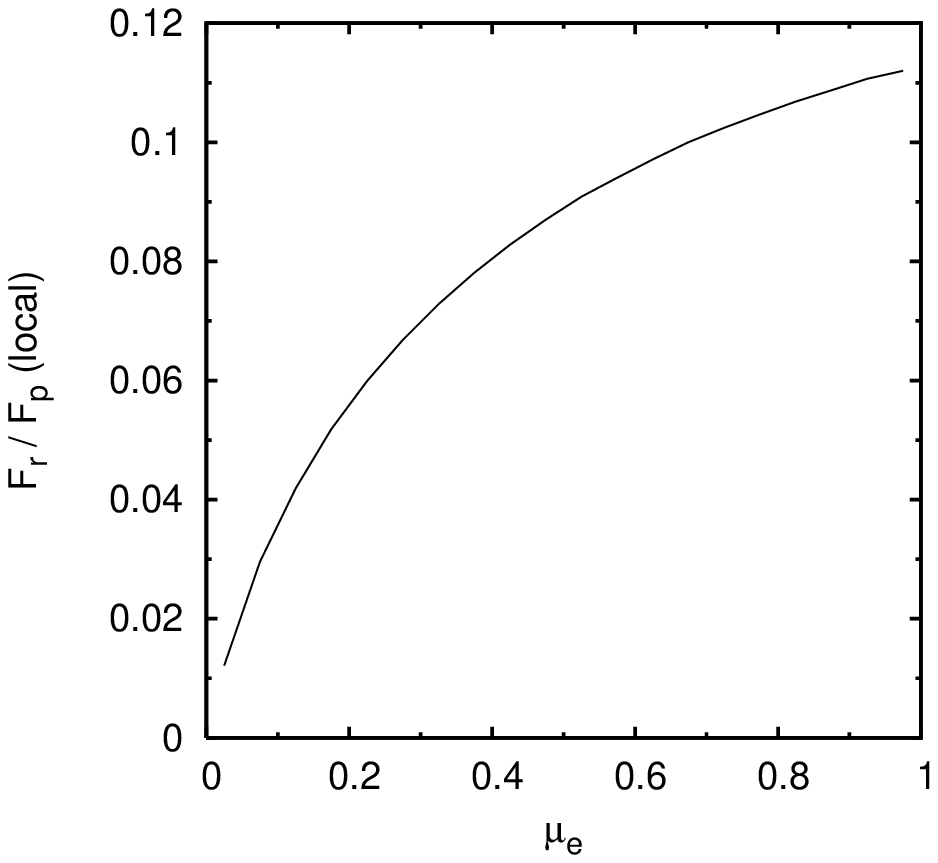}
  \includegraphics[width=4.1cm]{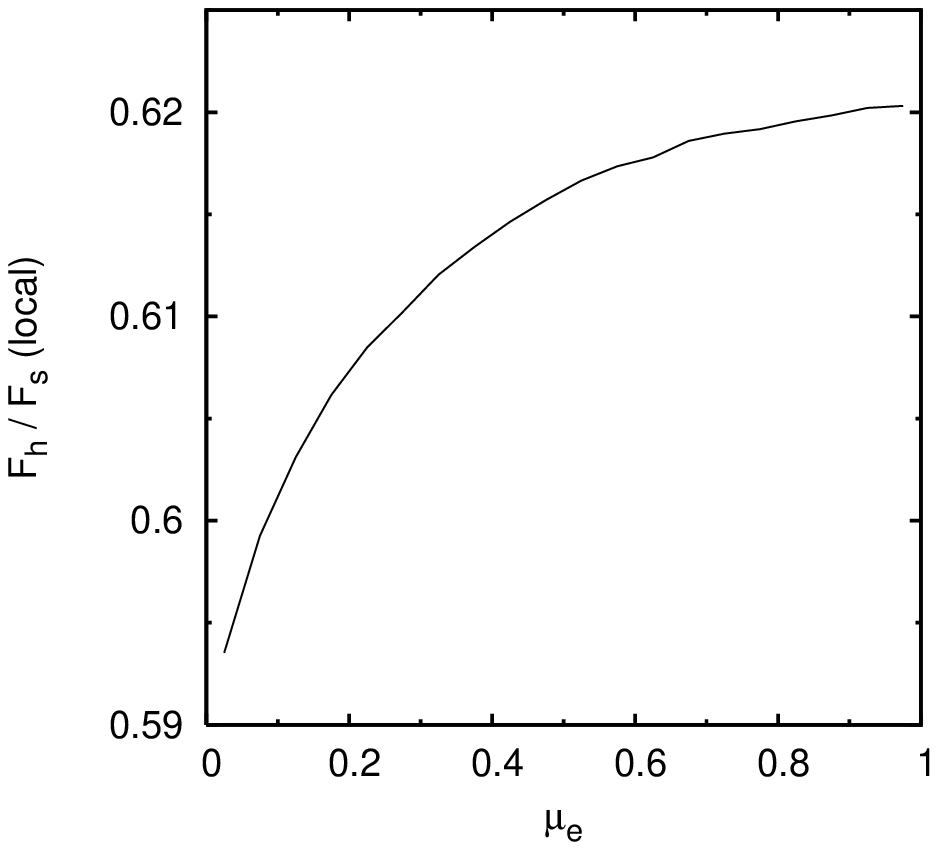}
  \caption{{\bf Left:} The ratio of the locally emitted energy flux in the
  direction   $\mu_{\rm e}$ to the primary flux. The fluxes are integrated in
  the energy   range 3--10~keV.
  {\bf Right:} The local hardness ratio of the fluxes in the ranges
  6.5--10~keV ($F_{\rm h}$) and 3--6.5~keV ($F_{\rm s}$).}
  \label{fig-ratio_loc}
\end{center}
\end{figure}

The intrinsic (local) spectra from the spot were computed by Monte Carlo
simulations considering multiple Compton scattering and iron line fluorescence
in a cold, neutral, constant density slab with solar iron abundance. We used
the NOAR code for these computations, see Section 5 of \citet{dum00} and
Chapter 5 of \citet{goo06}.
The local flux depends on the local incident and local emission angles, hence
the flux changes across the spot. Here and elsewhere in the text we
refer to the quantities measured in the local frame co-moving with the matter
in the disc as ``local''.

The local flux consists of only two components --- the flux from the
primary source (the flare) and the reflected flux from the spot. The latter one
consists of the reflection continuum (with the Compton hump and the iron edge
as the main features) and the neutral K$\alpha$ and K$\beta$ iron lines. No
other emission is taken into account. The spectral properties of the local
emission (the local EW, ratio of the reflected flux to the primary one and
hardness ratio) are shown in Figs.~\ref{fig-ew_loc} and \ref{fig-ratio_loc}.

As far as the photon trajectories from the spot to the observer are concerned,
all general relativistic effects --- energy shift, aberration, light bending,
lensing and relative time delays --- are taken into account. We assume that
only the gravity of the central black hole influences the photons on their
path from the disc to the observer. This allows us to define a relatively
simple scheme in which different intervening effects remain under full control
and can be well identified.

\section{Spectral characteristics of the observed signal}
\label{results}

\begin{figure*}
\begin{center}
  \includegraphics[width=4cm]{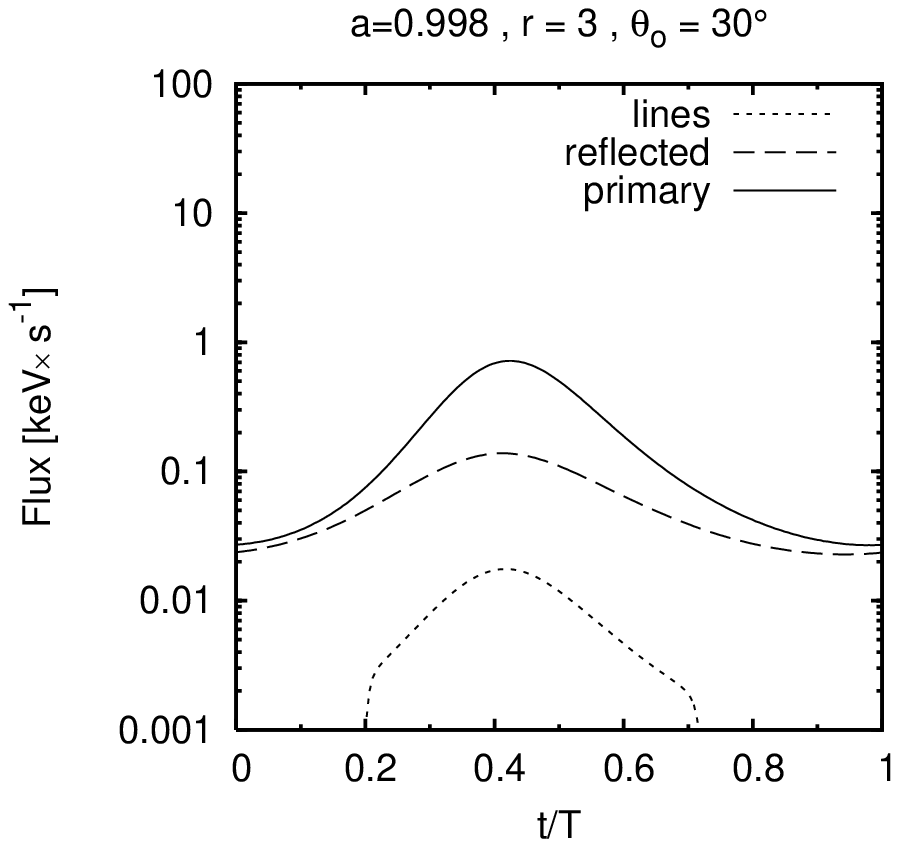}
  \includegraphics[width=4cm]{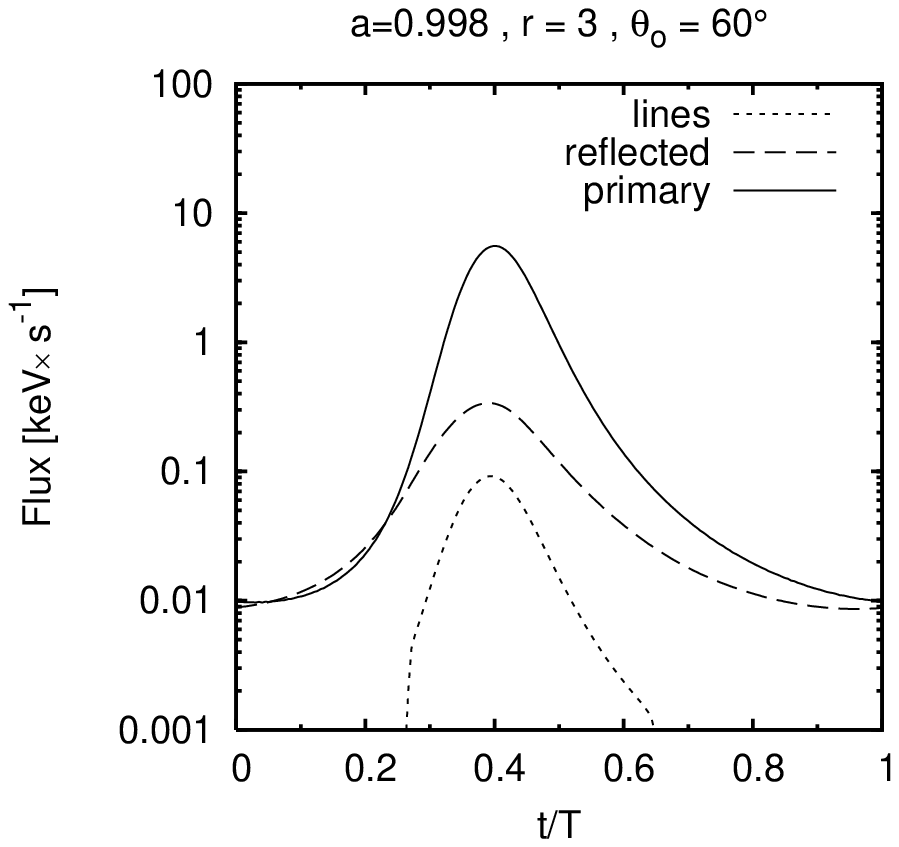}
  \includegraphics[width=4cm]{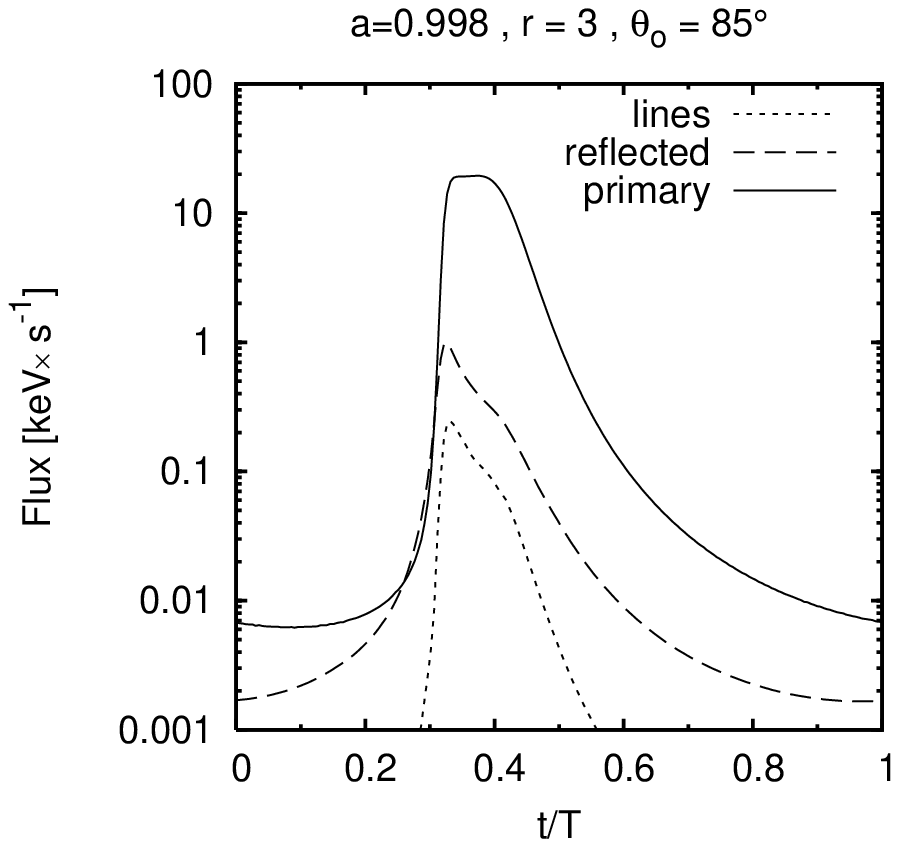}\\[4mm]
  \includegraphics[width=4cm]{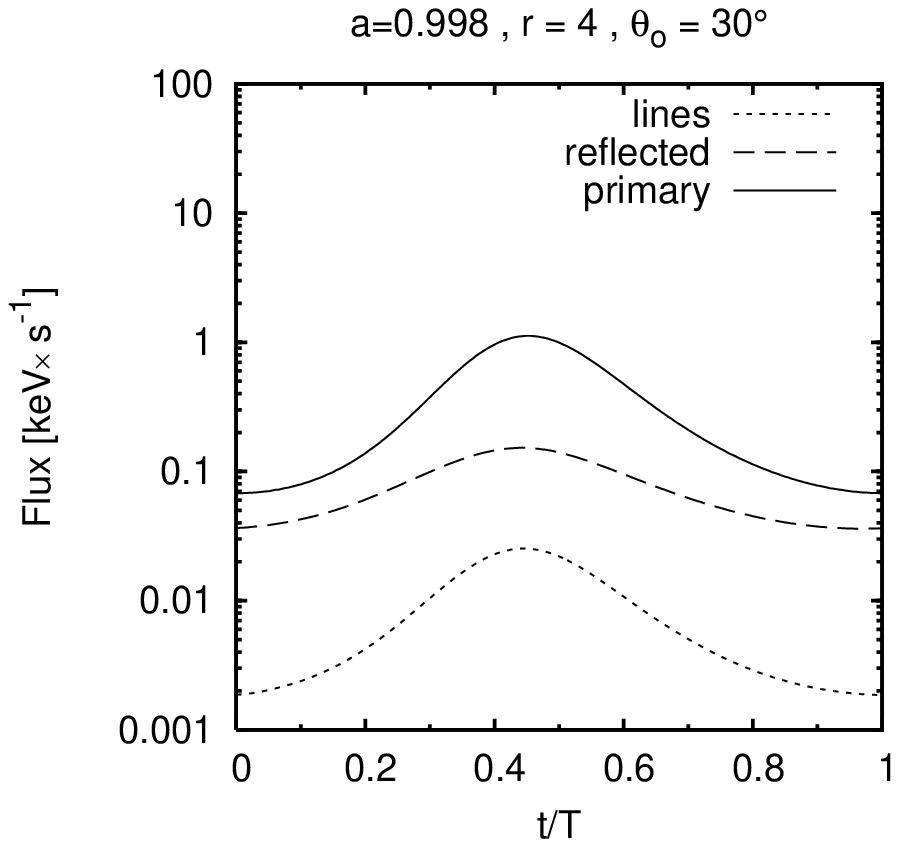}
  \includegraphics[width=4cm]{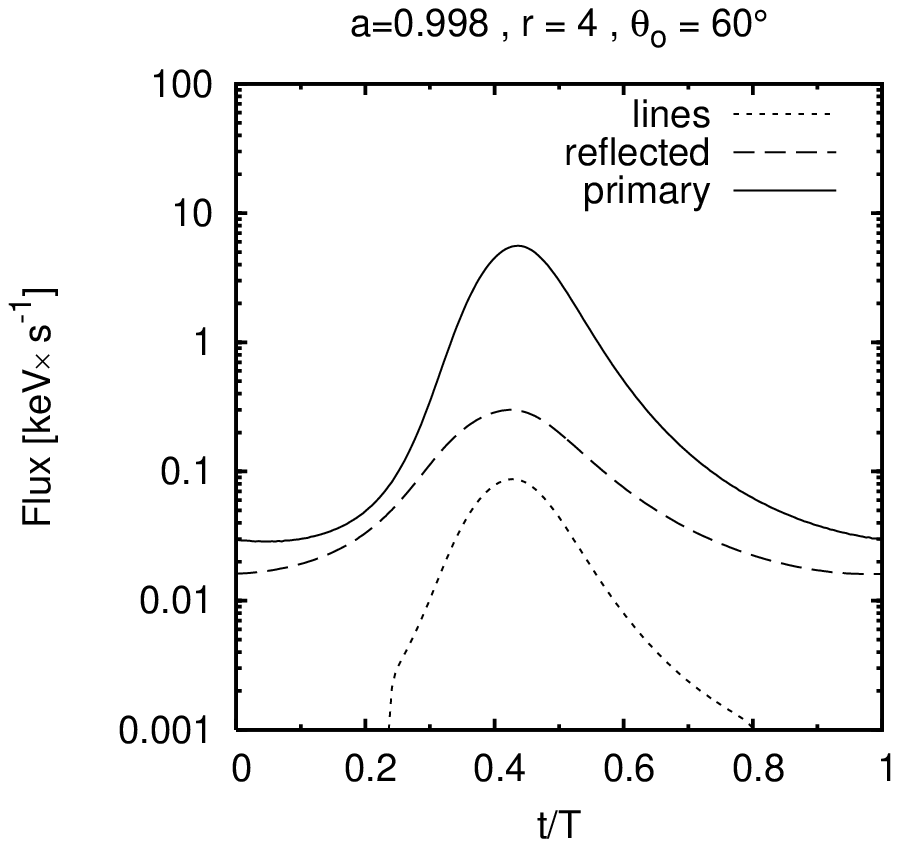}
  \includegraphics[width=4cm]{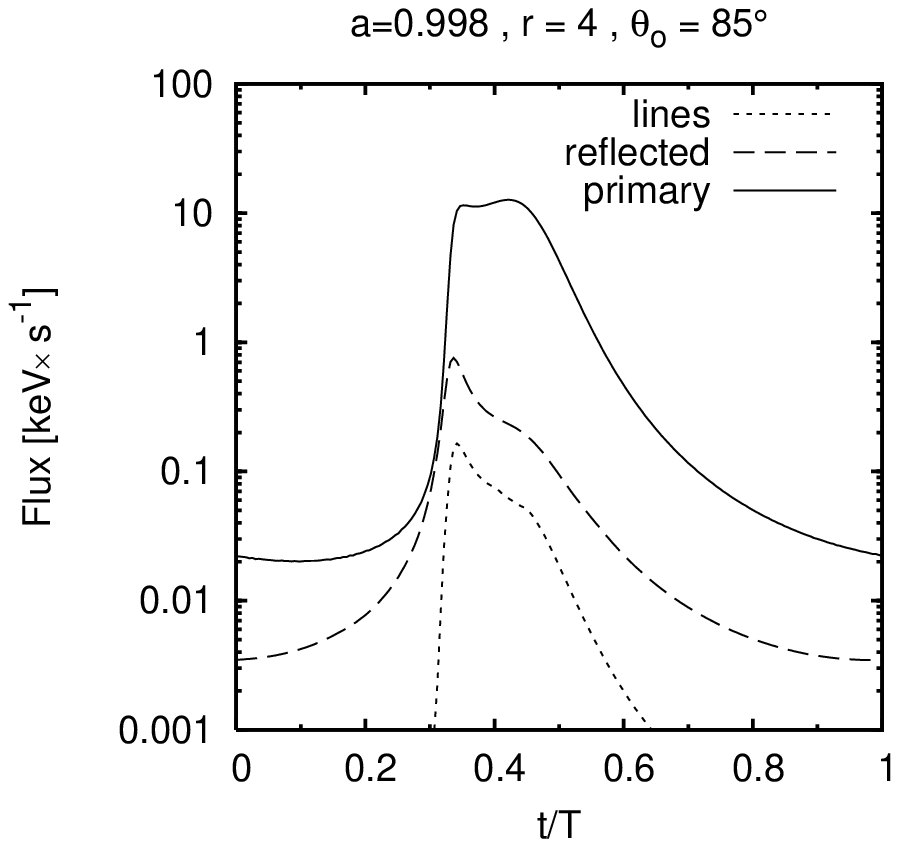}\\[4mm]
  \includegraphics[width=4cm]{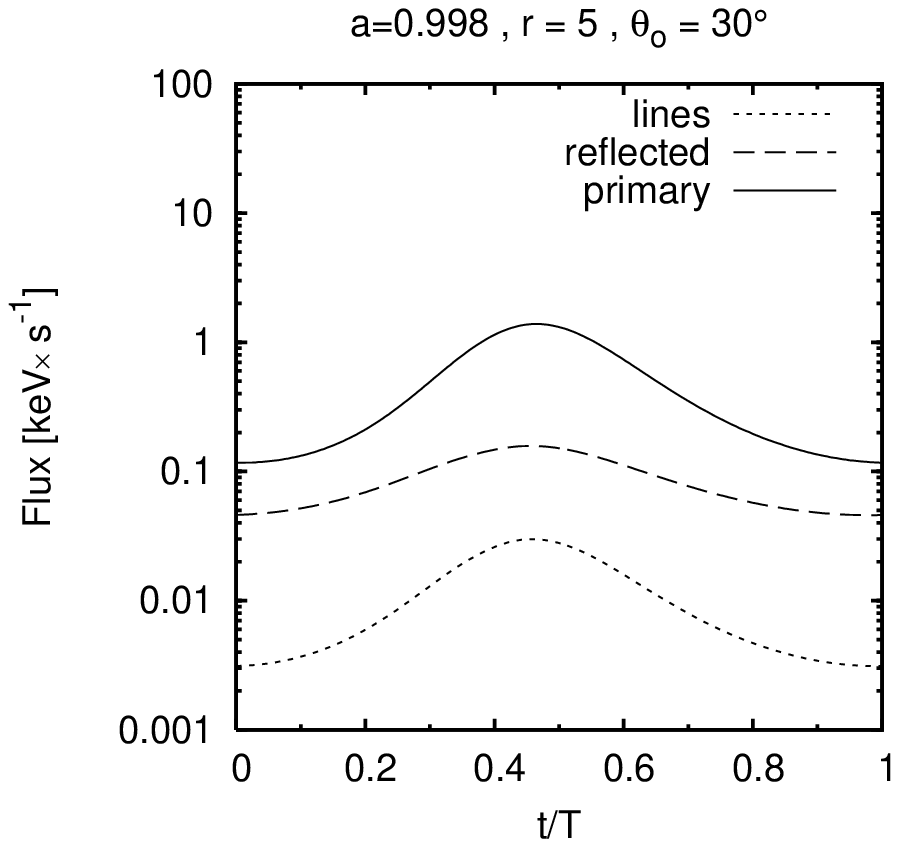}
  \includegraphics[width=4cm]{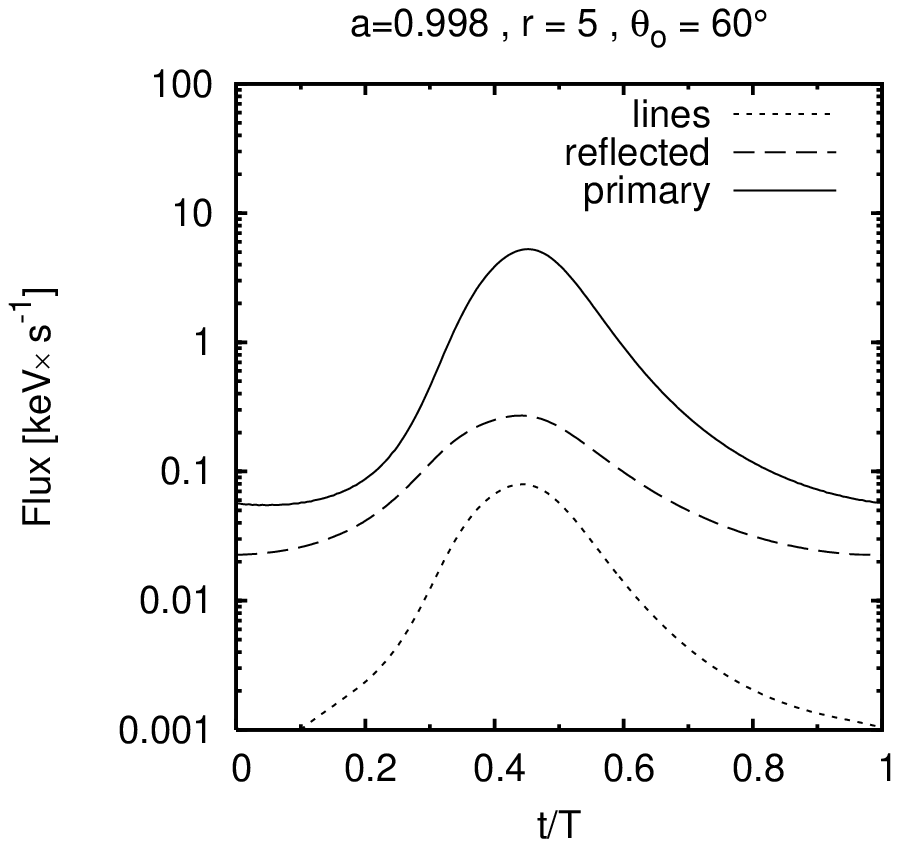}
  \includegraphics[width=4cm]{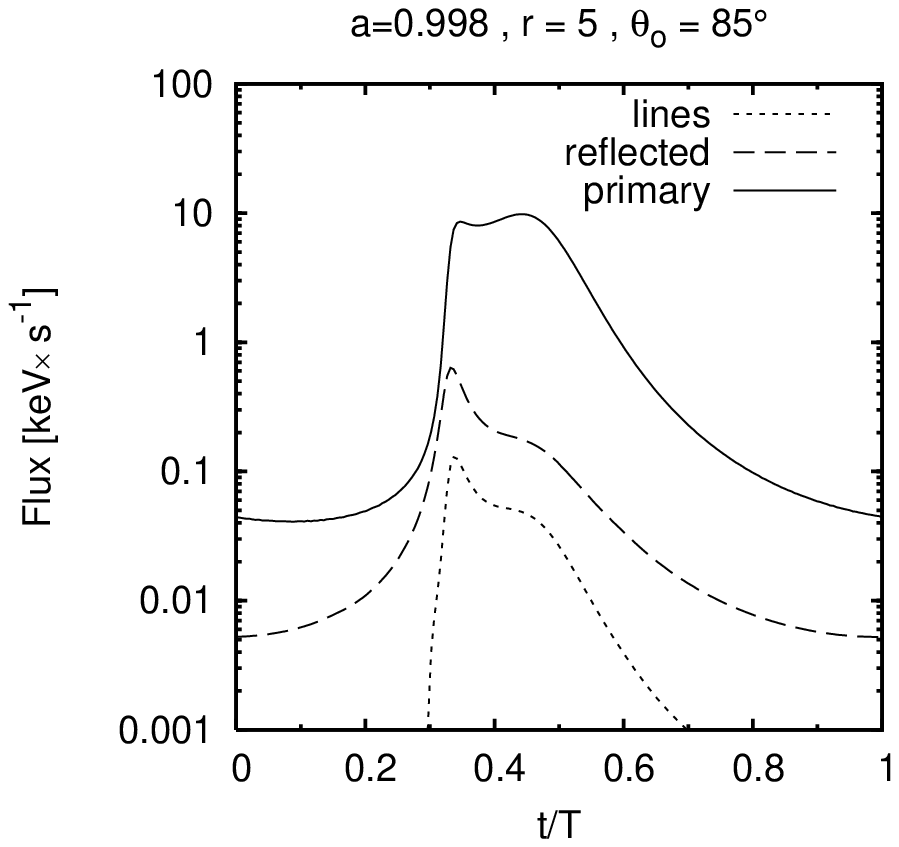}\\[4mm]
  \includegraphics[width=4cm]{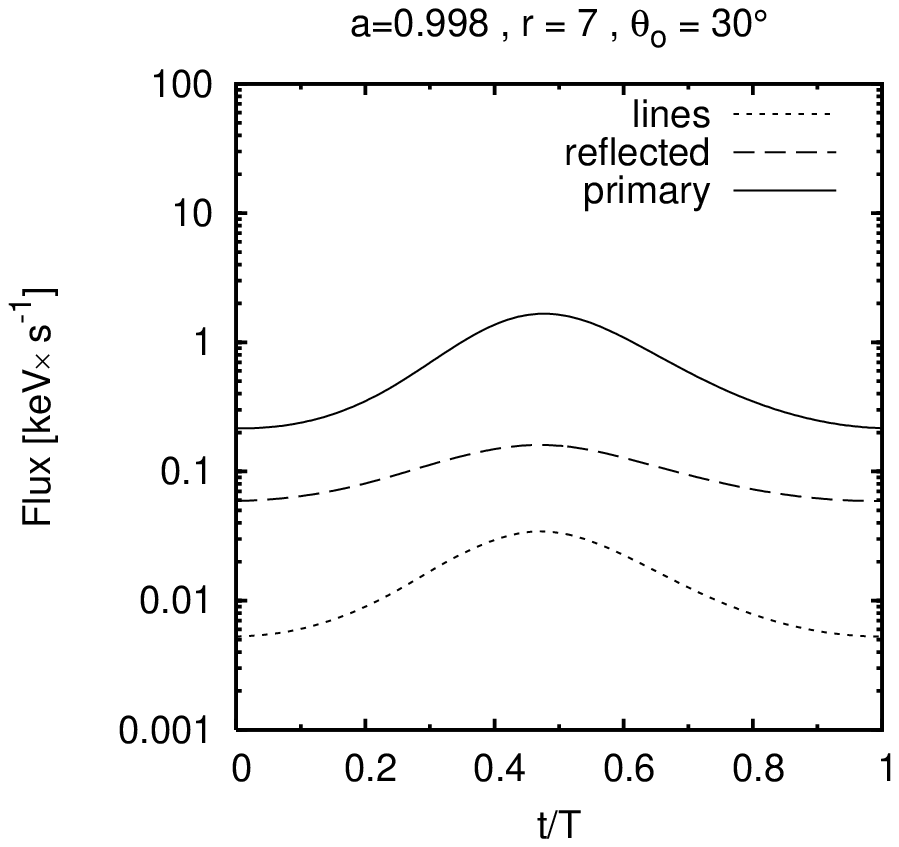}
  \includegraphics[width=4cm]{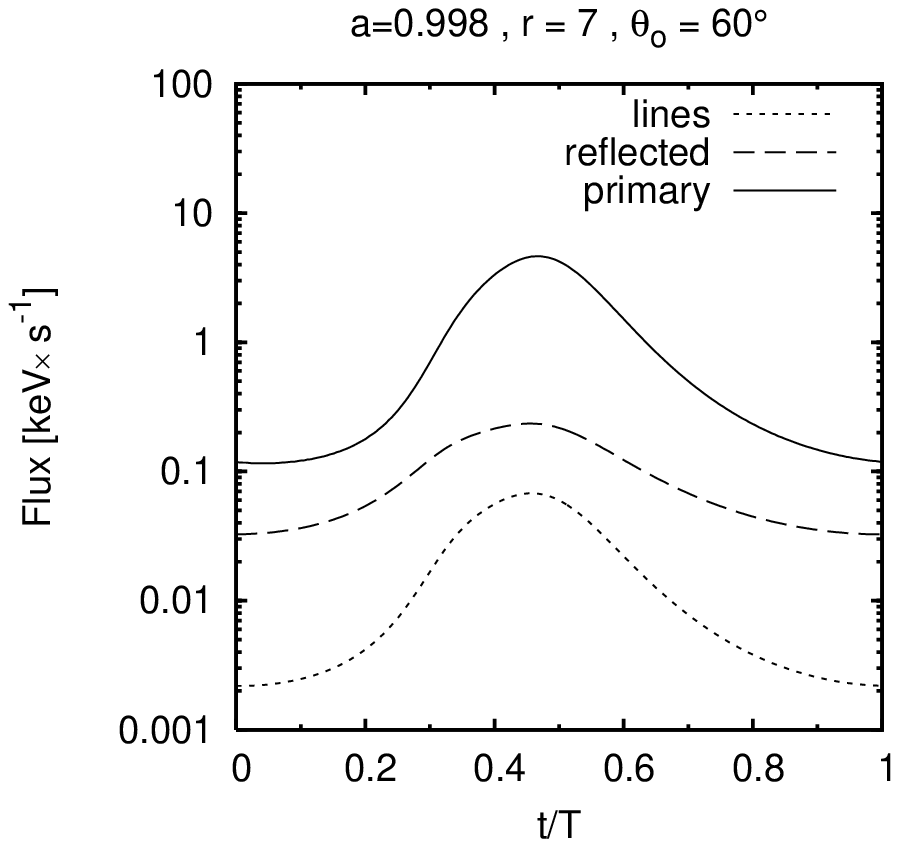}
  \includegraphics[width=4cm]{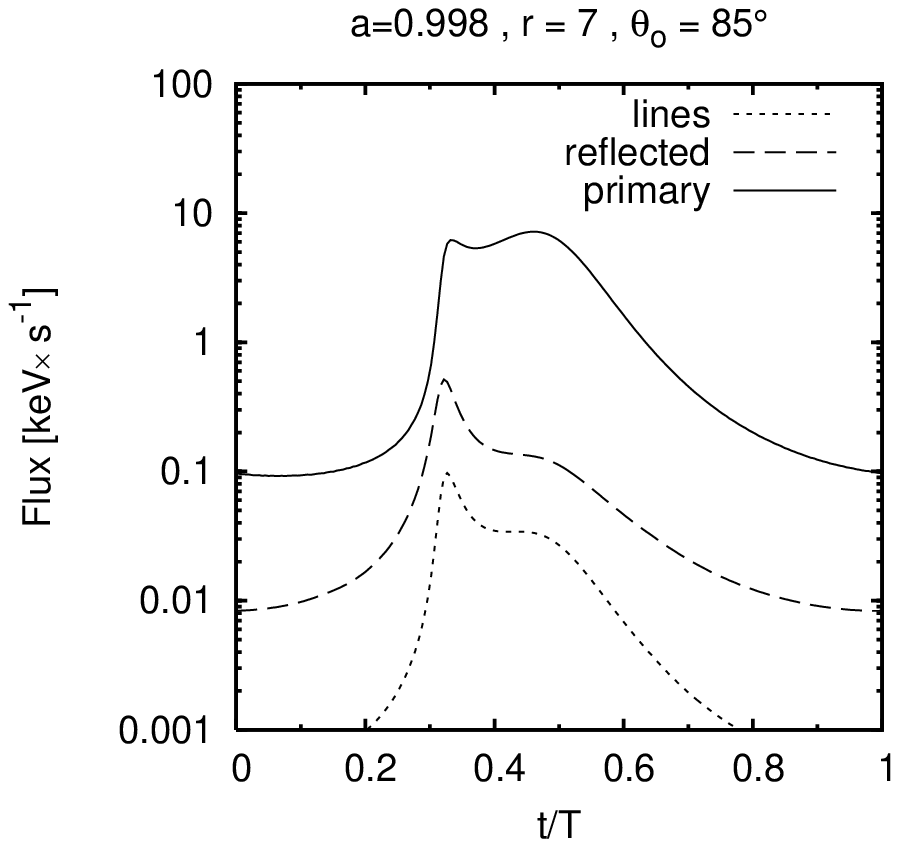}
  \caption{The light curves of the observed emission from the flare and the
  spot for the energy range 3--10~keV for the Kerr black hole with the spin
  $0.998\,GM/c^3$, the spot orbital radii 3, 4, 5 and 7 $GM/c^2$
  (from top to bottom) and the observer's inclination angles
  30$^\circ$, 60$^\circ$ and 85$^\circ$ (from left to right).
  The primary emission, spot's
  continuum   emission and spot's emission in K$\alpha$ and K$\beta$ lines are
  denoted by solid, dashed and dotted graphs, respectively.}
  \label{fig-light_curves1}
\end{center}
\end{figure*}

\begin{figure*}
\begin{center}
  \includegraphics[width=4cm]{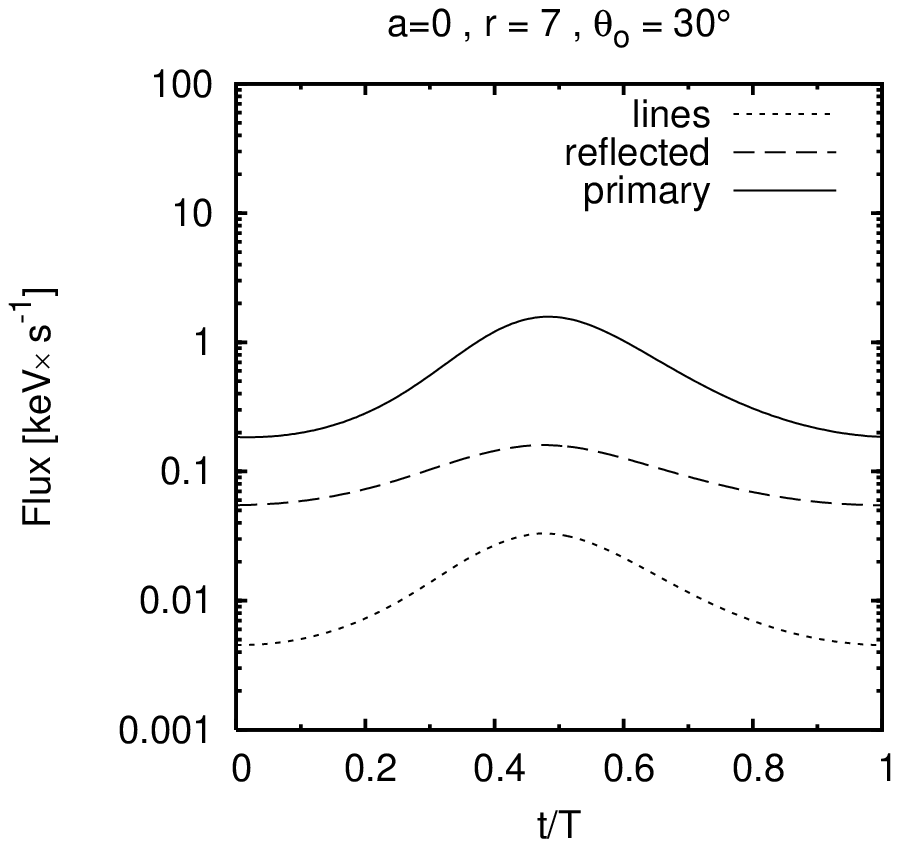}
  \includegraphics[width=4cm]{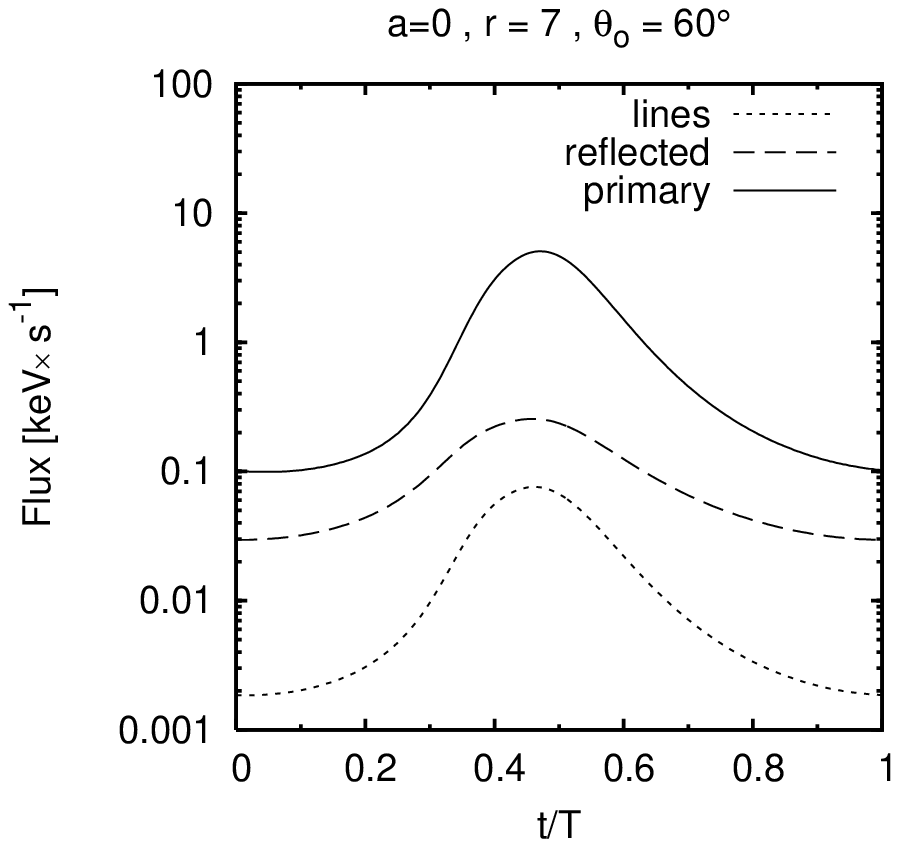}
  \includegraphics[width=4cm]{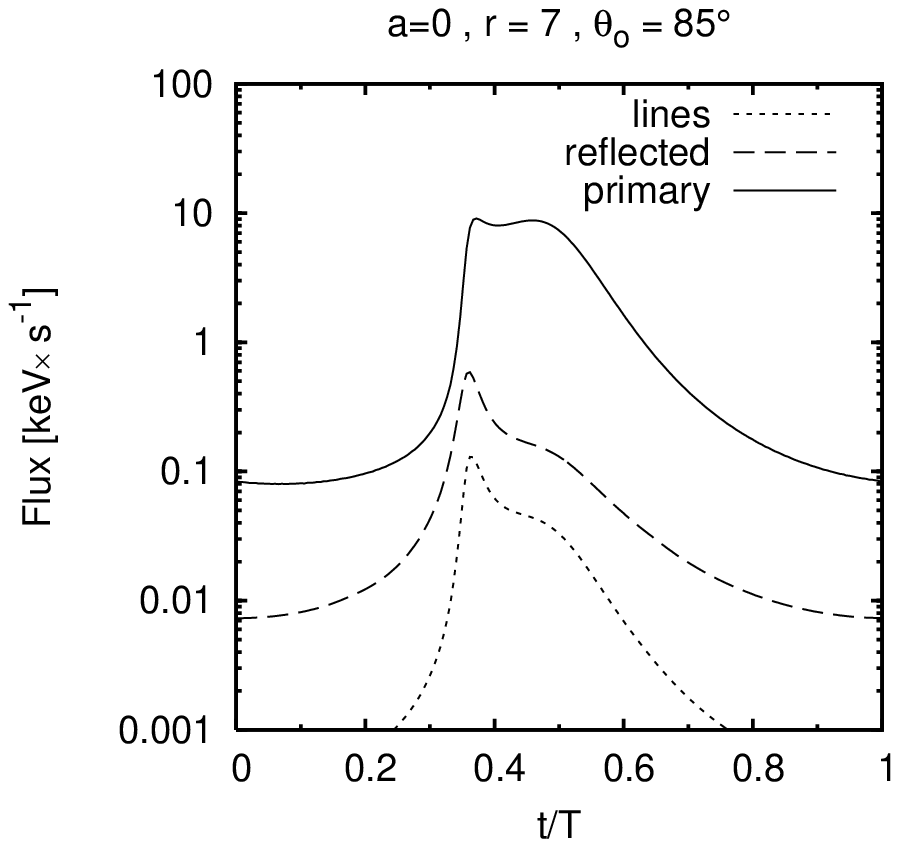}\\[4mm]
  \includegraphics[width=4cm]{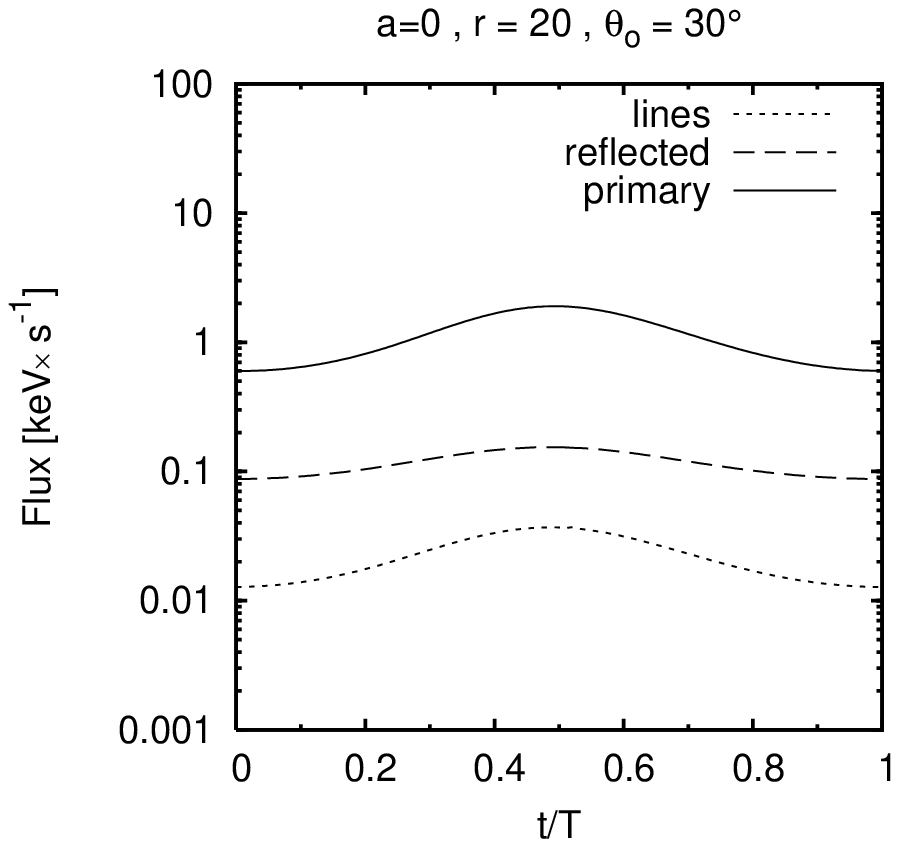}
  \includegraphics[width=4cm]{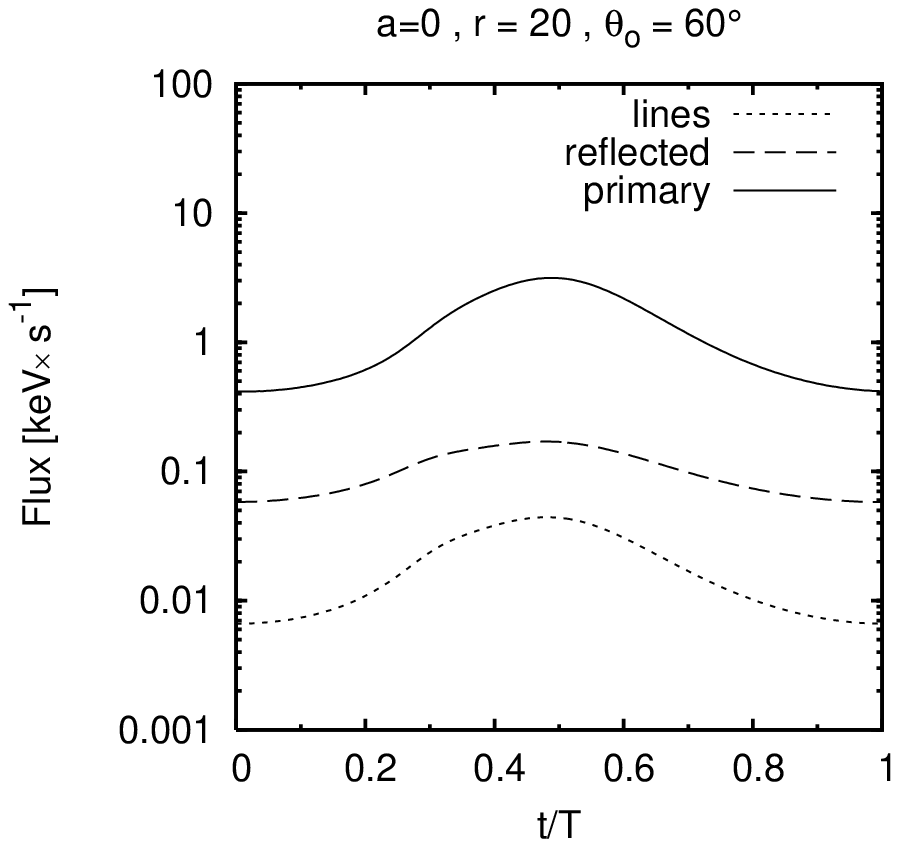}
  \includegraphics[width=4cm]{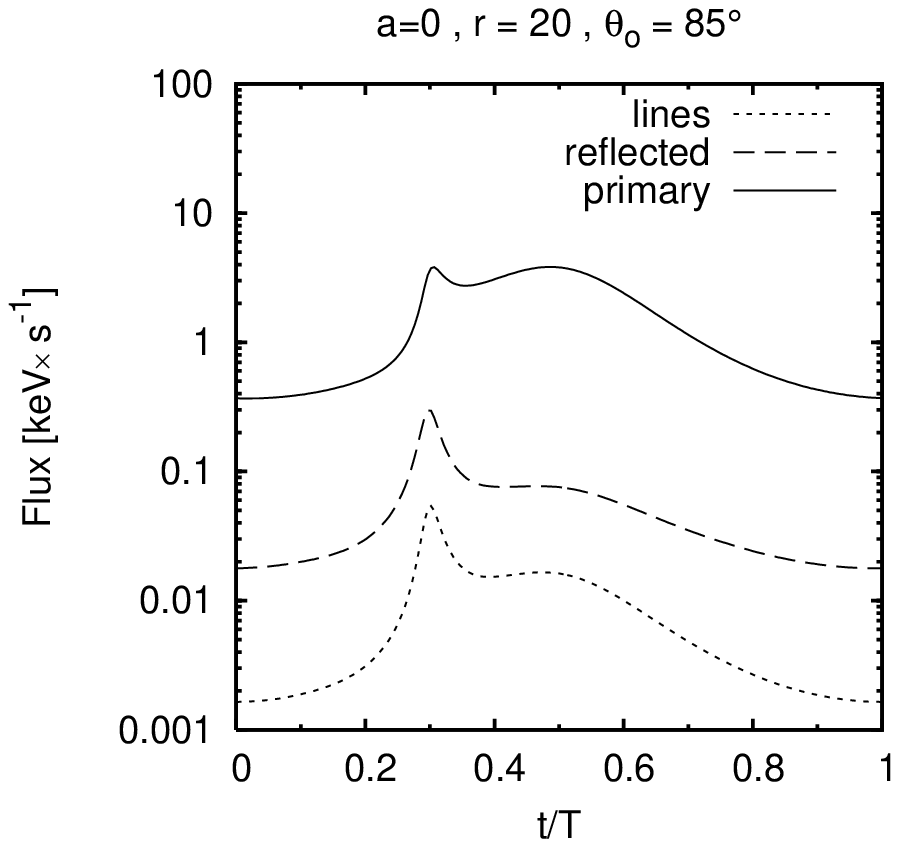}\\[4mm]
  \includegraphics[width=4cm]{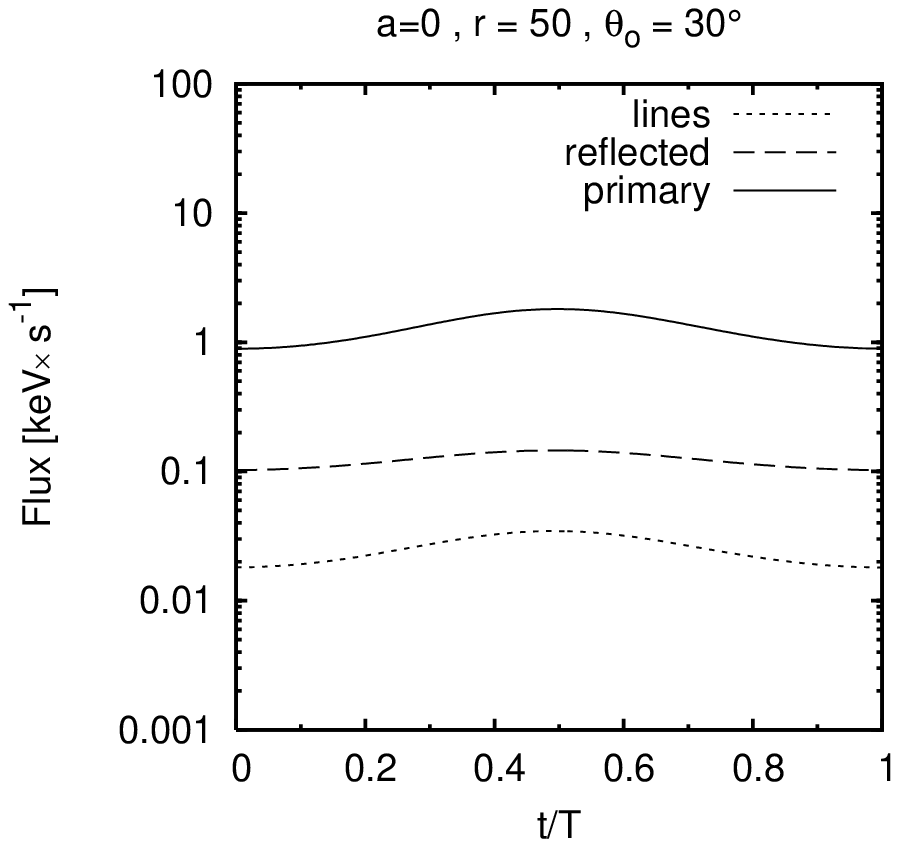}
  \includegraphics[width=4cm]{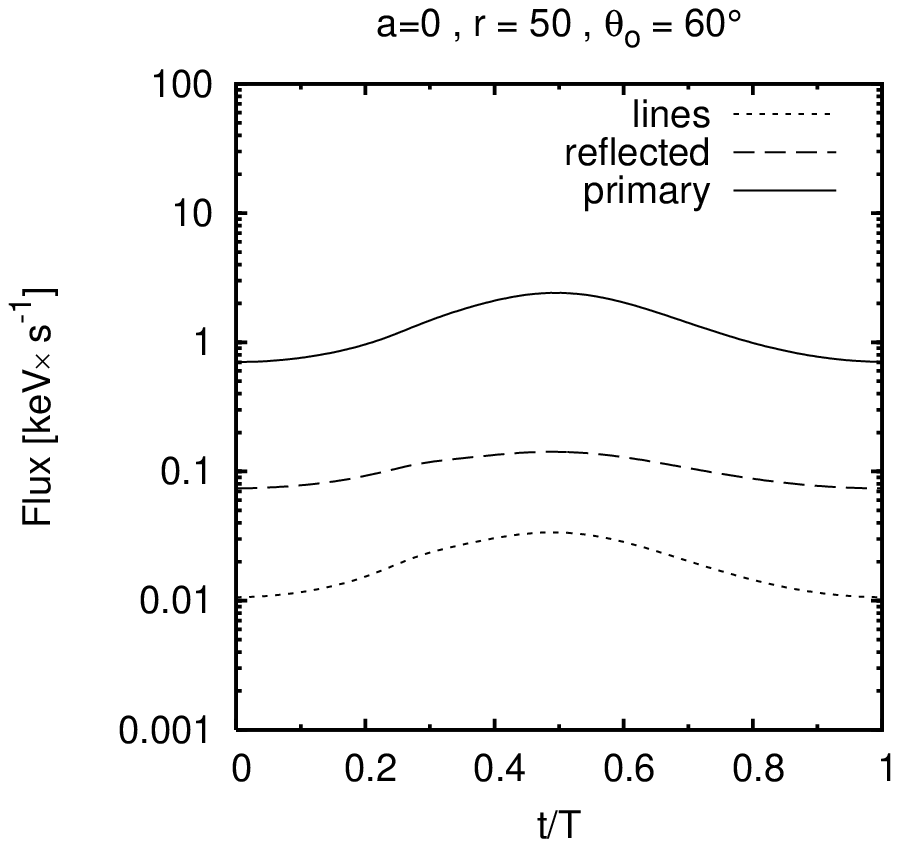}
  \includegraphics[width=4cm]{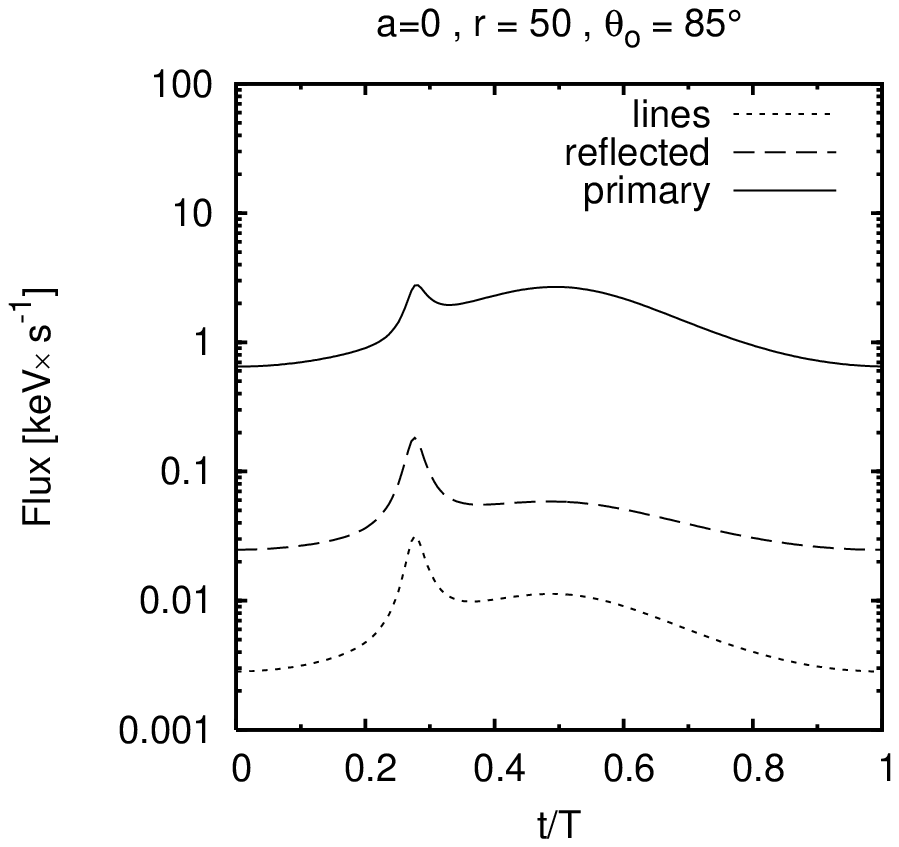}\\[4mm]
  \includegraphics[width=4cm]{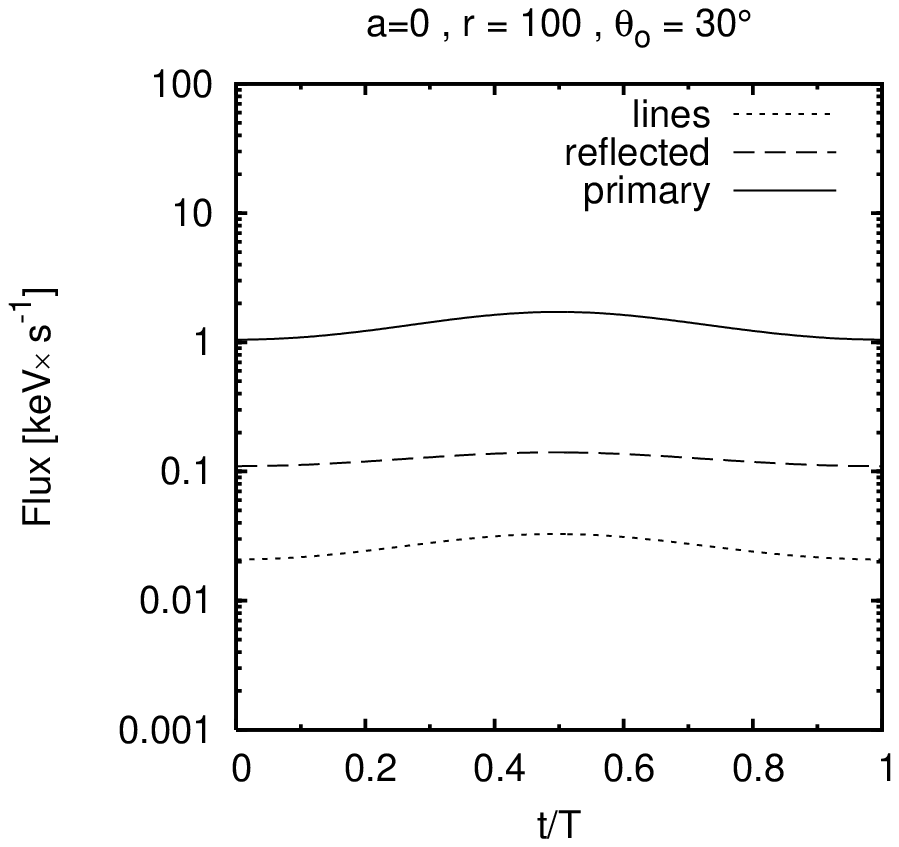}
  \includegraphics[width=4cm]{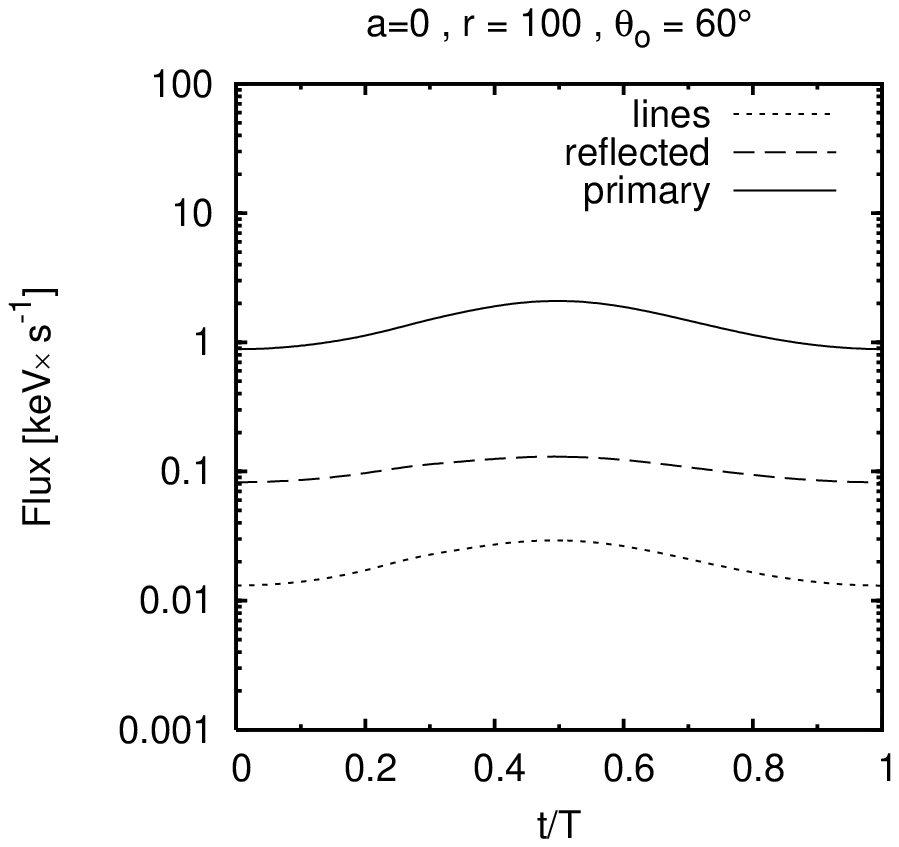}
  \includegraphics[width=4cm]{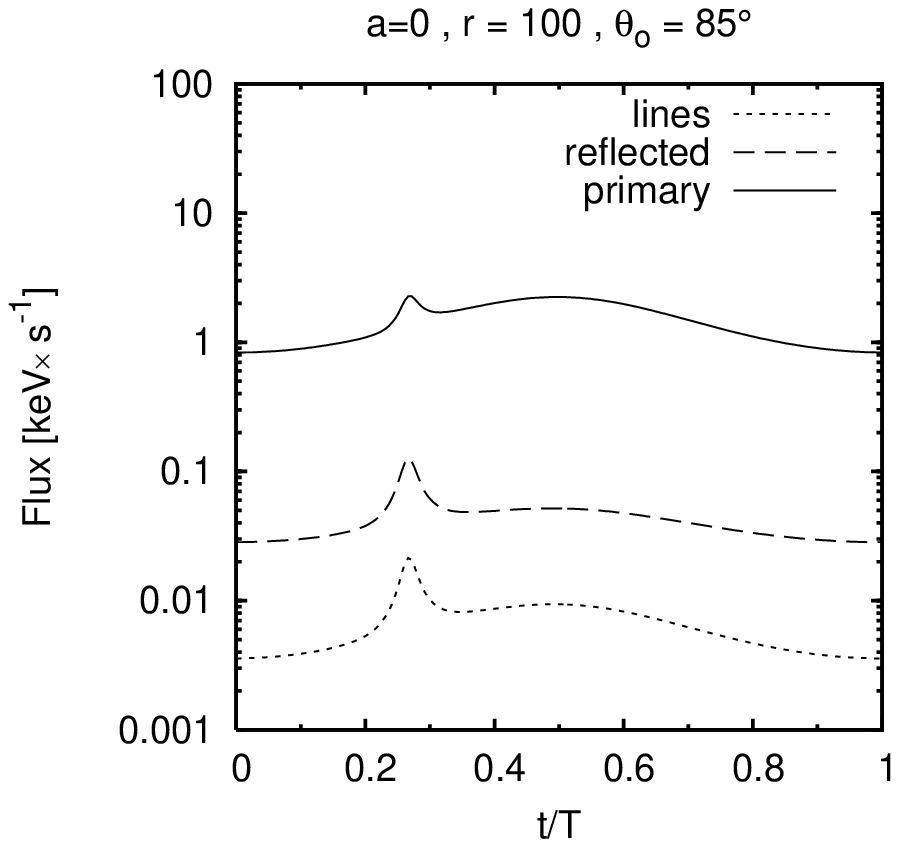}
  \caption{The same as in Fig.~\ref{fig-light_curves1} but for the
  Schwarzschild   black hole and the spot orbital radii 7, 20, 50 and 100
  $GM/c^2$ (from top to bottom).}
  \label{fig-light_curves2}
\end{center}
\end{figure*}

\begin{figure*}
\begin{center}
  \includegraphics[width=4cm]{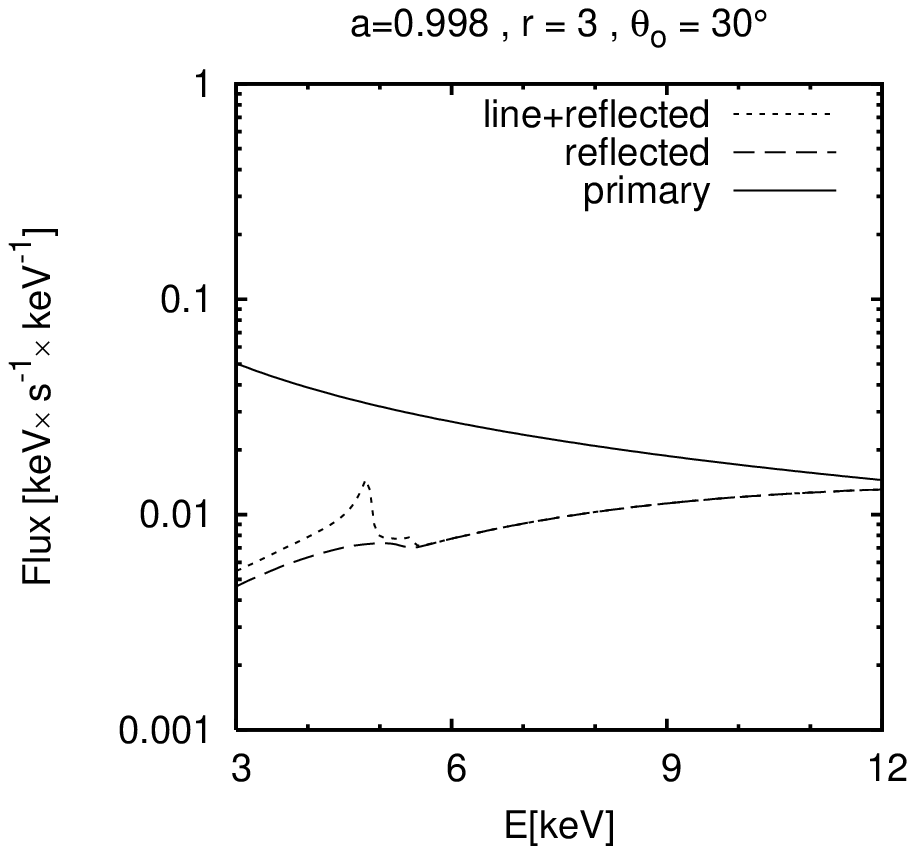}
  \includegraphics[width=4cm]{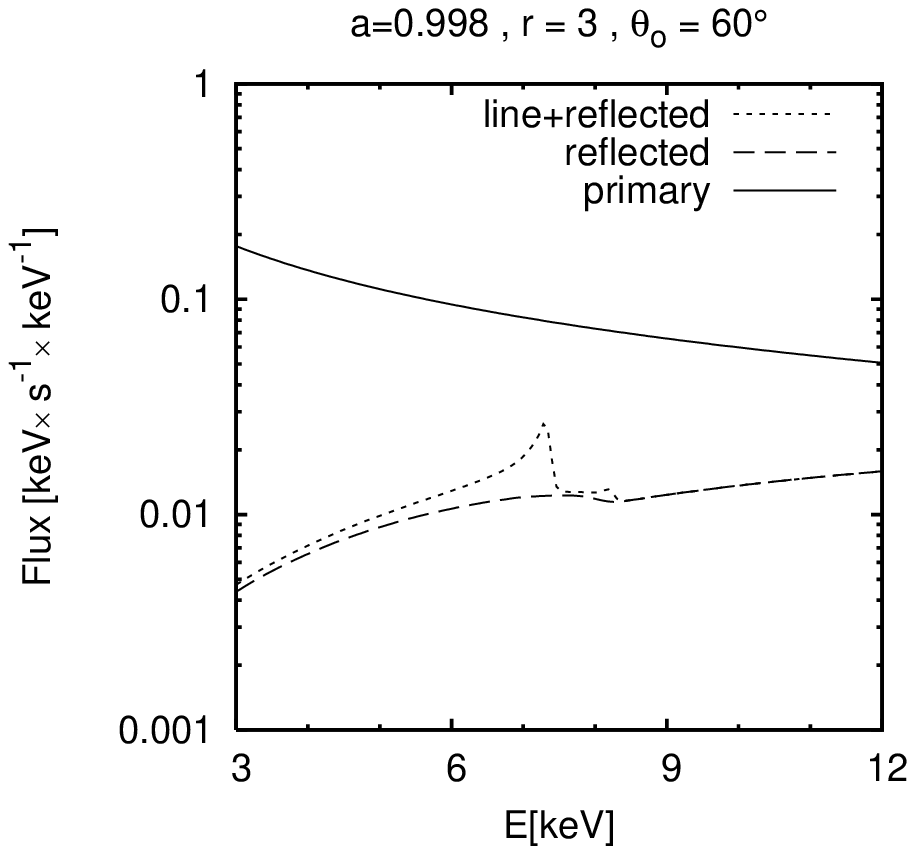}
  \includegraphics[width=4cm]{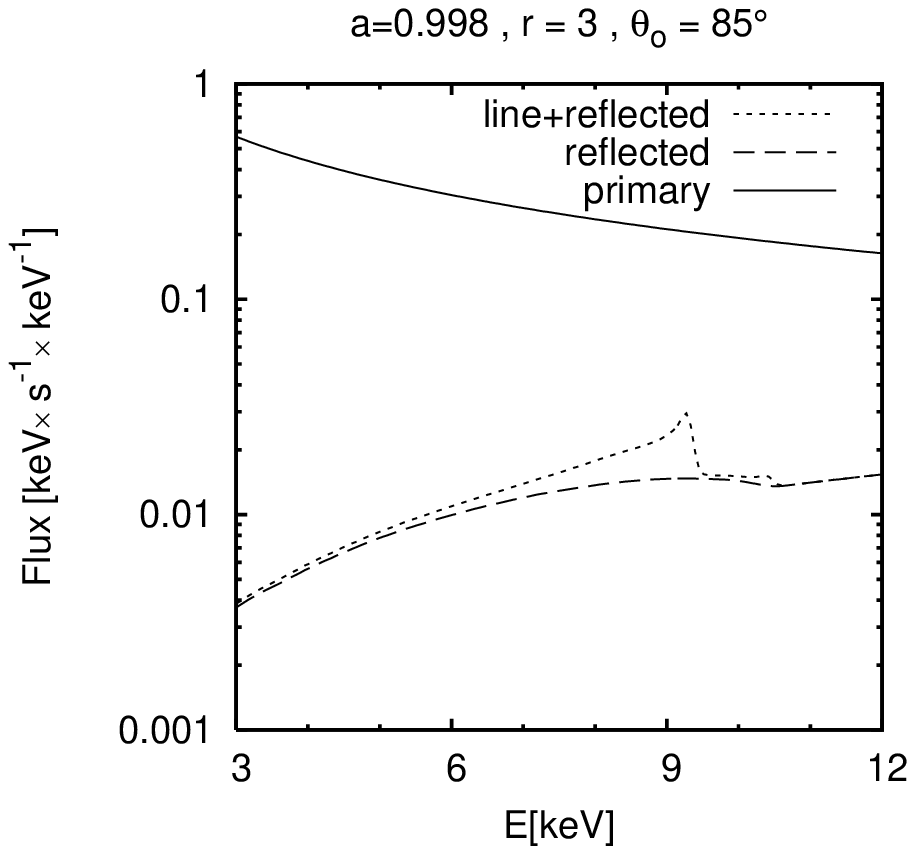}\\[4mm]
  \includegraphics[width=4cm]{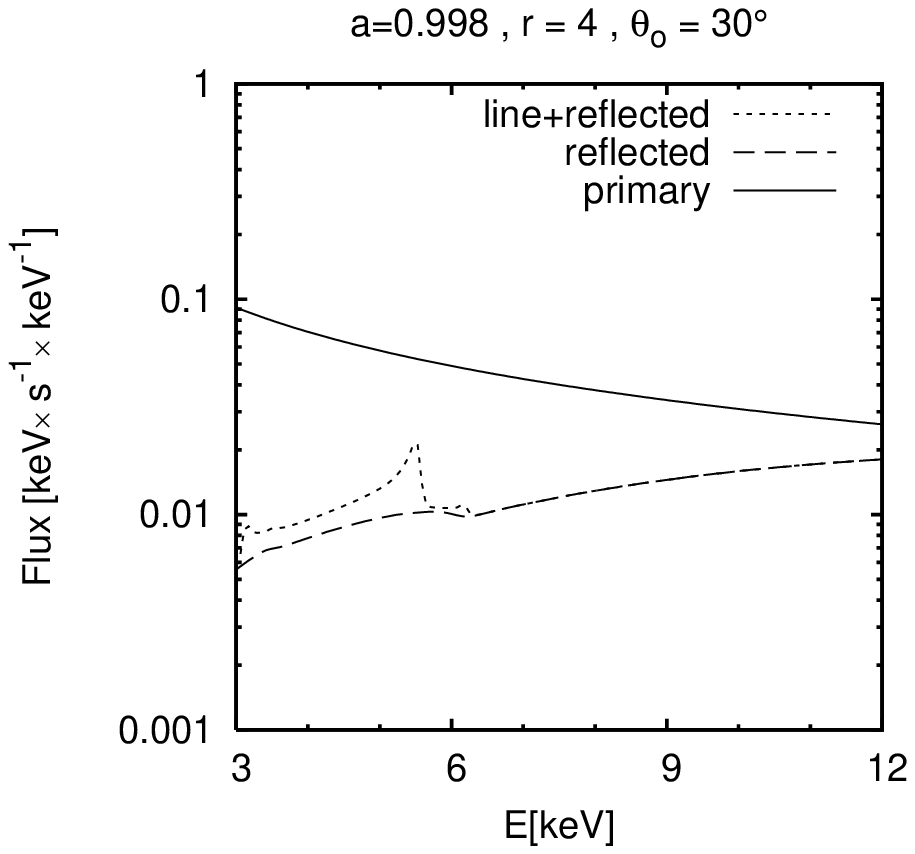}
  \includegraphics[width=4cm]{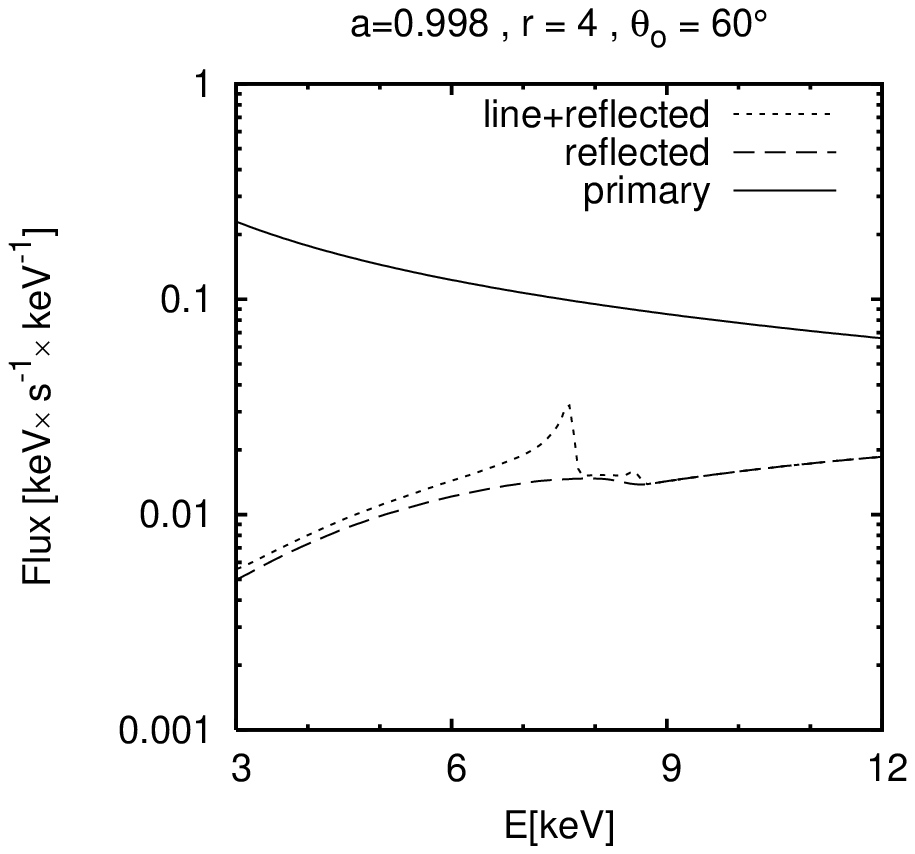}
  \includegraphics[width=4cm]{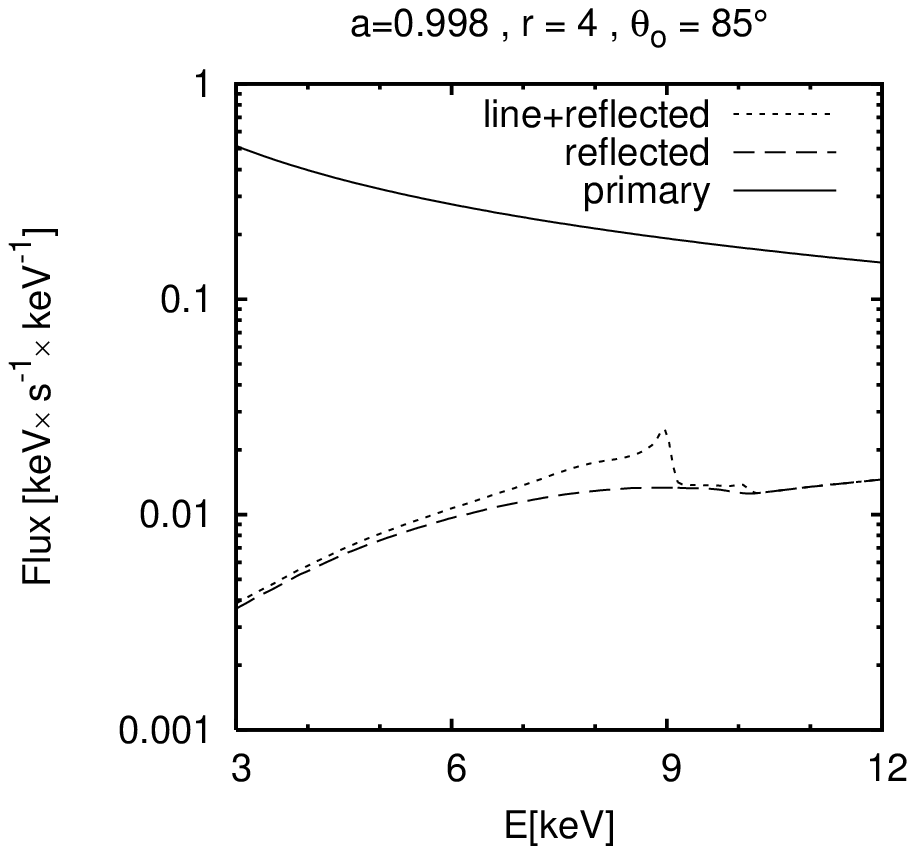}\\[4mm]
  \includegraphics[width=4cm]{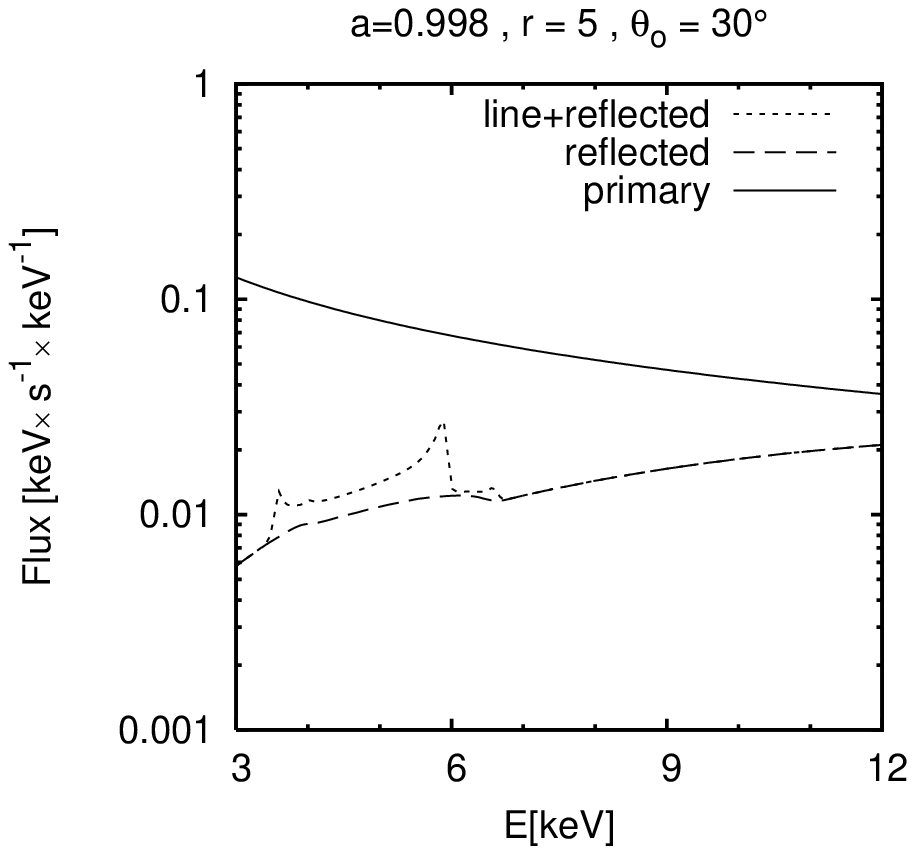}
  \includegraphics[width=4cm]{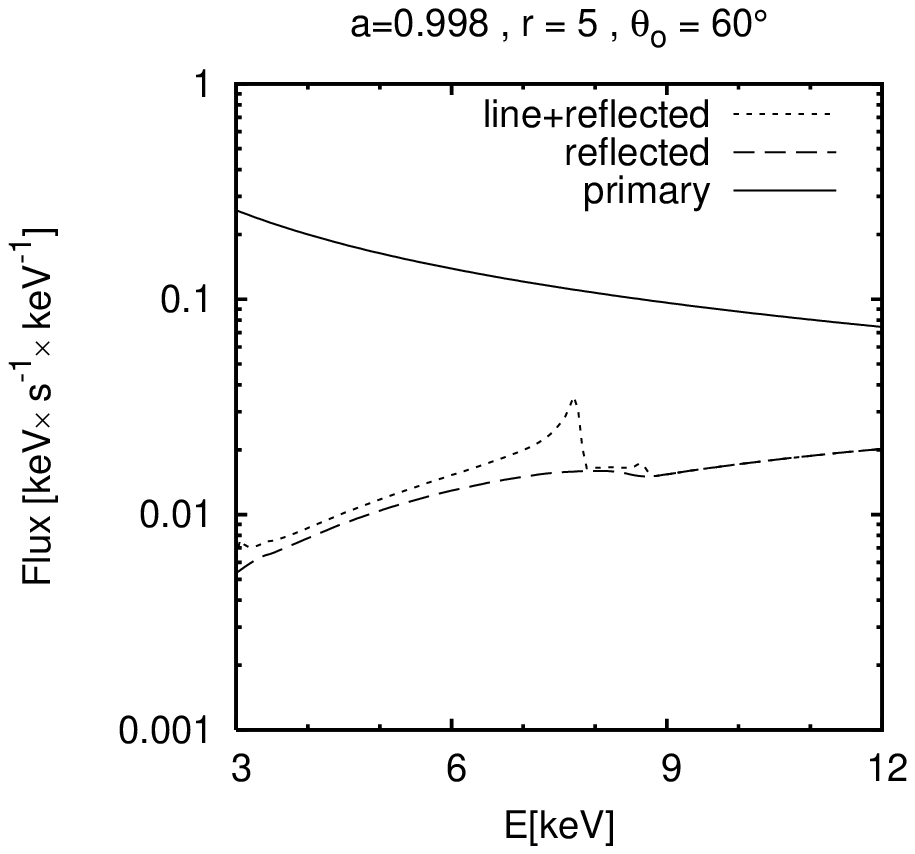}
  \includegraphics[width=4cm]{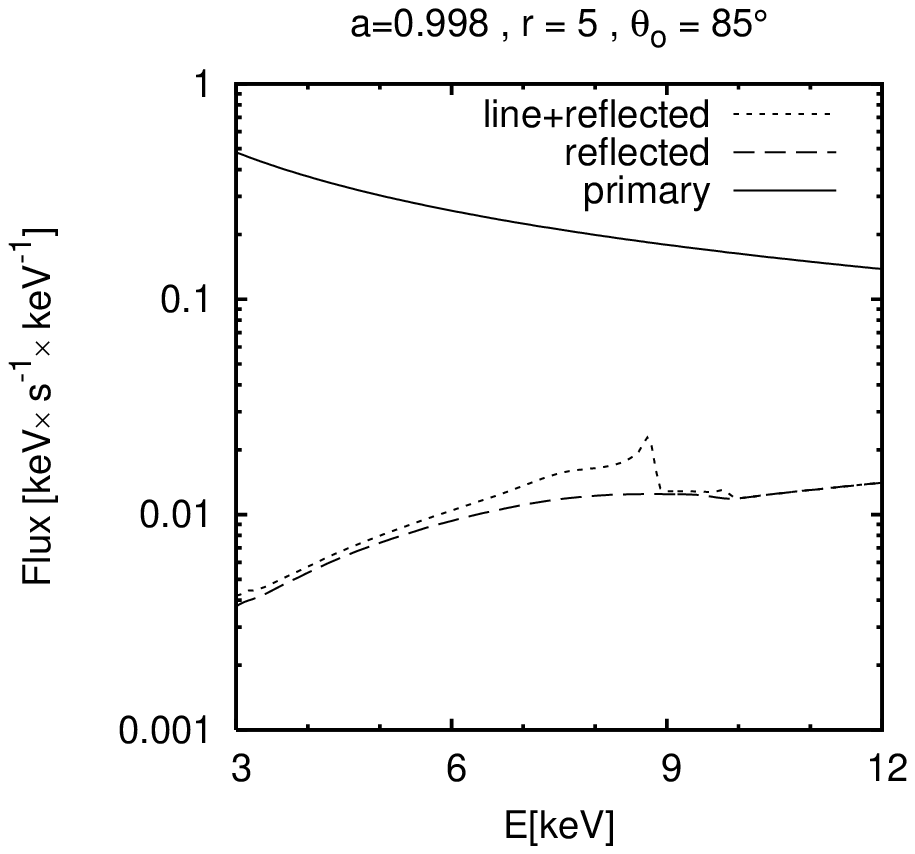}\\[4mm]
  \includegraphics[width=4cm]{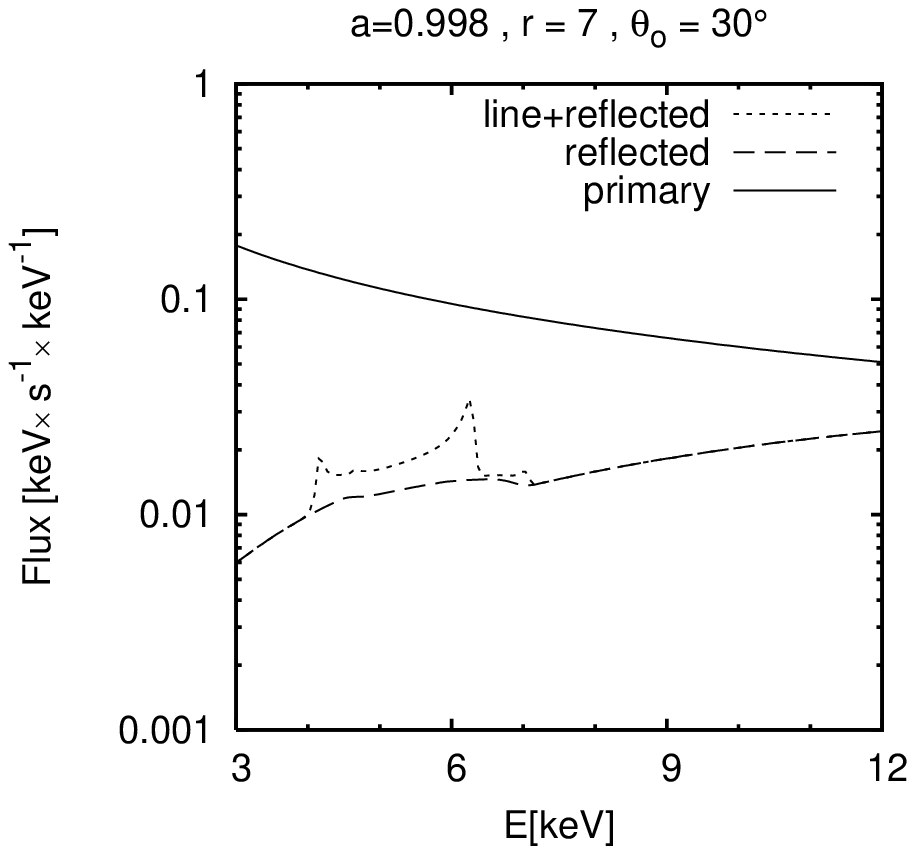}
  \includegraphics[width=4cm]{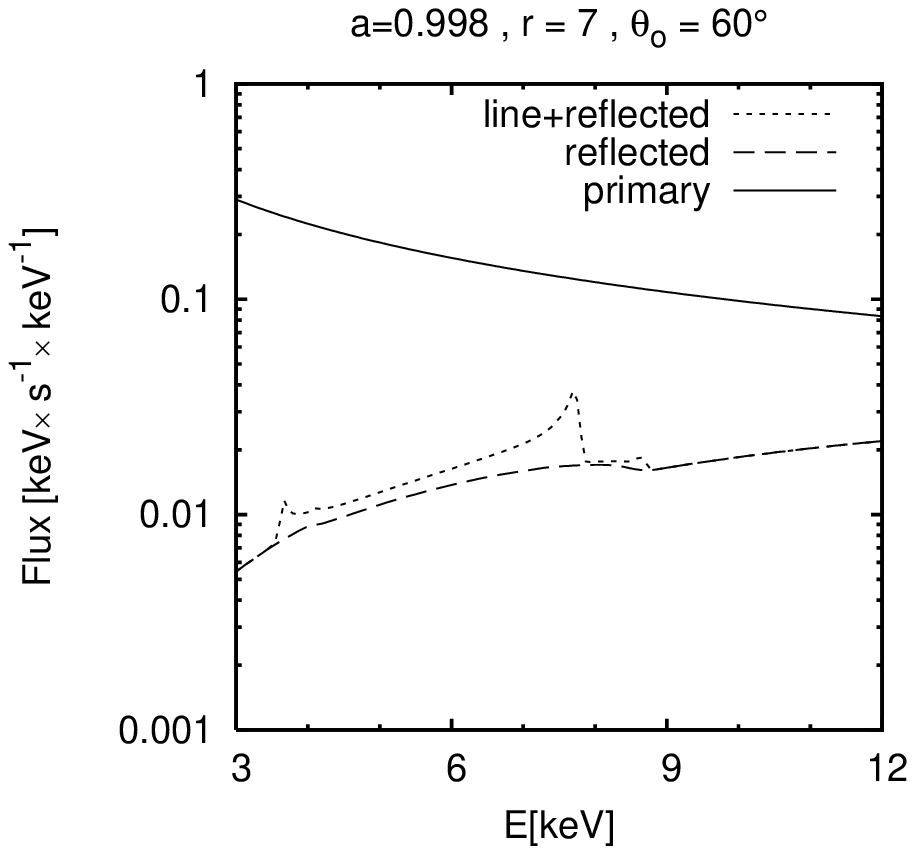}
  \includegraphics[width=4cm]{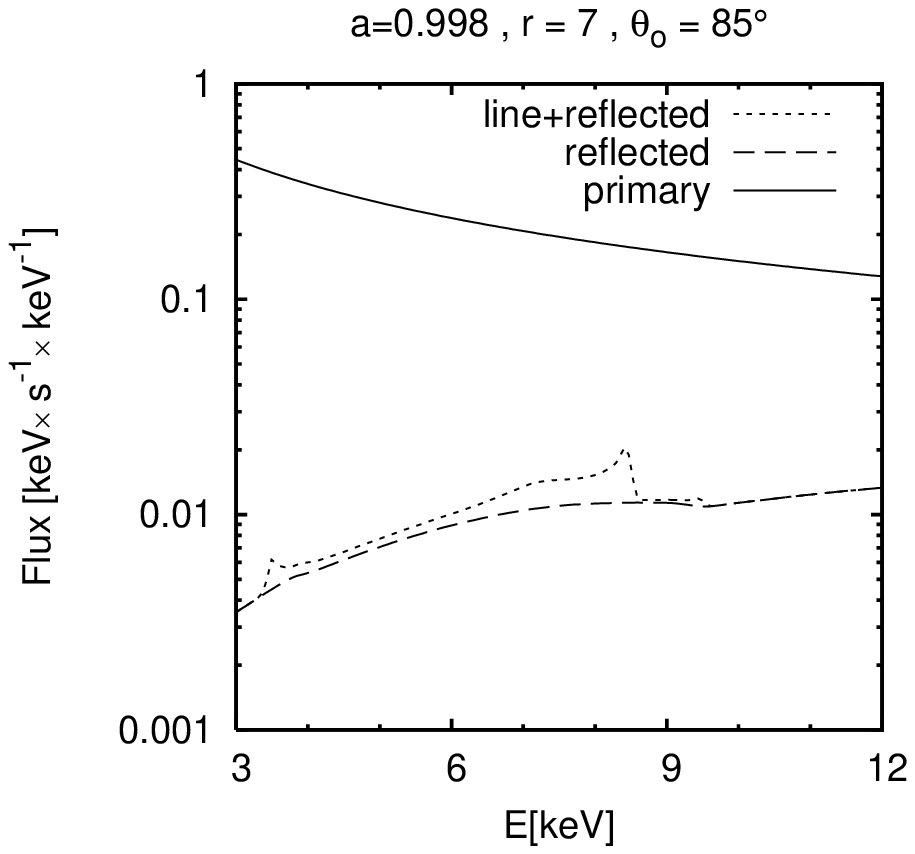}
  \caption{The observed spectra averaged over one orbit computed for the same
  set of parameters as in Fig.~\ref{fig-light_curves1}. Here, the observed line
  flux is shown on top of the spot continuum emission.}
  \label{fig-spectrum1}
\end{center}
\end{figure*}

\begin{figure*}
\begin{center}
  \includegraphics[width=4cm]{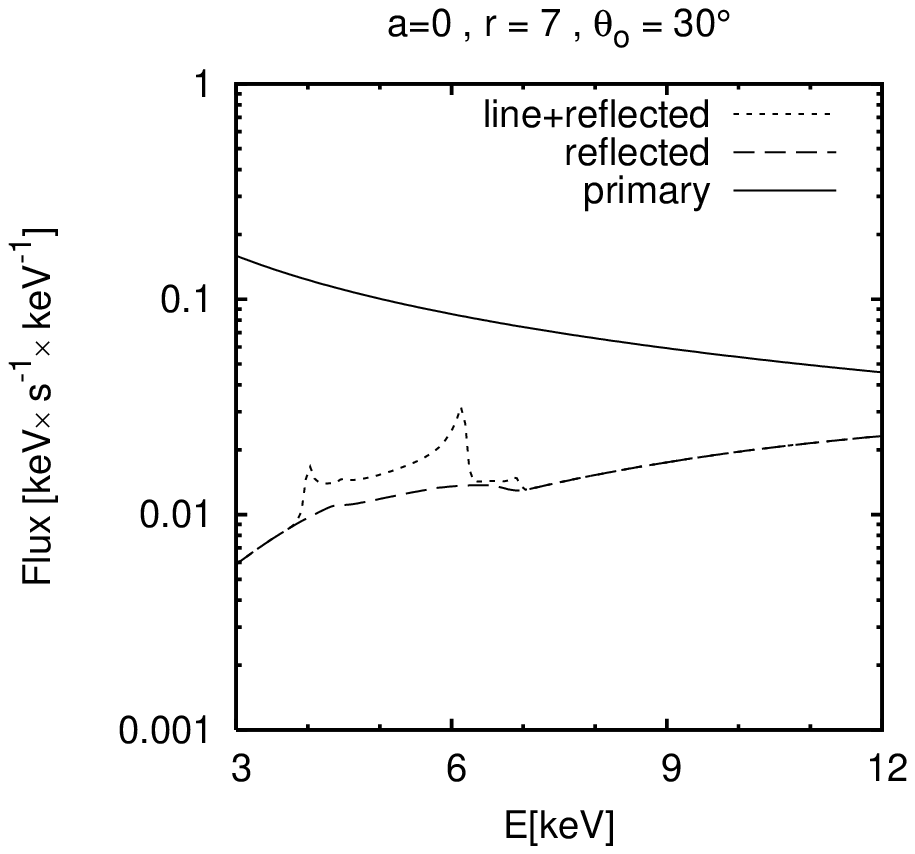}
  \includegraphics[width=4cm]{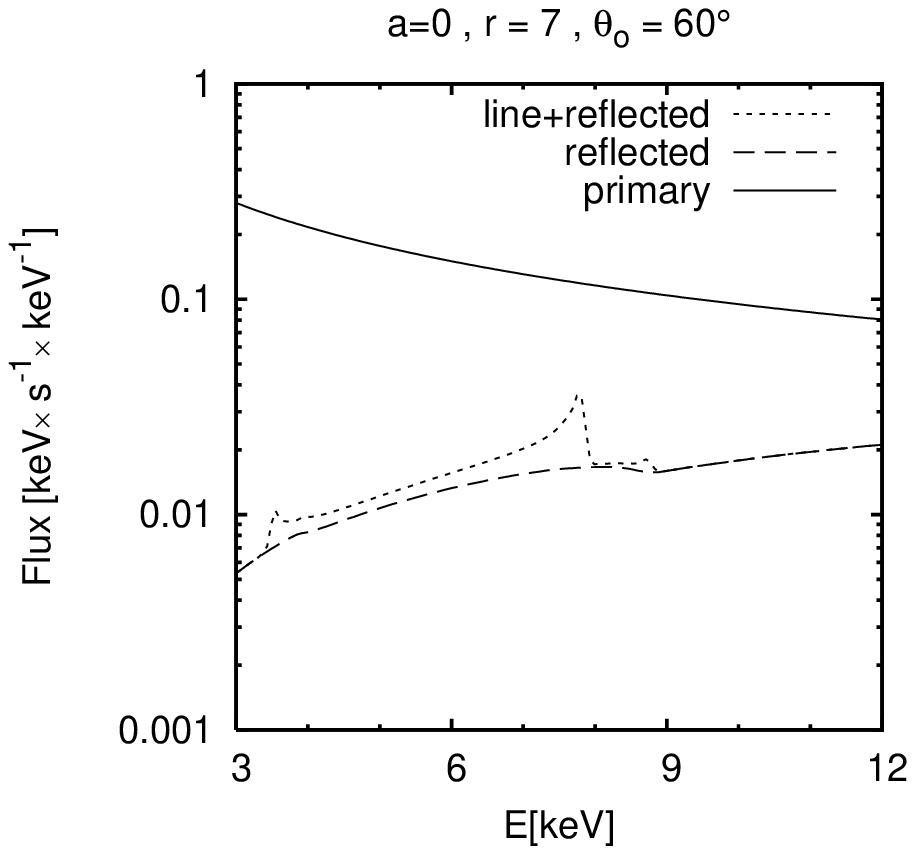}
  \includegraphics[width=4cm]{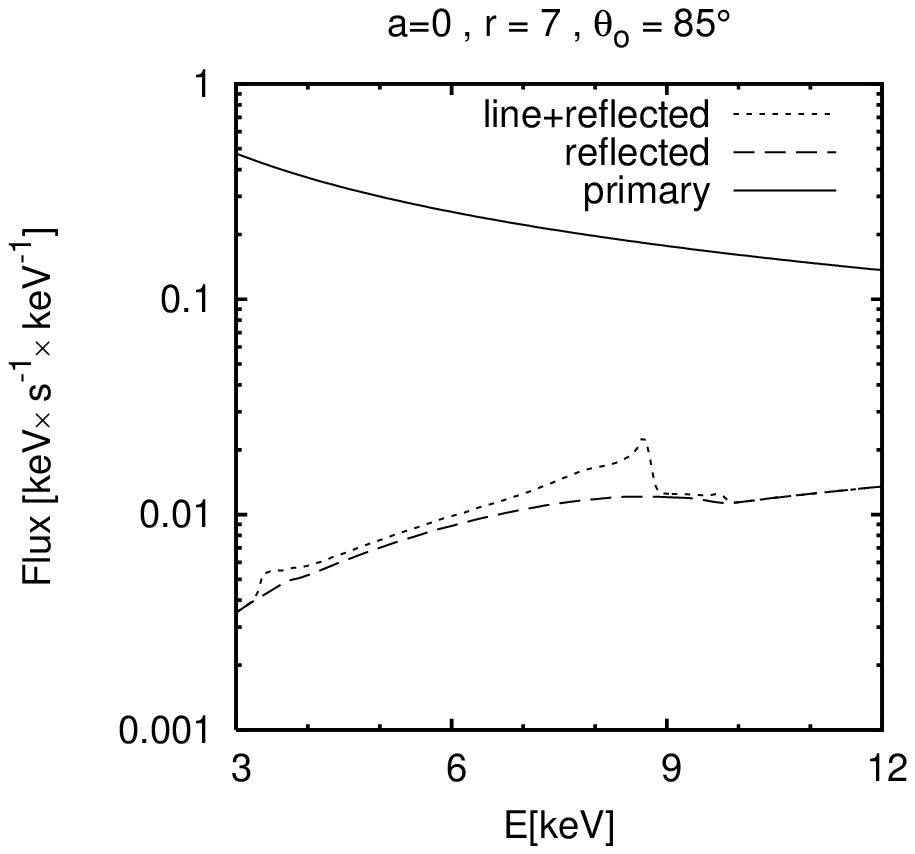}\\[4mm]
  \includegraphics[width=4cm]{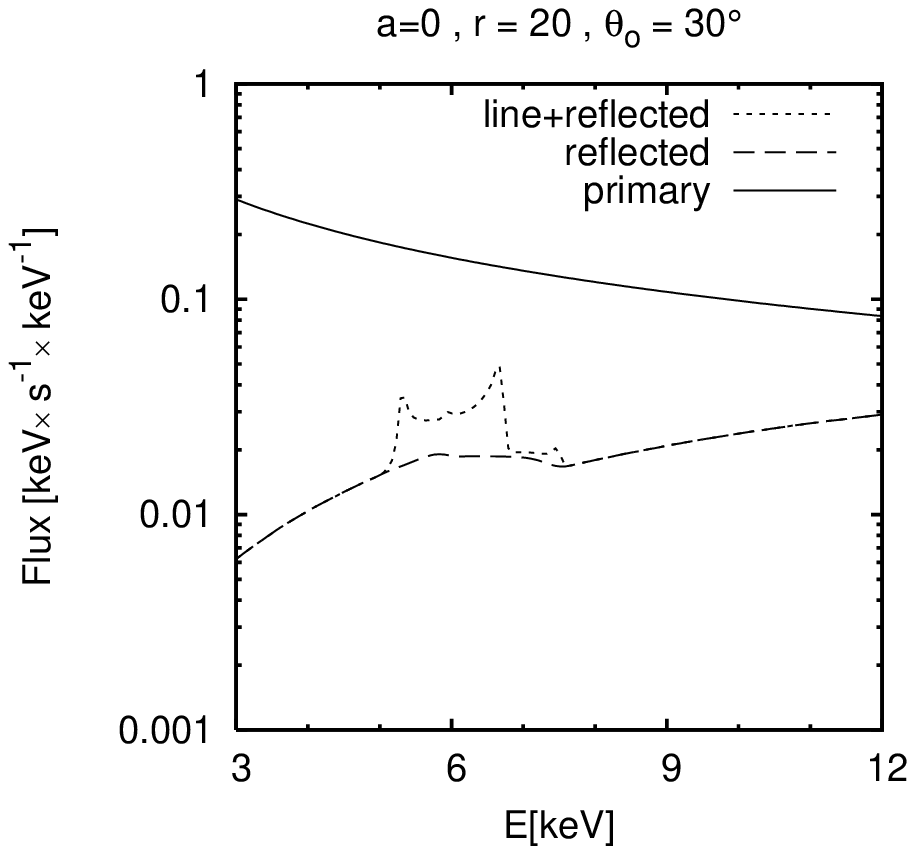}
  \includegraphics[width=4cm]{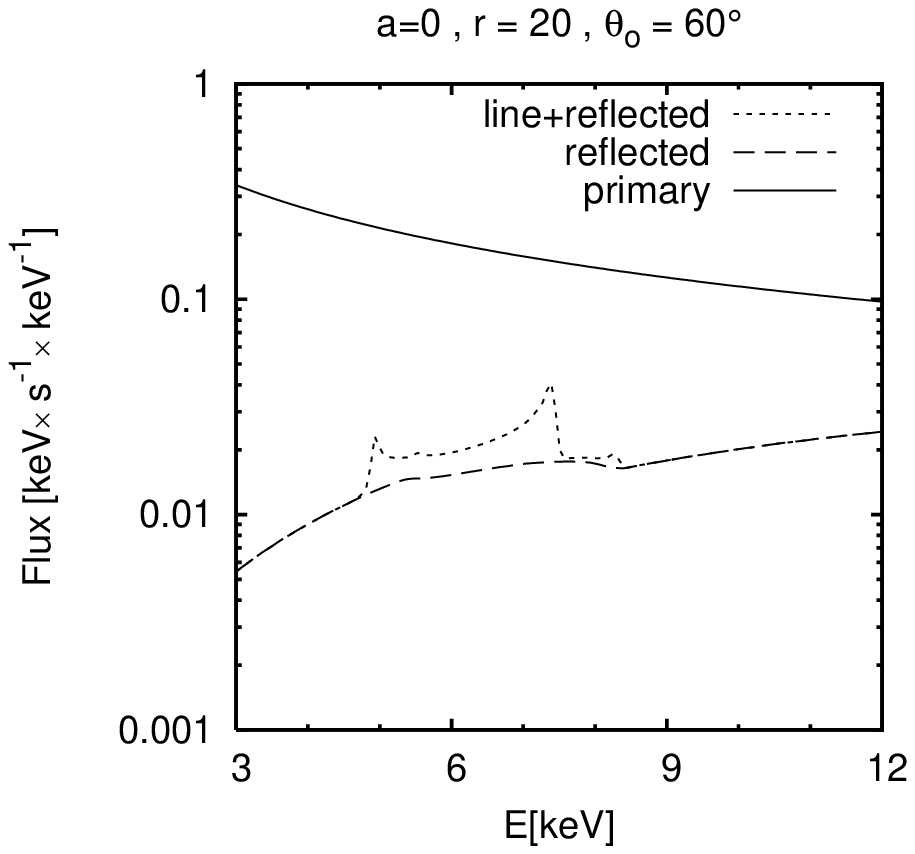}
  \includegraphics[width=4cm]{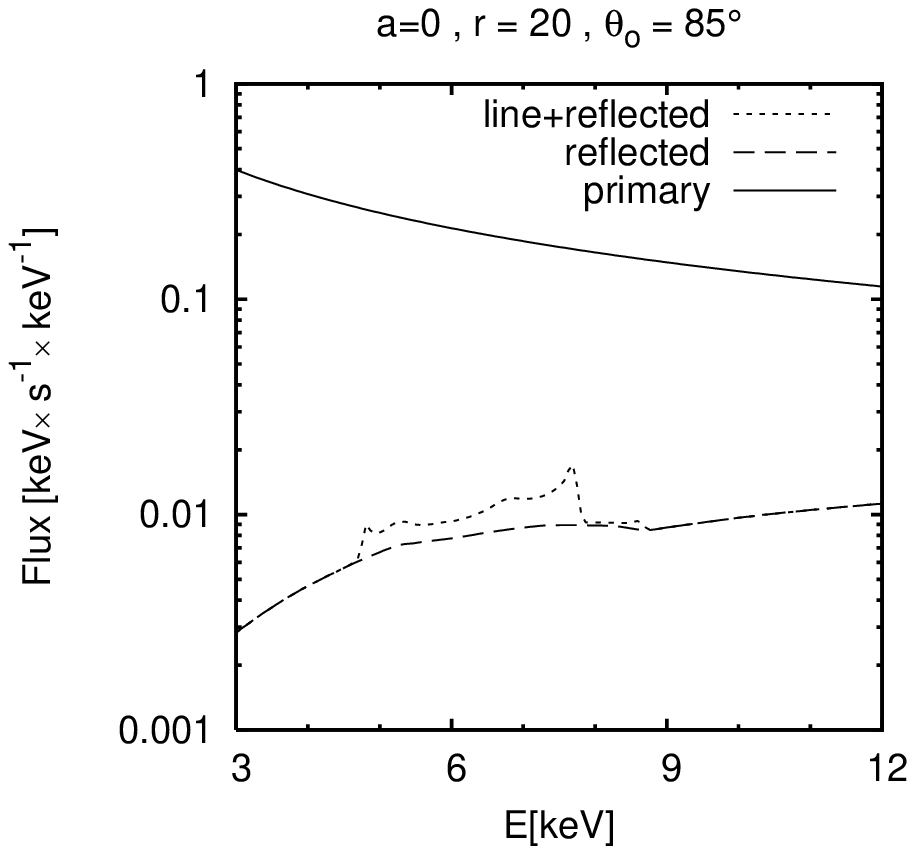}\\[4mm]
  \includegraphics[width=4cm]{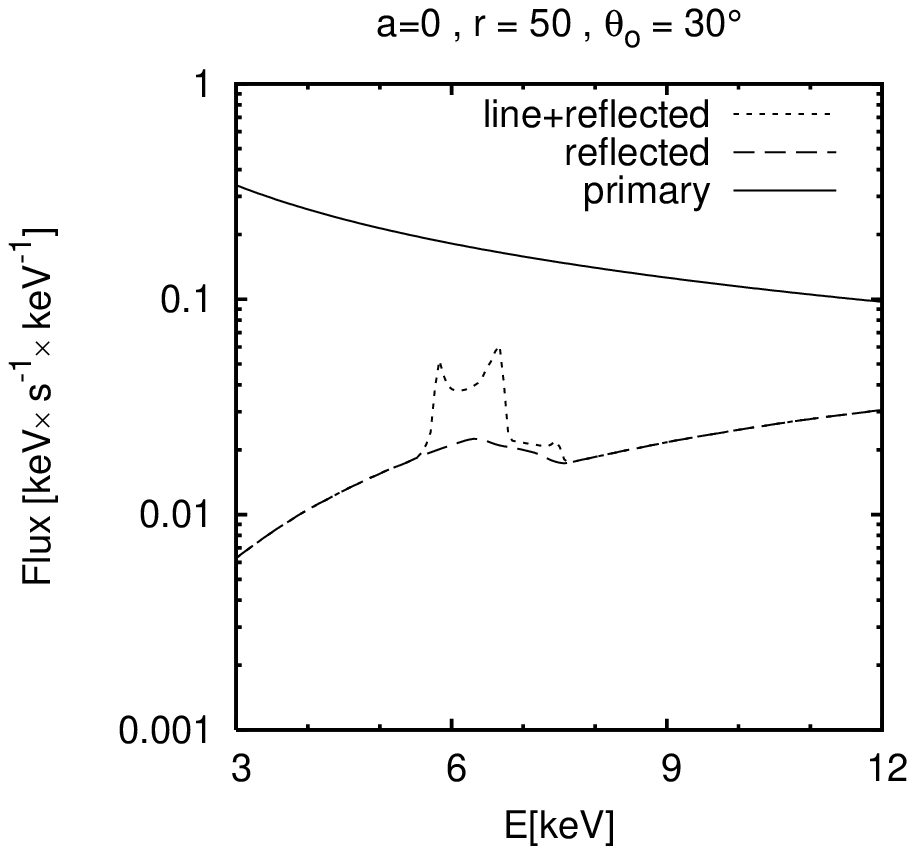}
  \includegraphics[width=4cm]{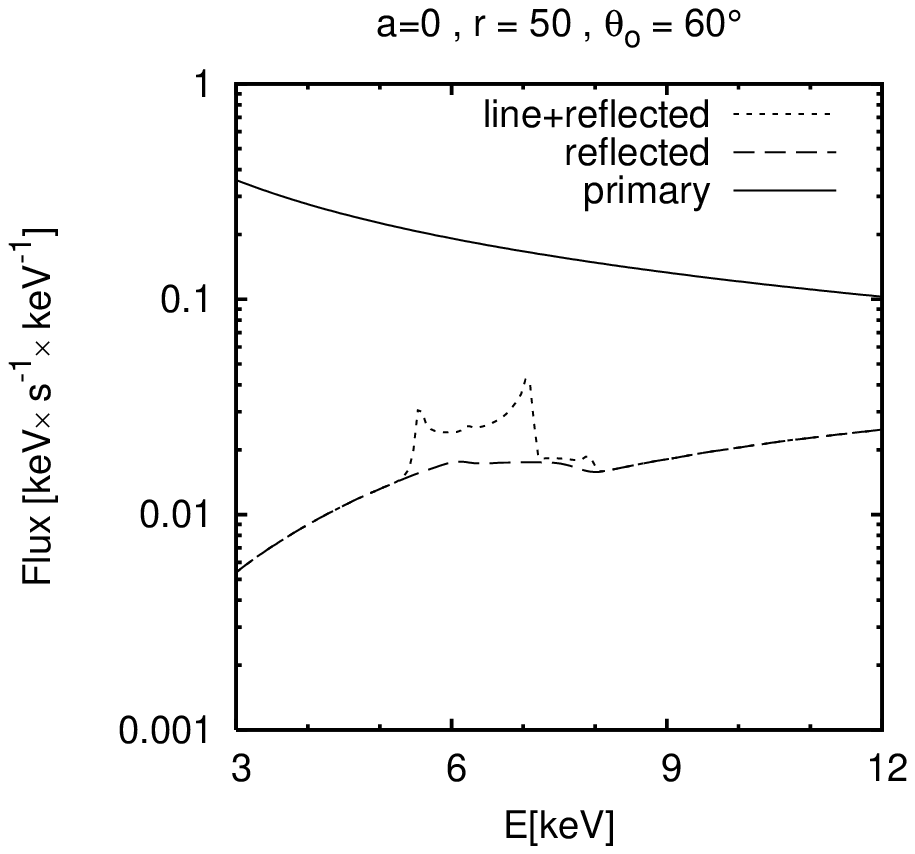}
  \includegraphics[width=4cm]{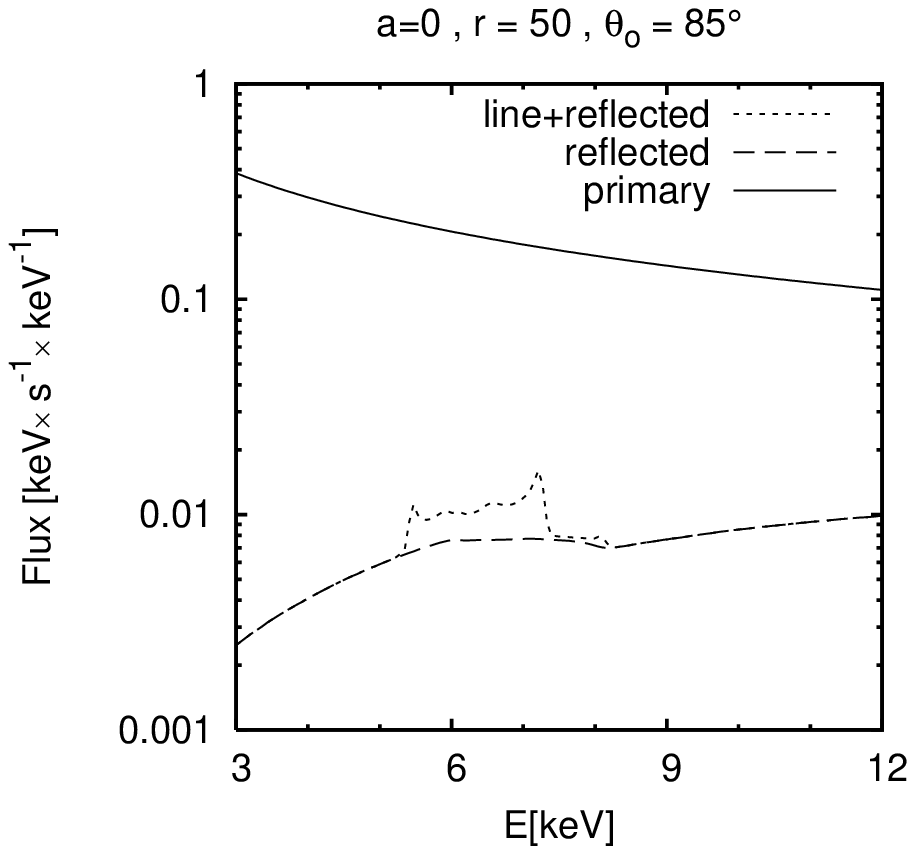}\\[4mm]
  \includegraphics[width=4cm]{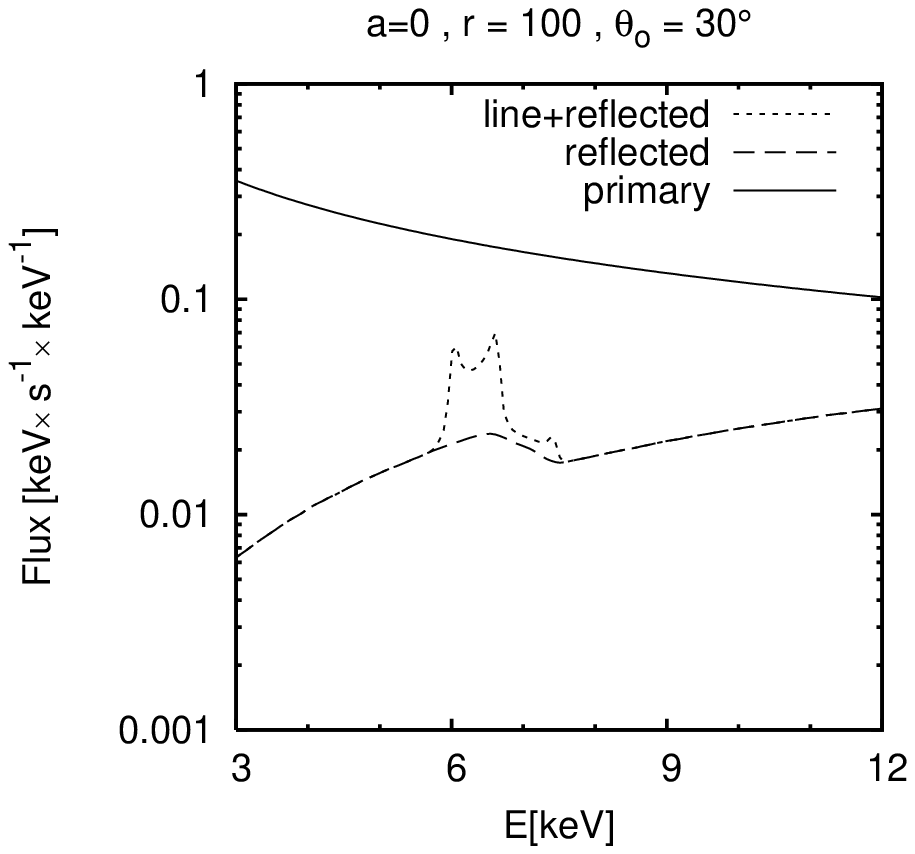}
  \includegraphics[width=4cm]{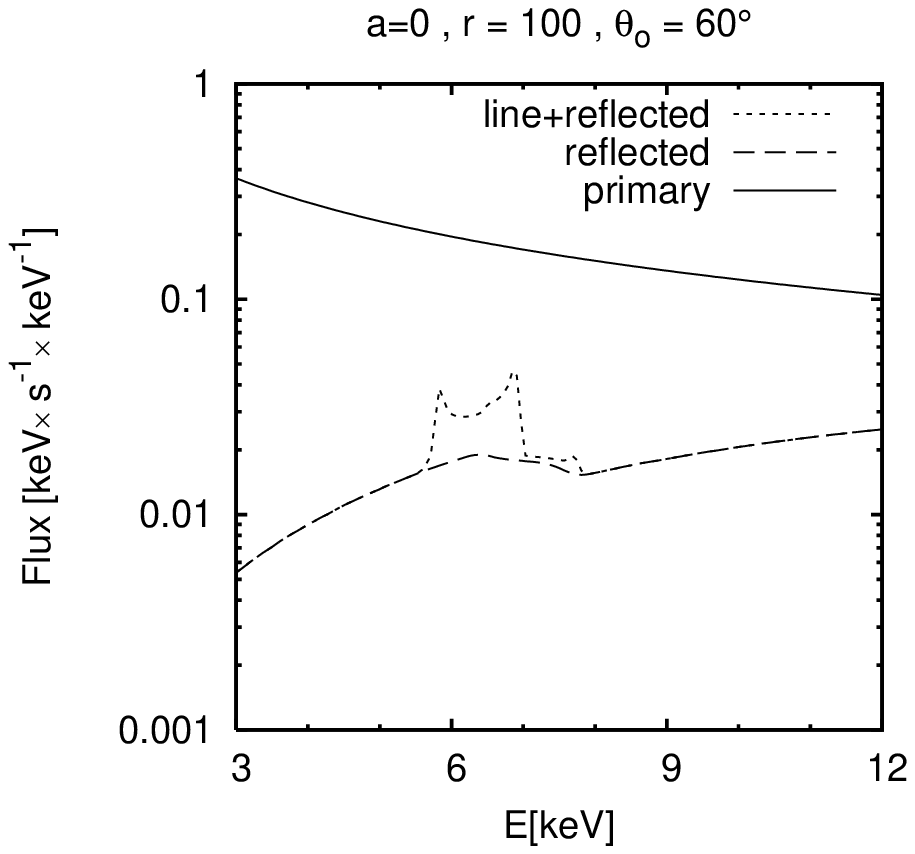}
  \includegraphics[width=4cm]{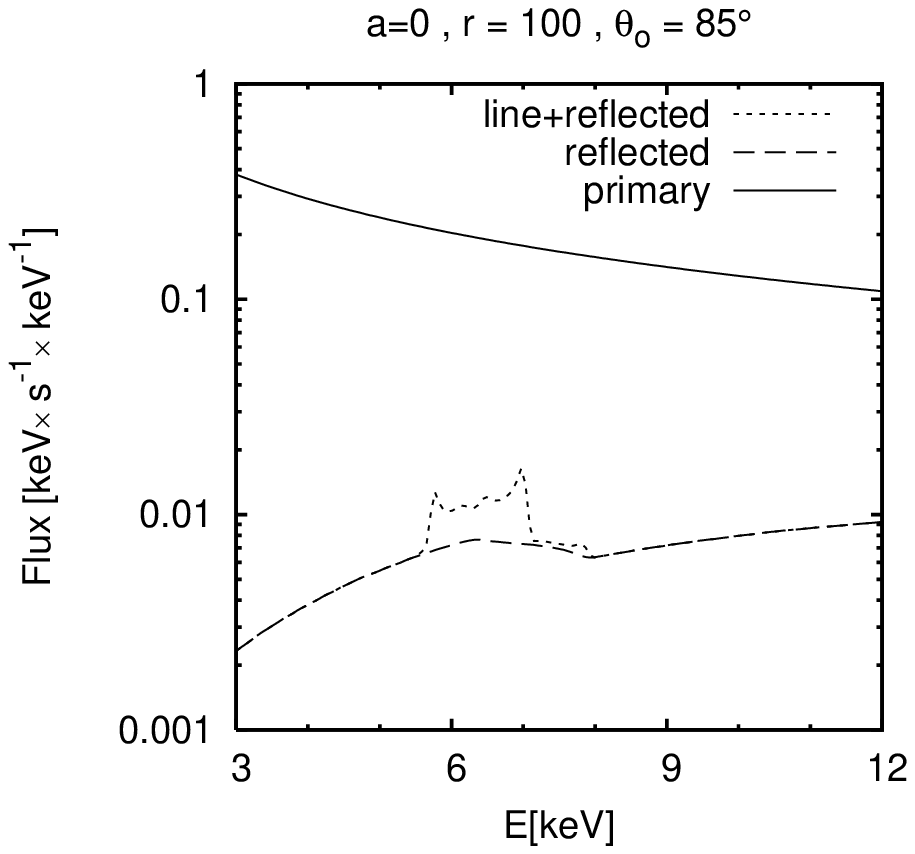}
  \caption{The observed spectra averaged over one orbit computed for the same
  set of parameters as in Fig.~\ref{fig-light_curves2}. Here, the observed line
  flux is shown on top of the spot continuum emission.}
  \label{fig-spectrum2}
\end{center}
\end{figure*}

\begin{figure*}
\begin{center}
  \vspace*{-2mm}
  \includegraphics[width=3.75cm]{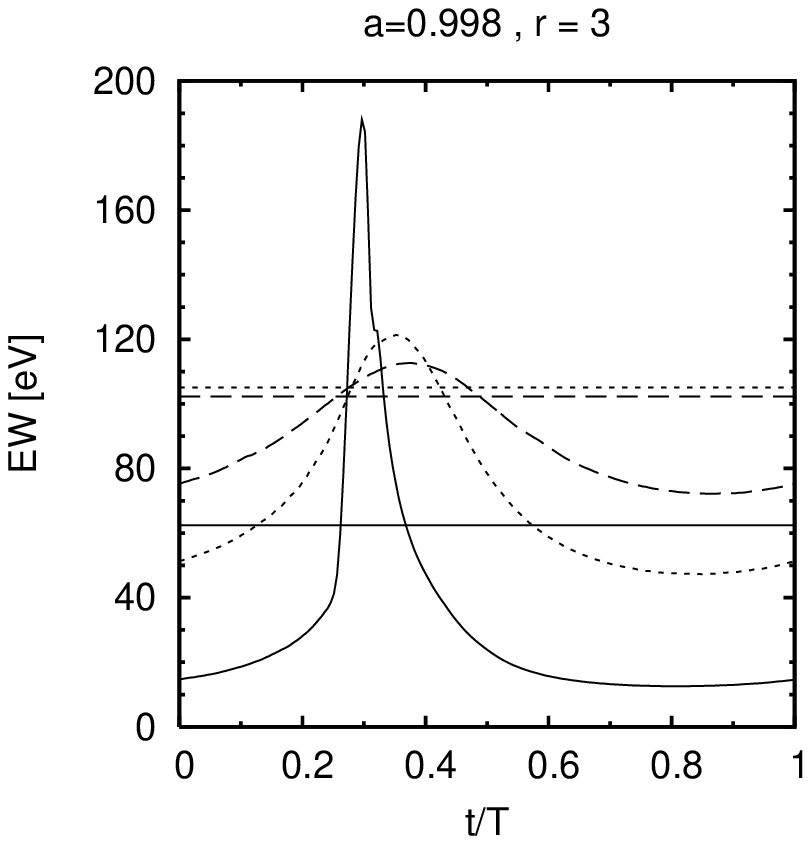}
  \includegraphics[width=3.77cm]{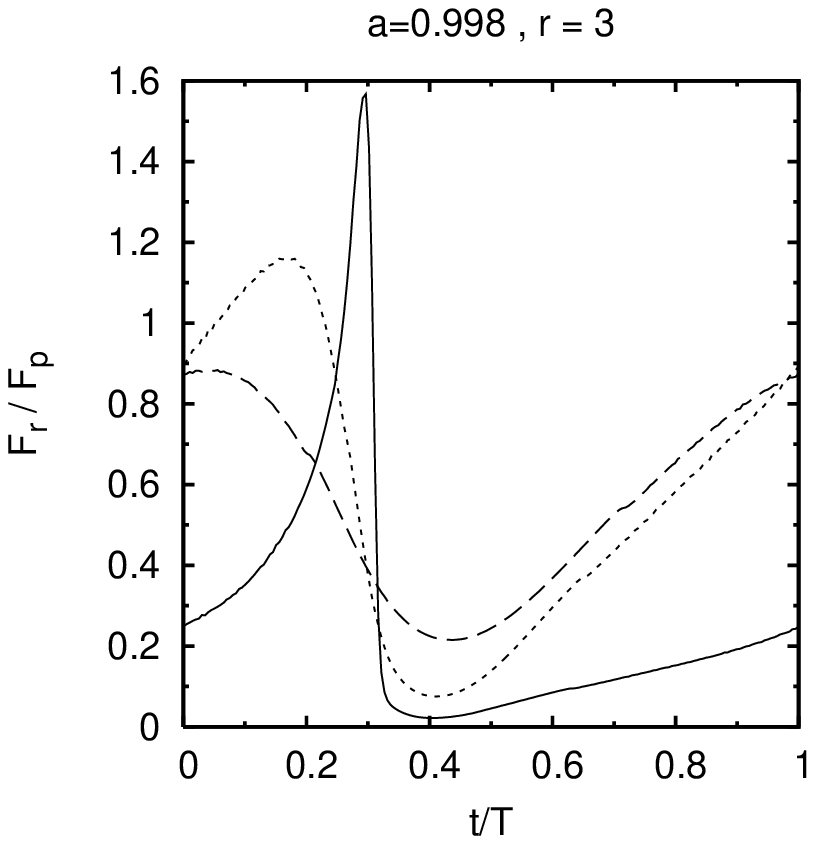}
  \includegraphics[width=3.9cm]{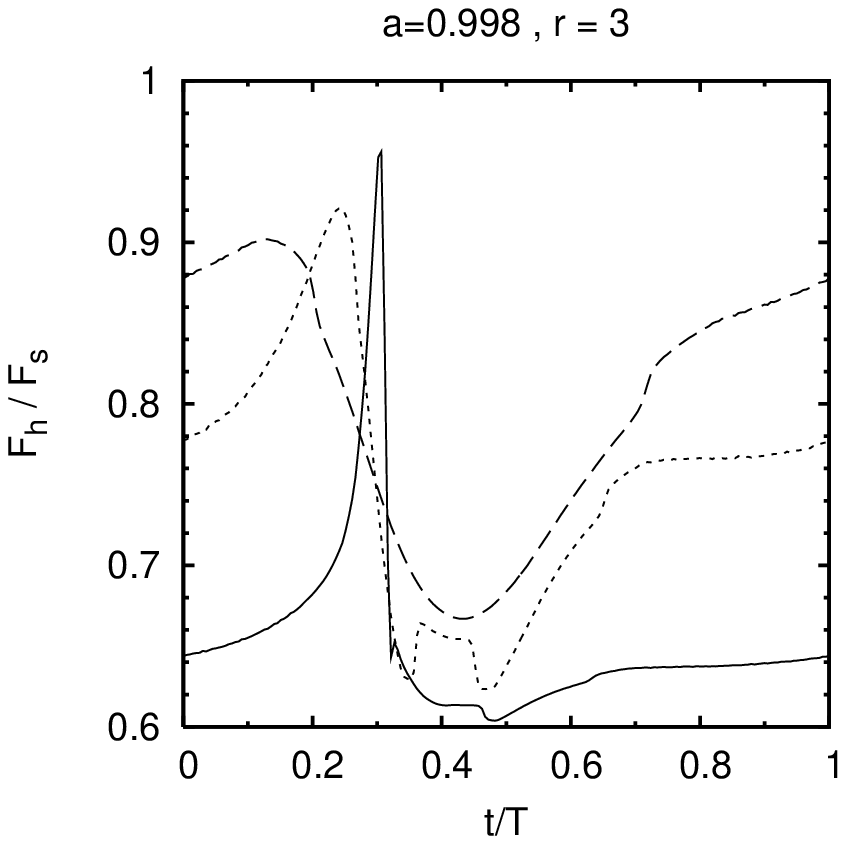}\\[4mm]
  \vspace*{-2mm}
  \includegraphics[width=3.75cm]{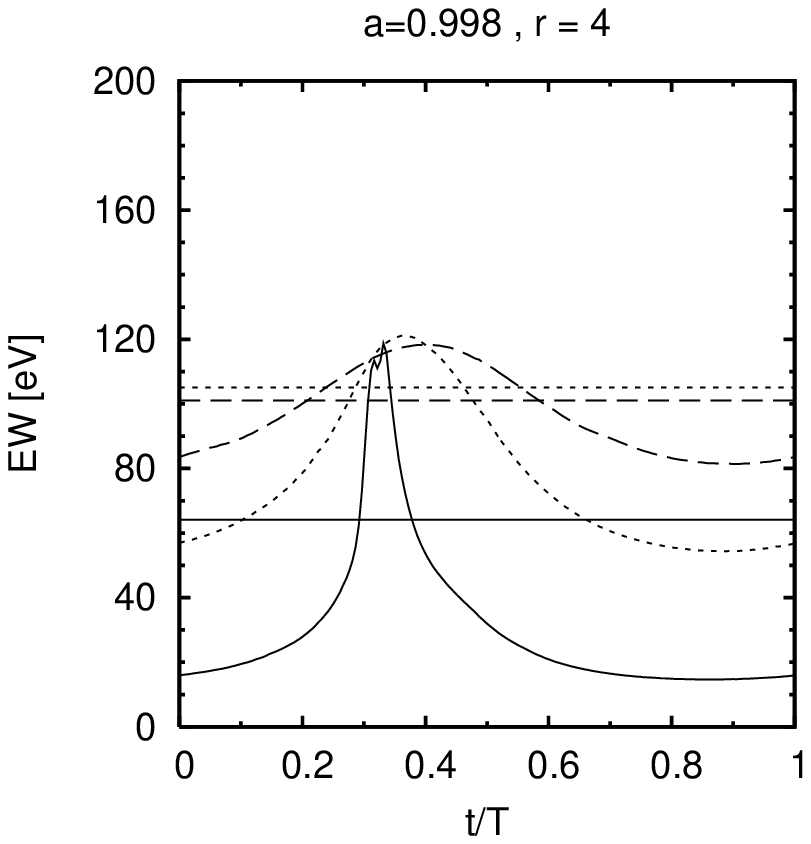}
  \includegraphics[width=3.77cm]{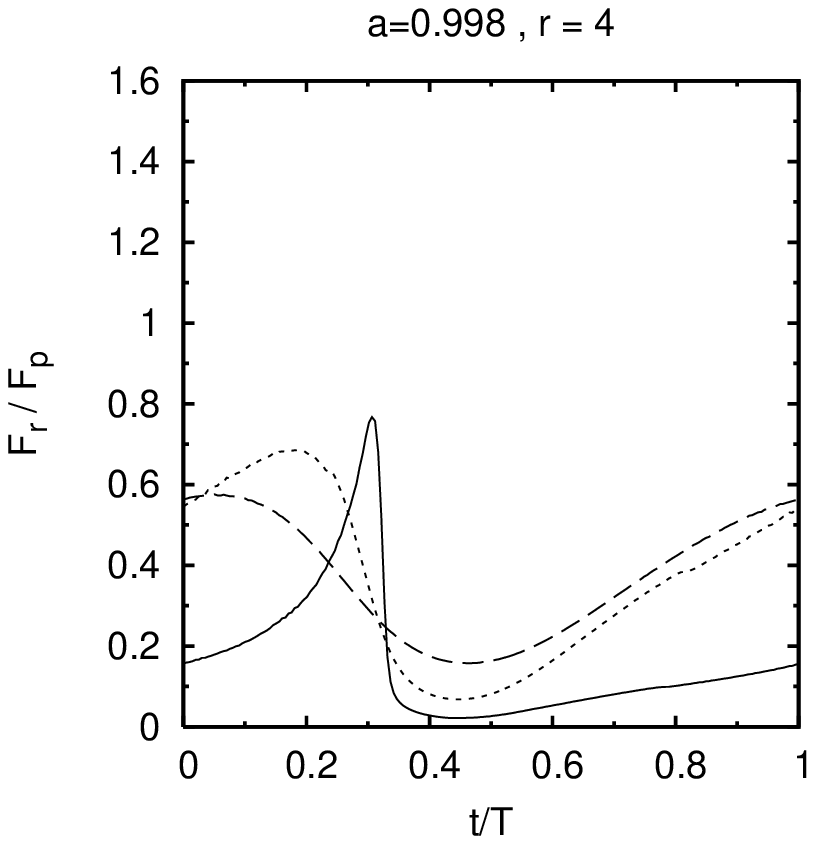}
  \includegraphics[width=3.9cm]{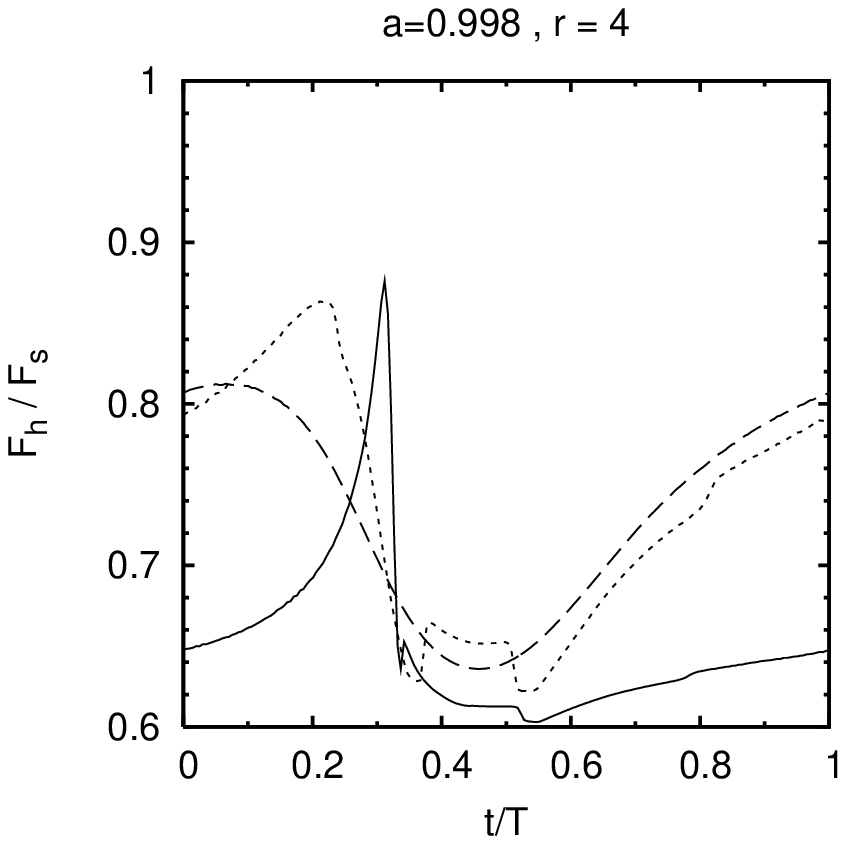}\\[4mm]
  \vspace*{-2mm}
  \includegraphics[width=3.75cm]{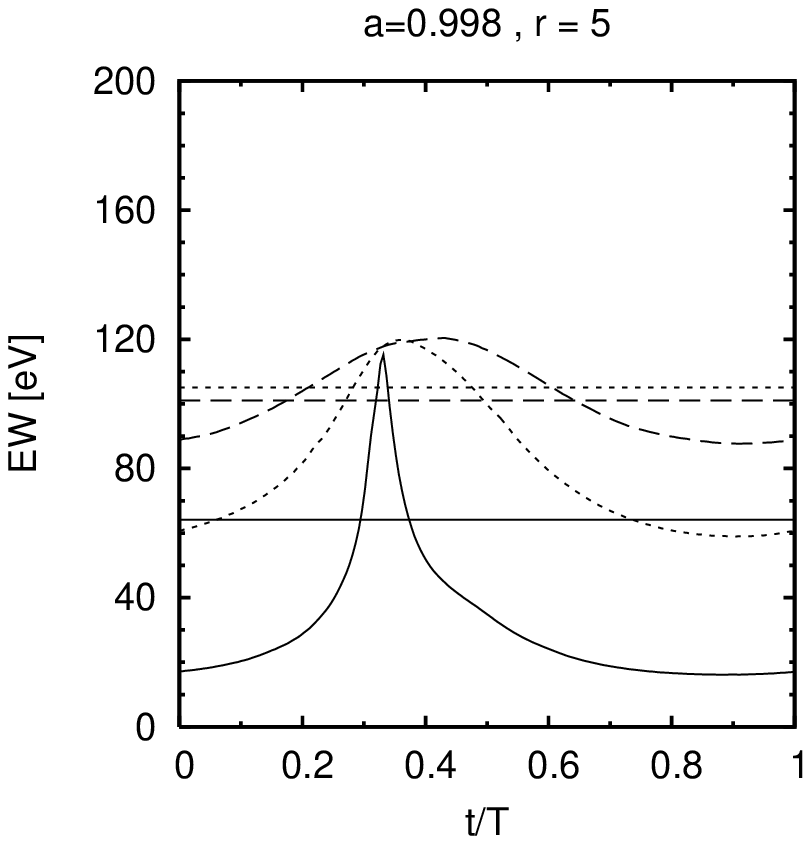}
  \includegraphics[width=3.77cm]{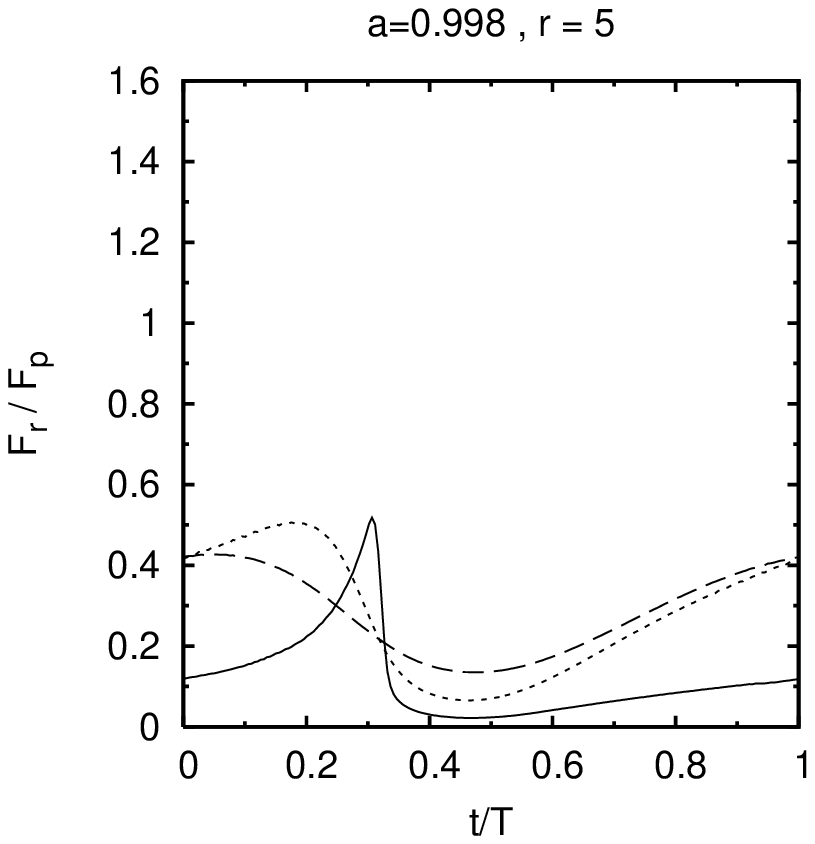}
  \includegraphics[width=3.9cm]{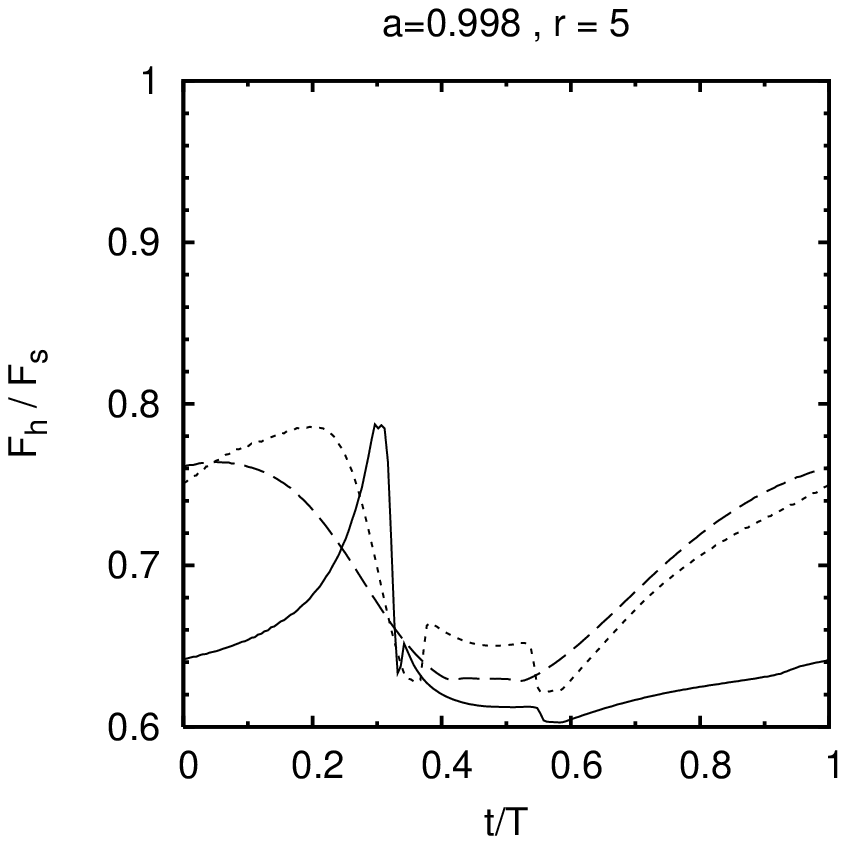}\\[4mm]
  \vspace*{-2mm}
  \hspace*{0mm}
  \includegraphics[width=3.75cm]{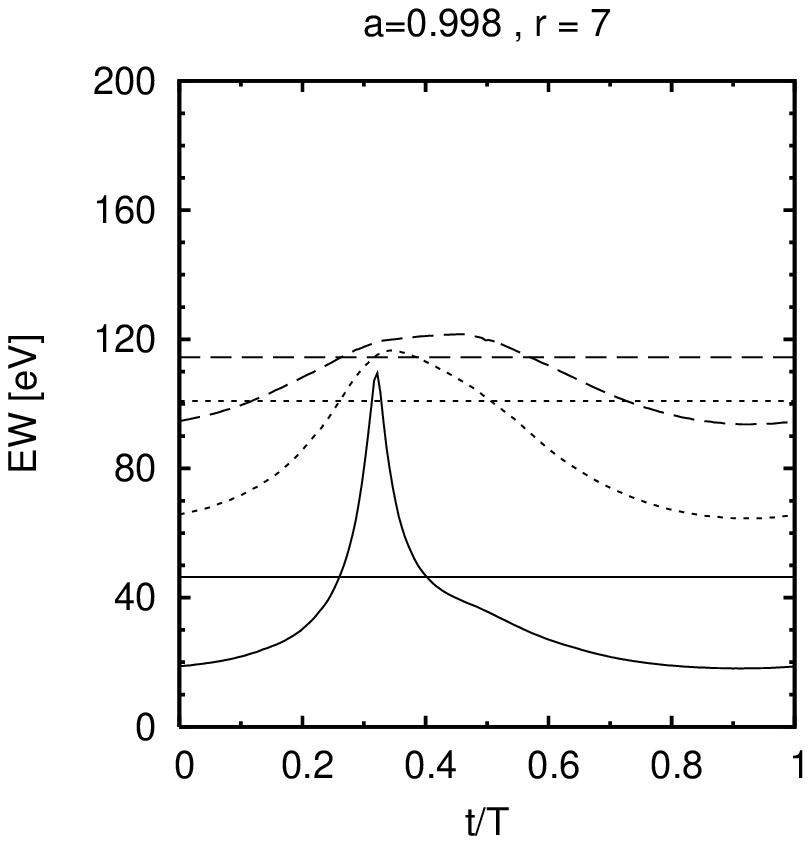}
  \includegraphics[width=3.77cm]{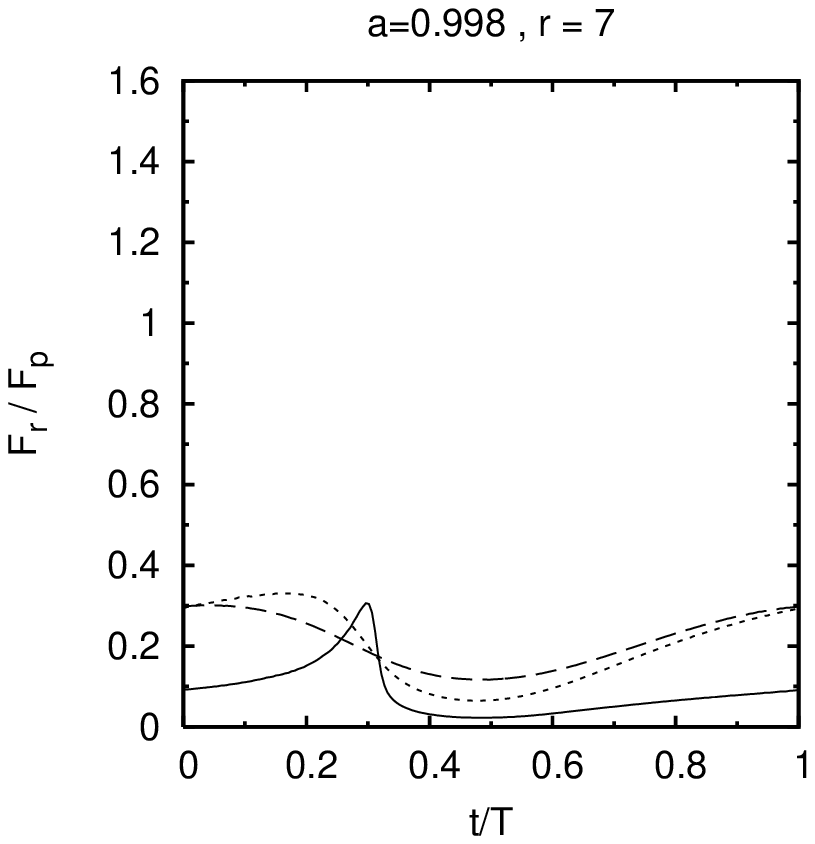}
  \includegraphics[width=3.9cm]{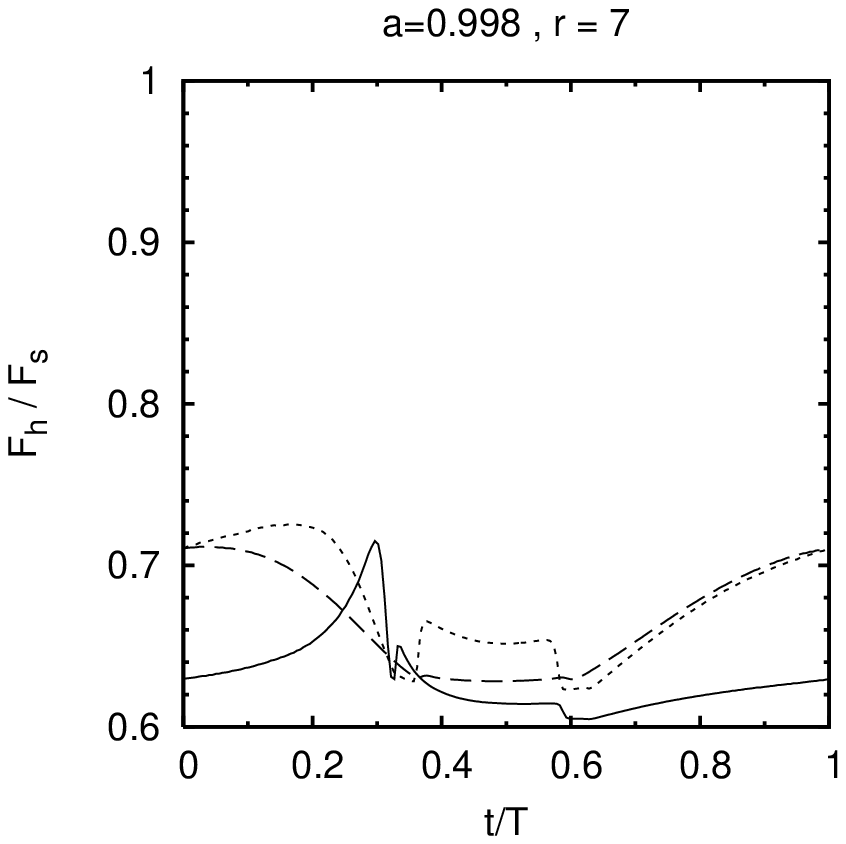}
  \vspace*{-1mm}
  \caption{{\bf Left:}
   The time variation of the observed EW of the K$\alpha$ line.
   The integrated EW is shown in horizontal lines.
  {\bf Middle:} The ratio of the observed reflected emission to the
  observed primary emission. The fluxes are integrated in the 3--10~keV energy
  range.
  {\bf Right:} The hardness ratio of the hard flux $F_{\rm h}$
  (6.5--10~keV) to the soft flux $F_{\rm s}$(3--6.5~keV).
  The flux in the Fe lines is also included.
  The dashed, dotted and solid lines correspond to the inclinations 30$^\circ$,
  60$^\circ$ and 85$^\circ$.}
  \label{fig-ew1}
\end{center}
\end{figure*}

\begin{figure*}
\begin{center}
  \vspace*{-2mm}
  \includegraphics[width=4cm]{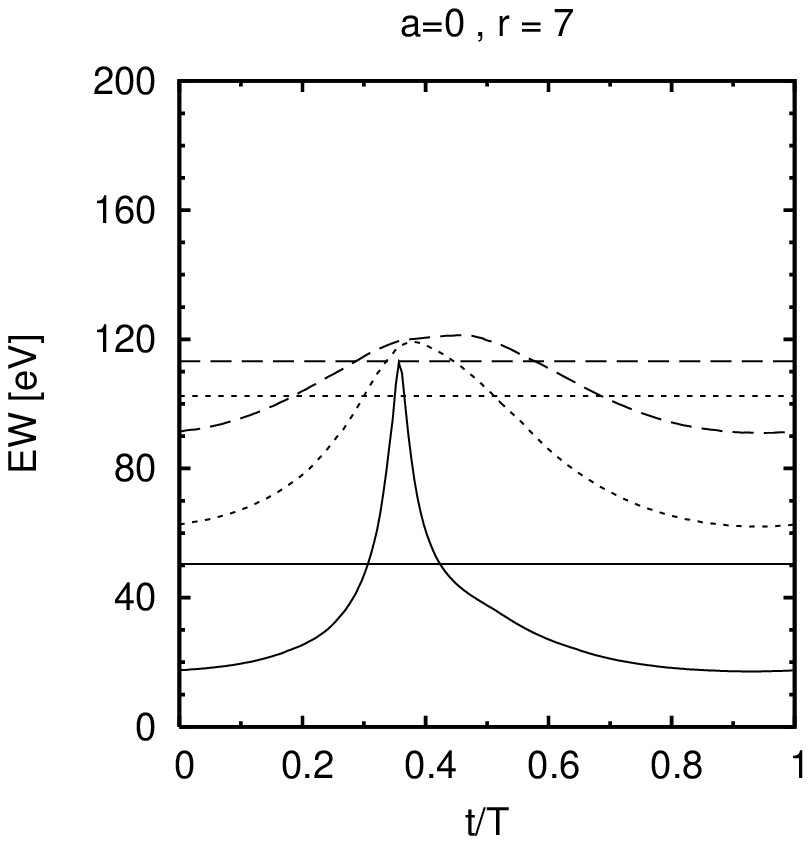}
  \includegraphics[width=4.02cm]{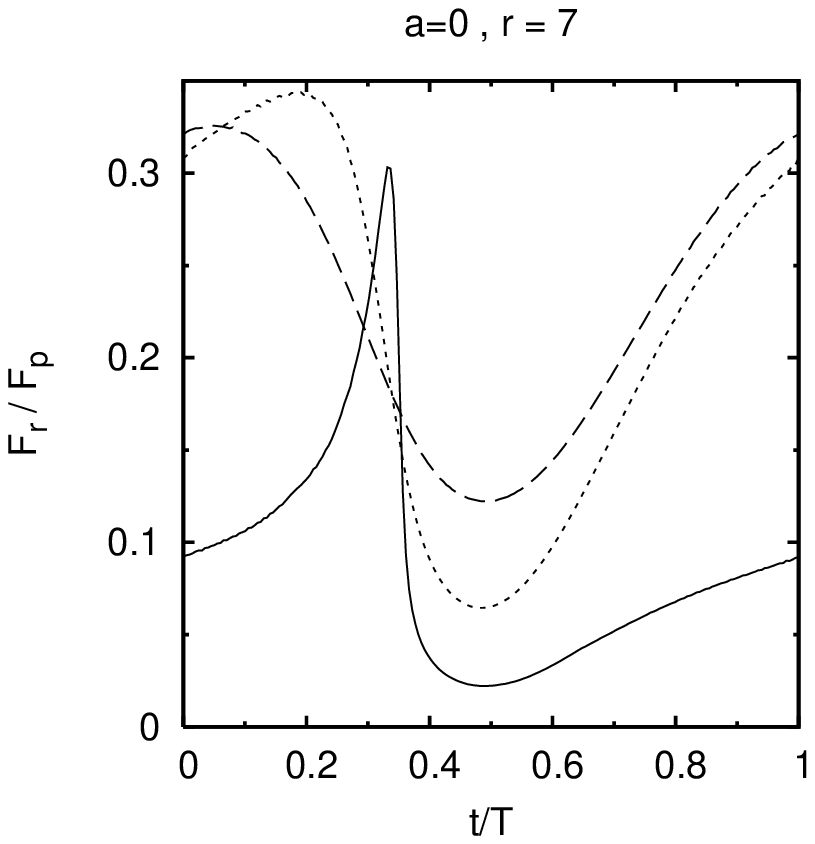}
  \includegraphics[width=4.28cm]{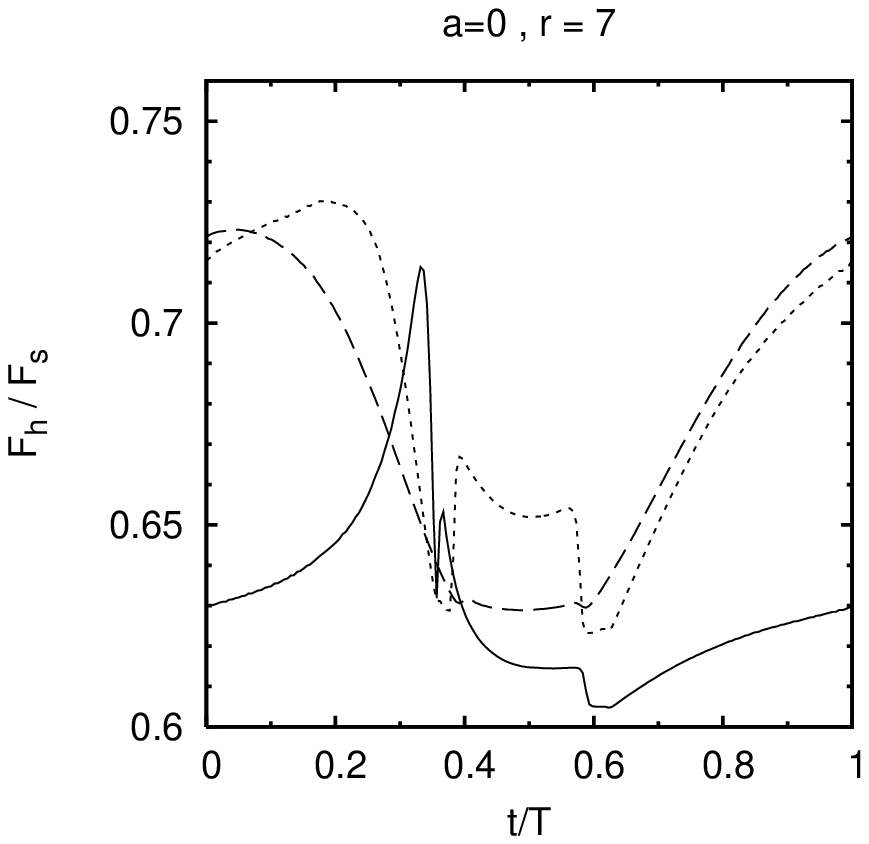}\\[4mm]
  \vspace*{-2mm}
  \includegraphics[width=4cm]{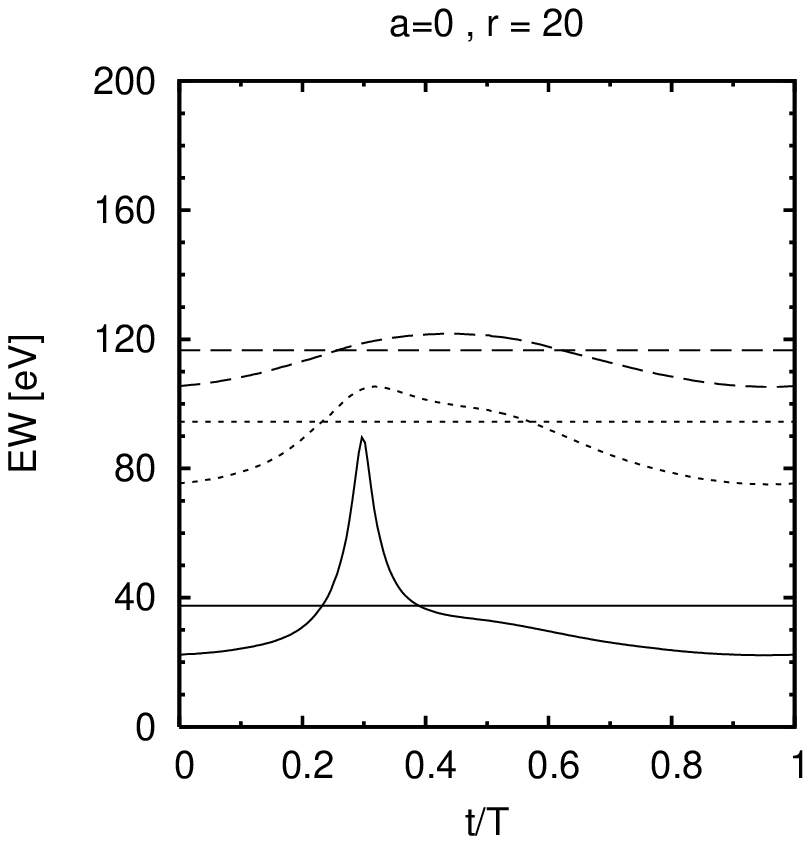}
  \includegraphics[width=4.02cm]{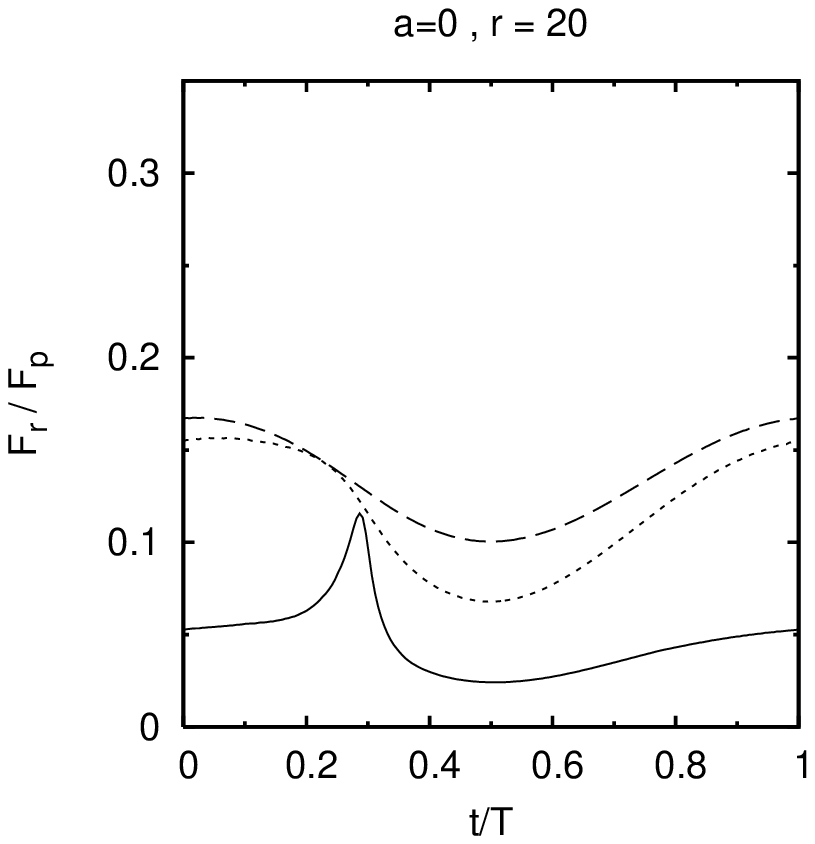}
  \includegraphics[width=4.28cm]{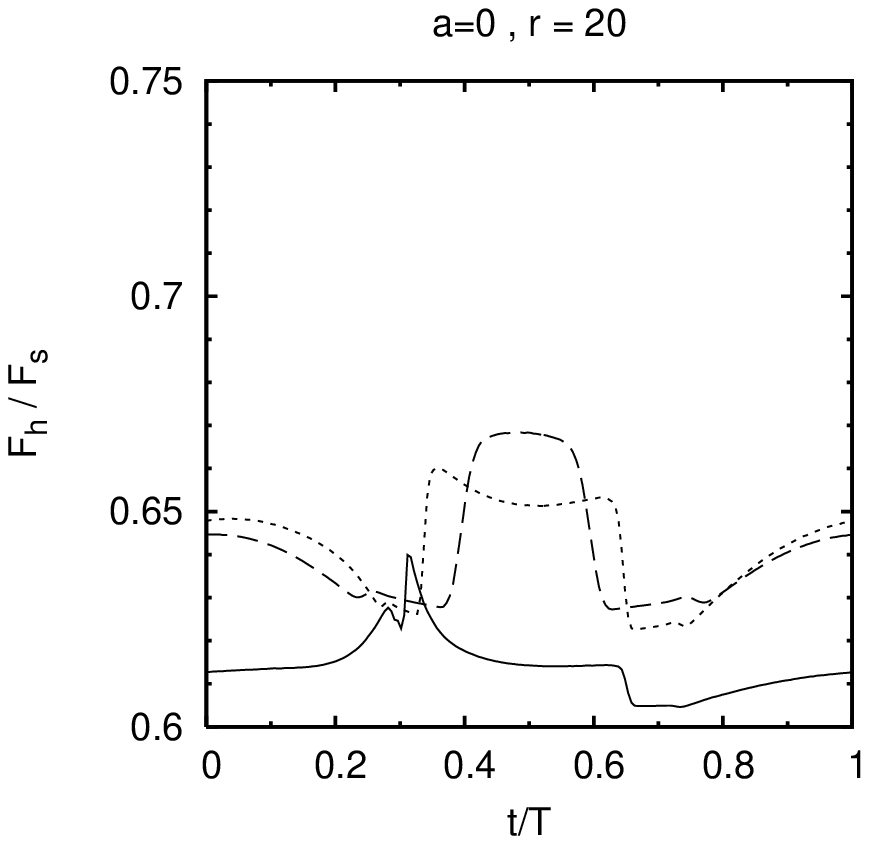}\\[4mm]
  \vspace*{-2mm}
  \includegraphics[width=4cm]{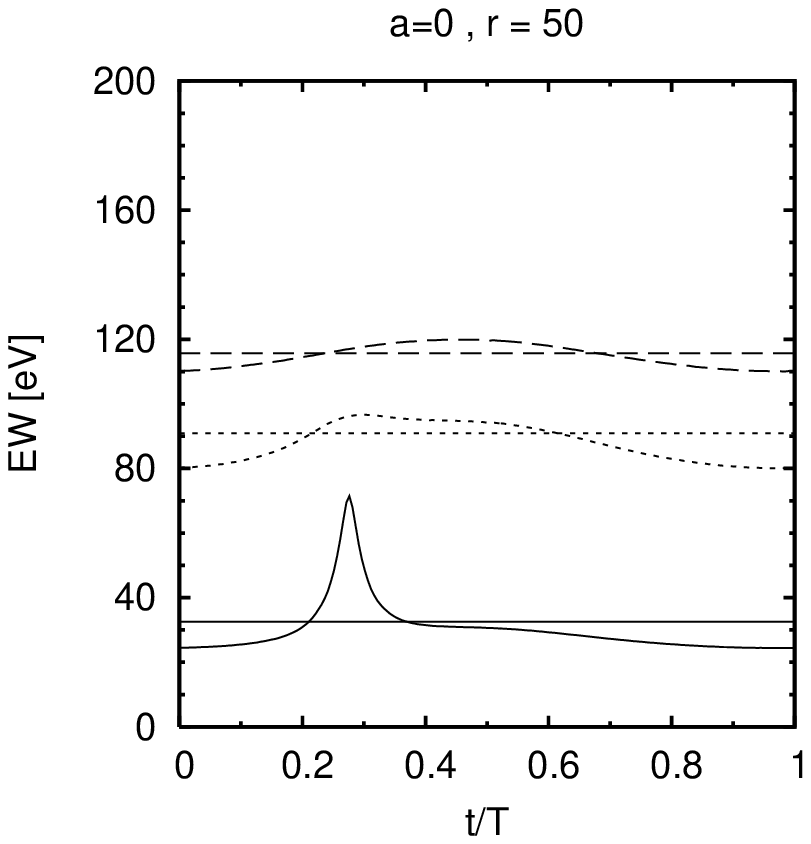}
  \includegraphics[width=4.02cm]{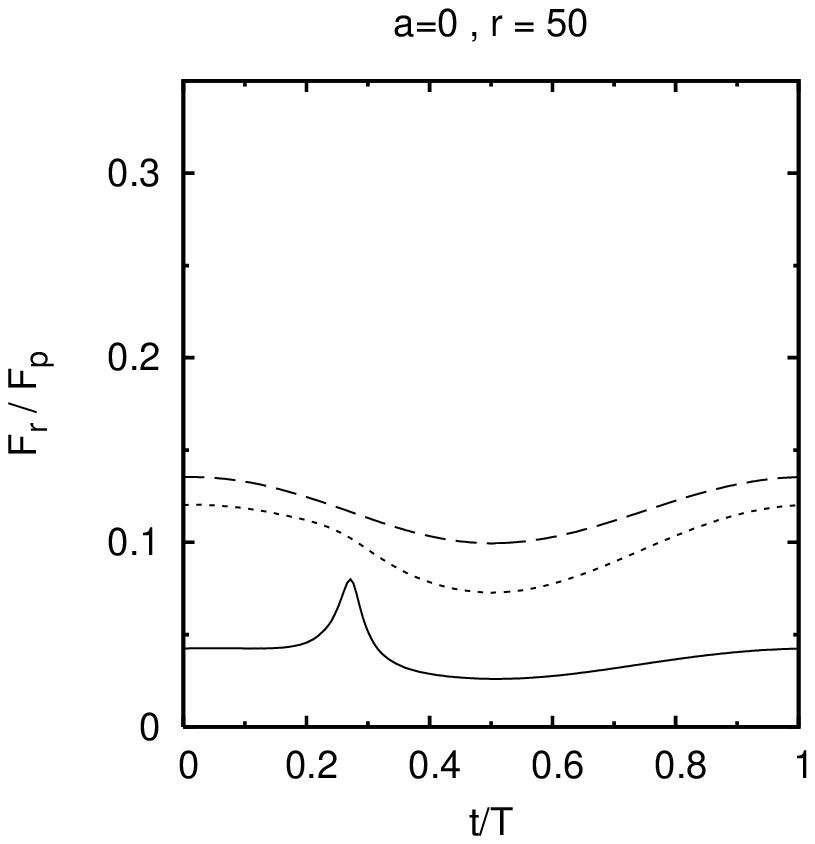}
  \includegraphics[width=4.28cm]{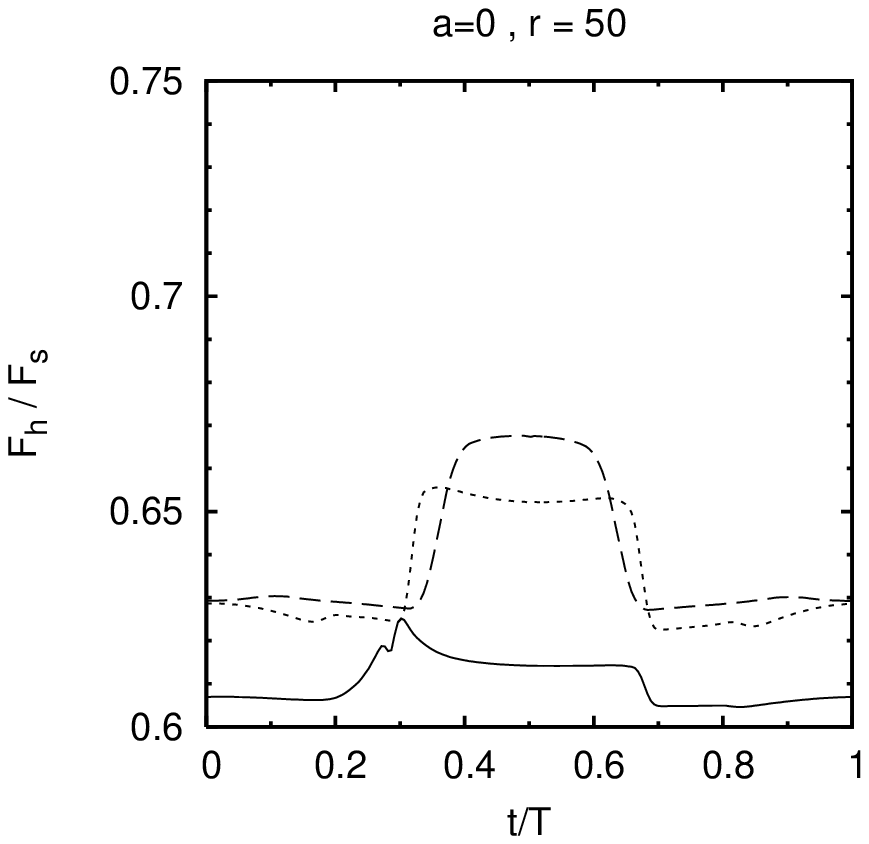}\\[4mm]
  \vspace*{-2mm}
  \includegraphics[width=4cm]{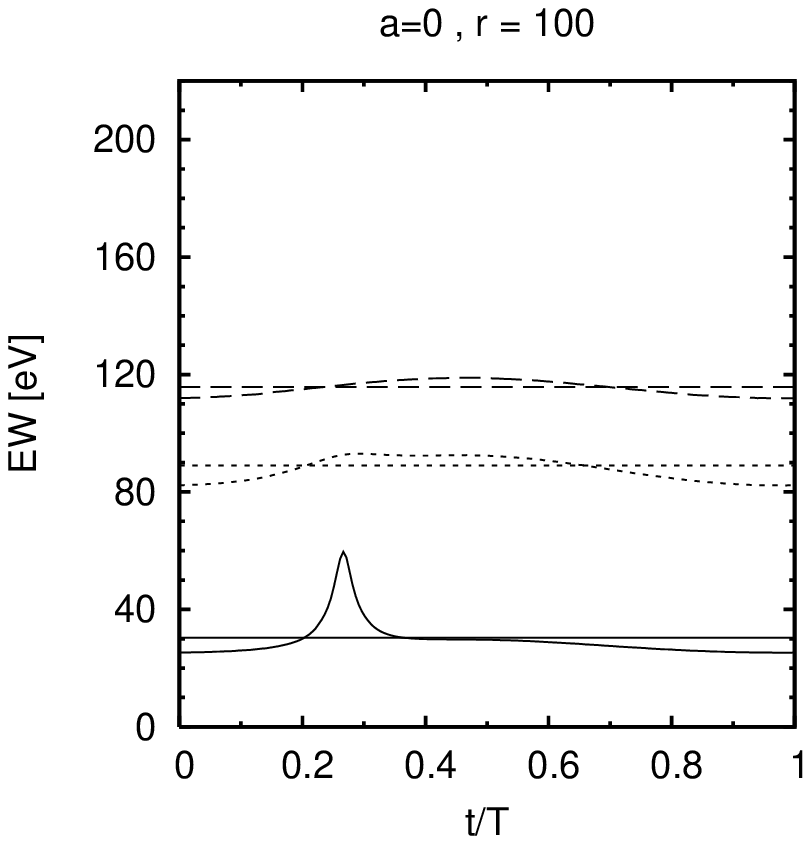}
  \includegraphics[width=4.02cm]{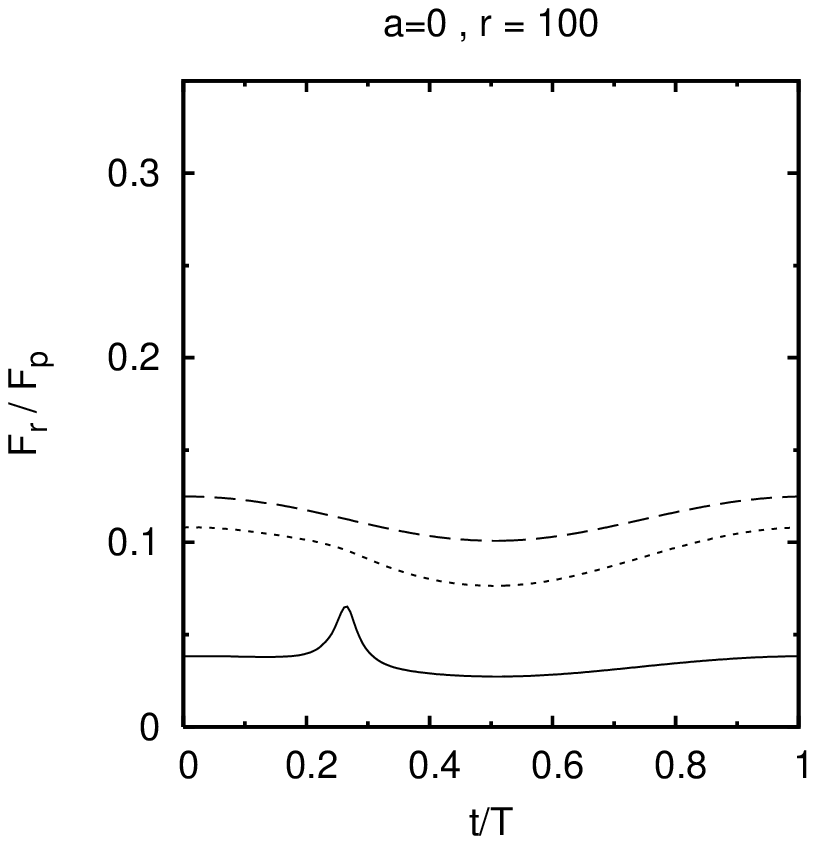}
  \includegraphics[width=4.28cm]{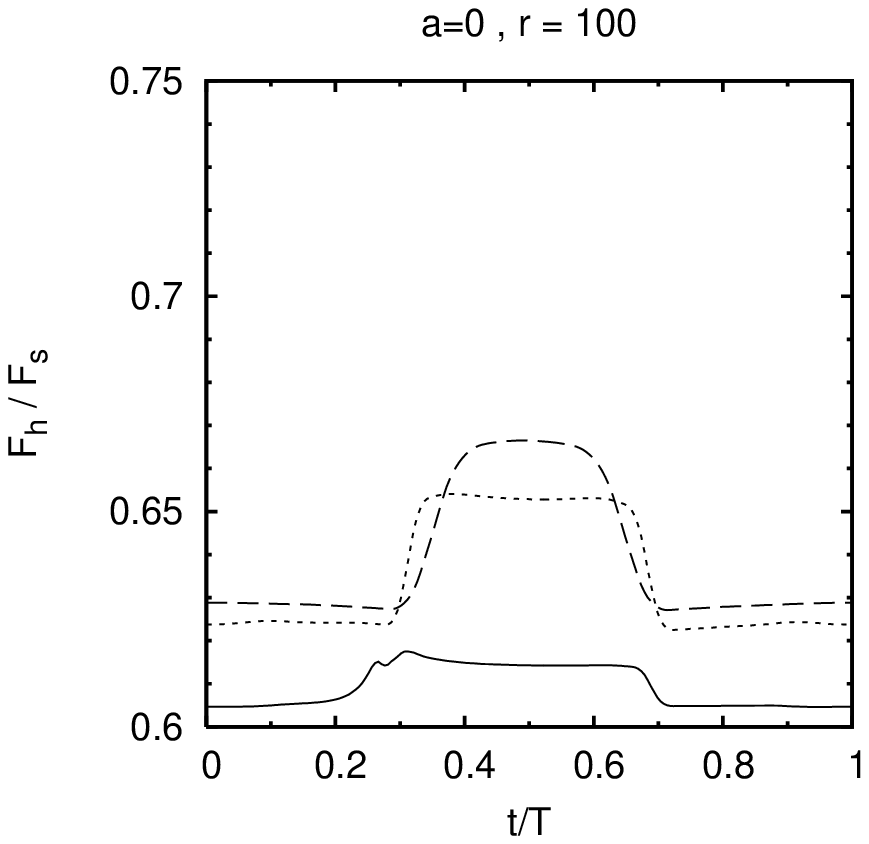}
  \caption{The same as in Fig.~\ref{fig-ew1} but for the
  Schwarzschild   black hole and the spot orbital radii 7, 20, 50 and 100
  $GM/c^2$ (from top to bottom).}
  \label{fig-ew2}
\end{center}
\end{figure*}

\begin{figure*}
\begin{center}
  \vspace*{-2mm}
  \hspace*{-3.5mm}
  \includegraphics[width=3.9cm]{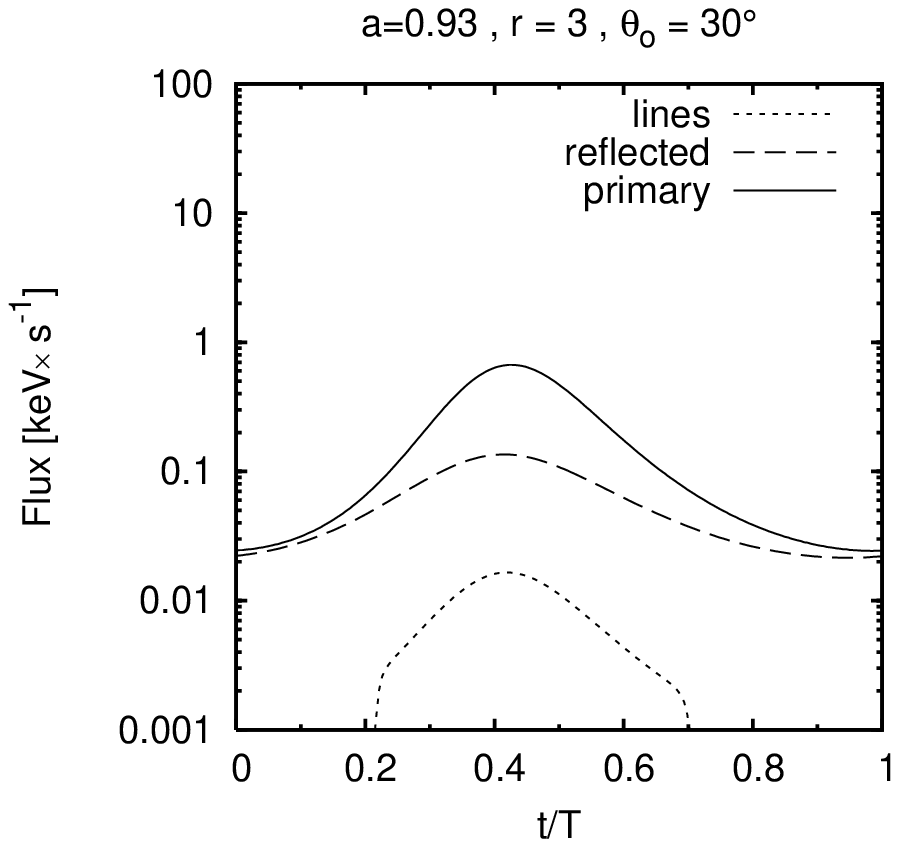}
  \hspace*{0mm}
  \includegraphics[width=3.9cm]{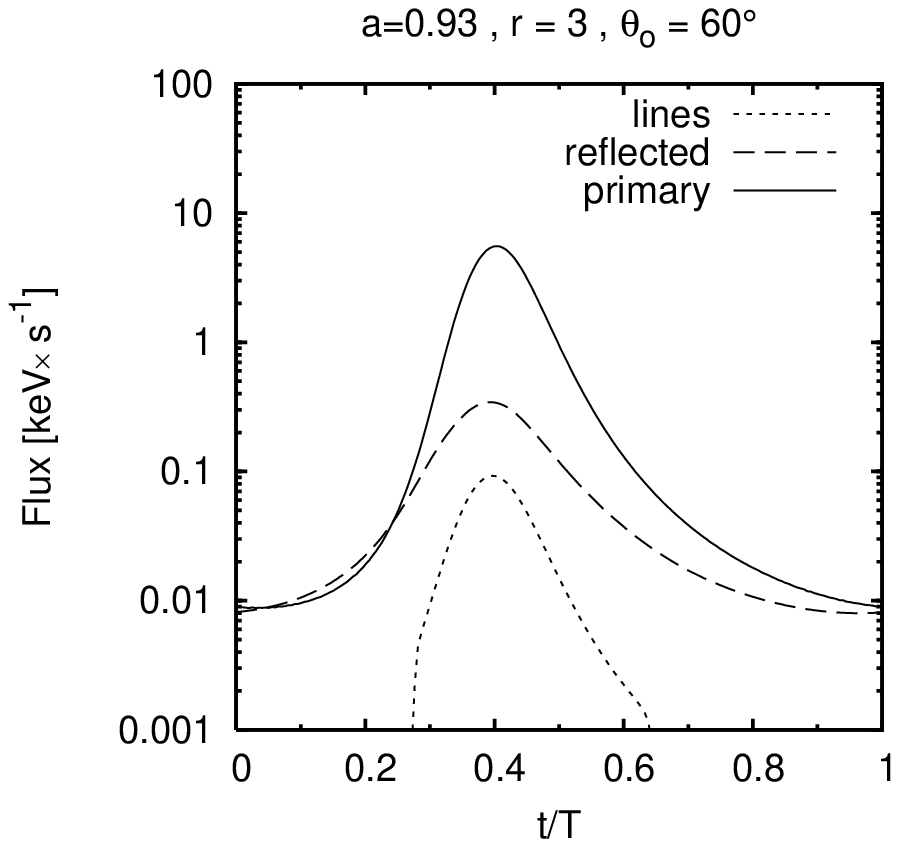}
  \hspace*{-1mm}
  \includegraphics[width=3.9cm]{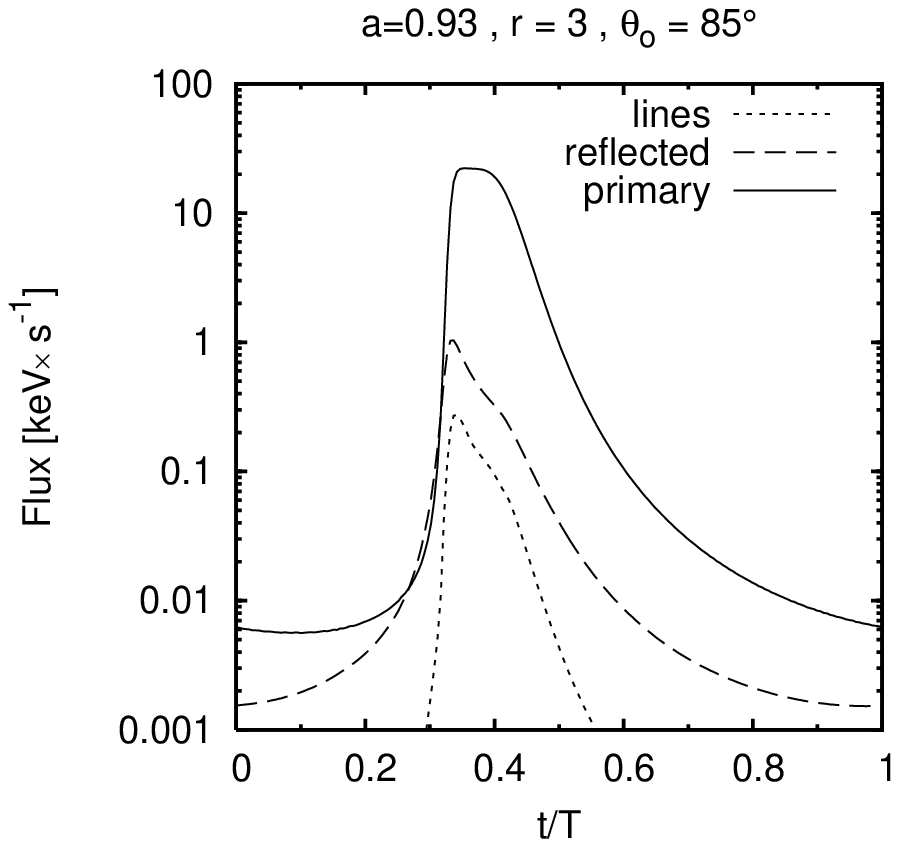}\\[4mm]
  \vspace*{-2mm}
  \hspace*{-3mm}
  \includegraphics[width=4cm]{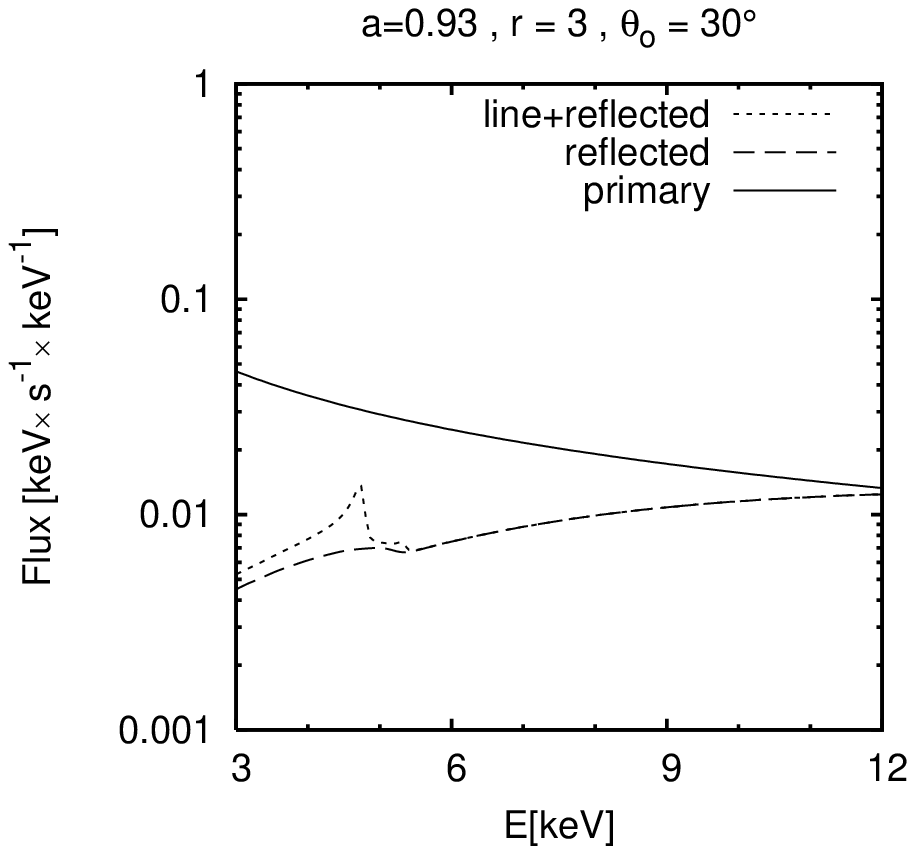}
  \hspace*{-1mm}
  \includegraphics[width=4cm]{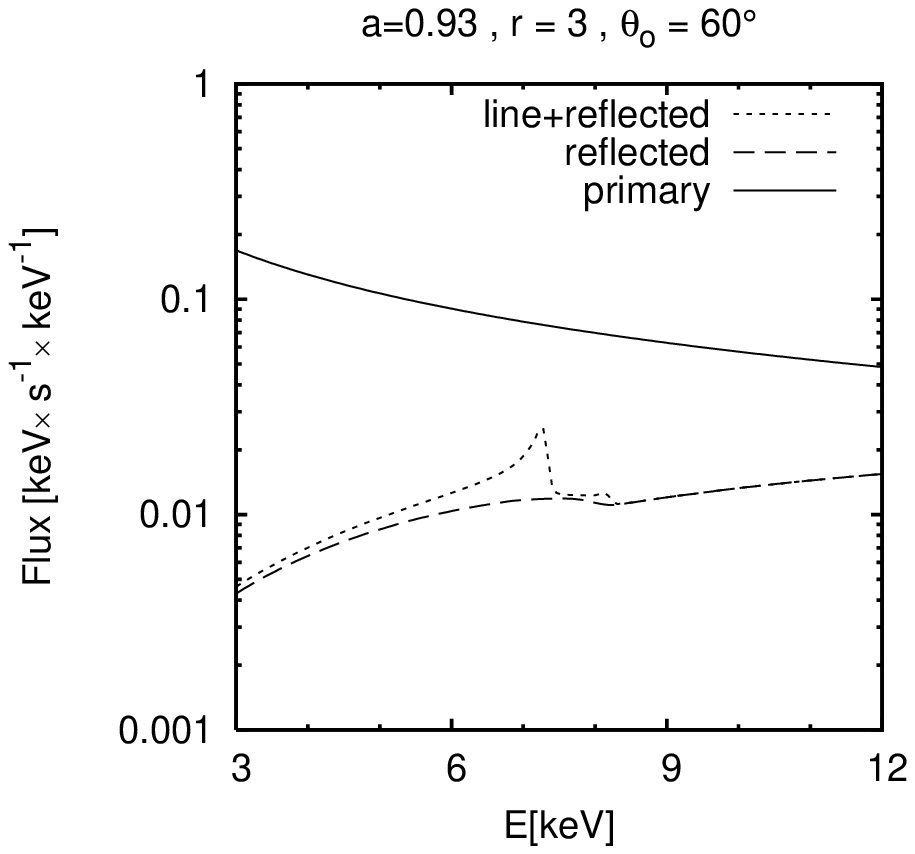}
  \hspace*{-2mm}
  \includegraphics[width=4cm]{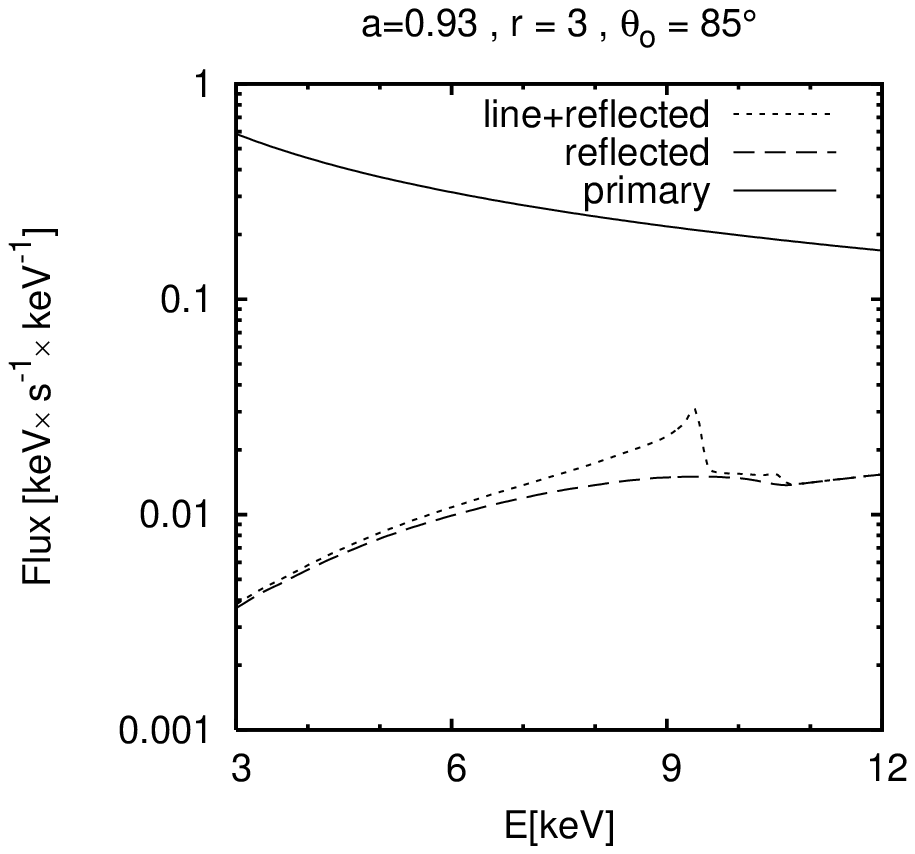}\\[4mm]
  \vspace*{-2mm}
  \hspace*{-0.8mm}
  \includegraphics[width=3.8cm]{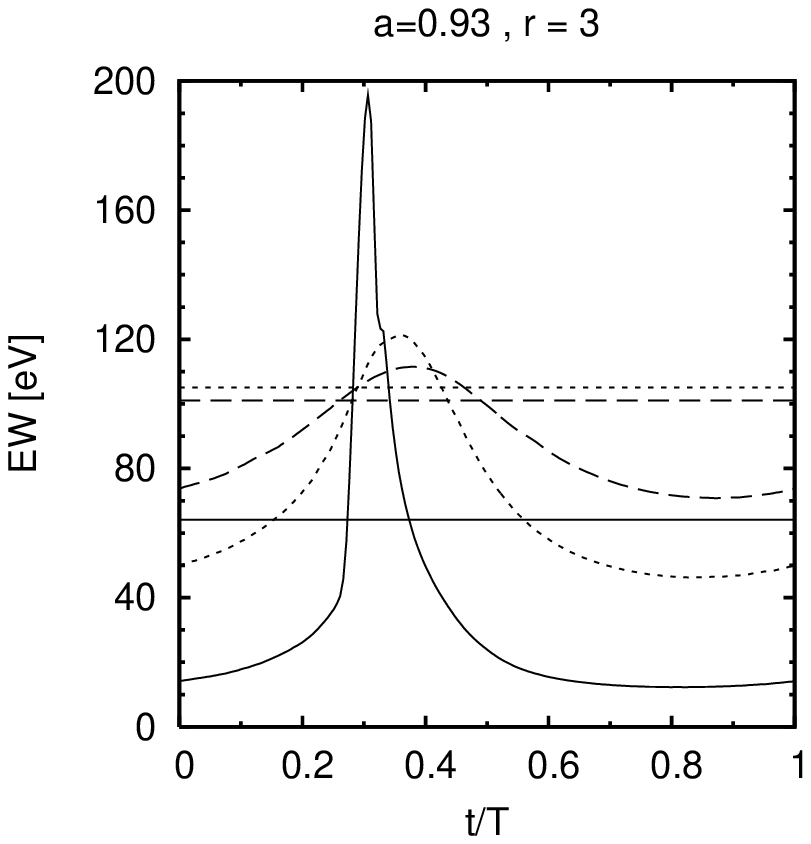}
  \hspace*{0.3mm}
  \includegraphics[width=3.85cm]{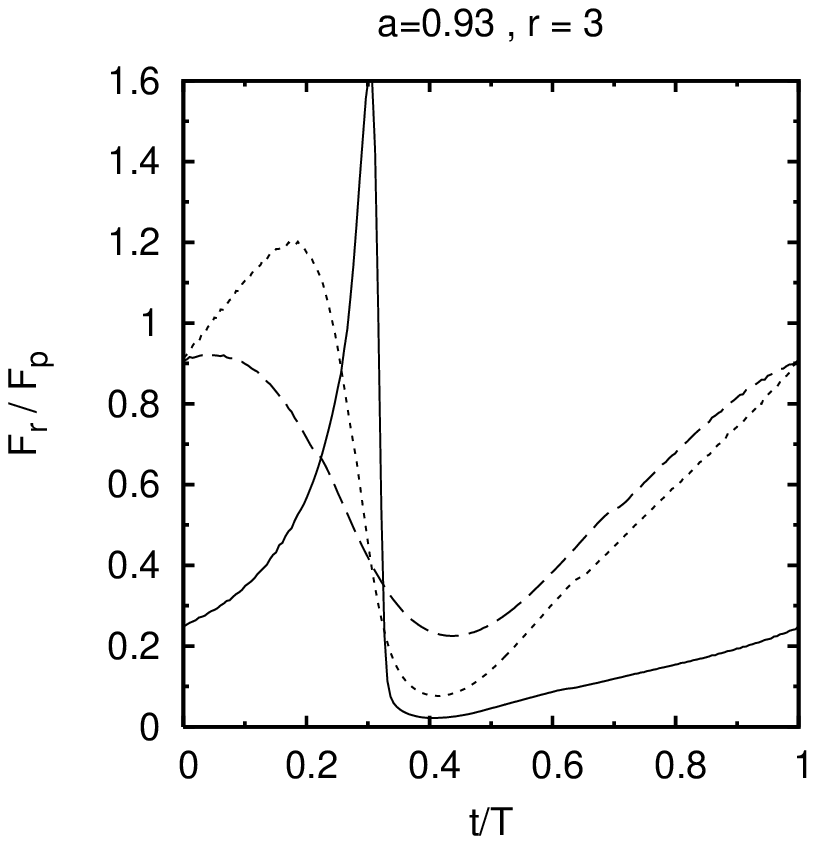}
  \hspace*{-2mm}
  \includegraphics[width=3.98cm]{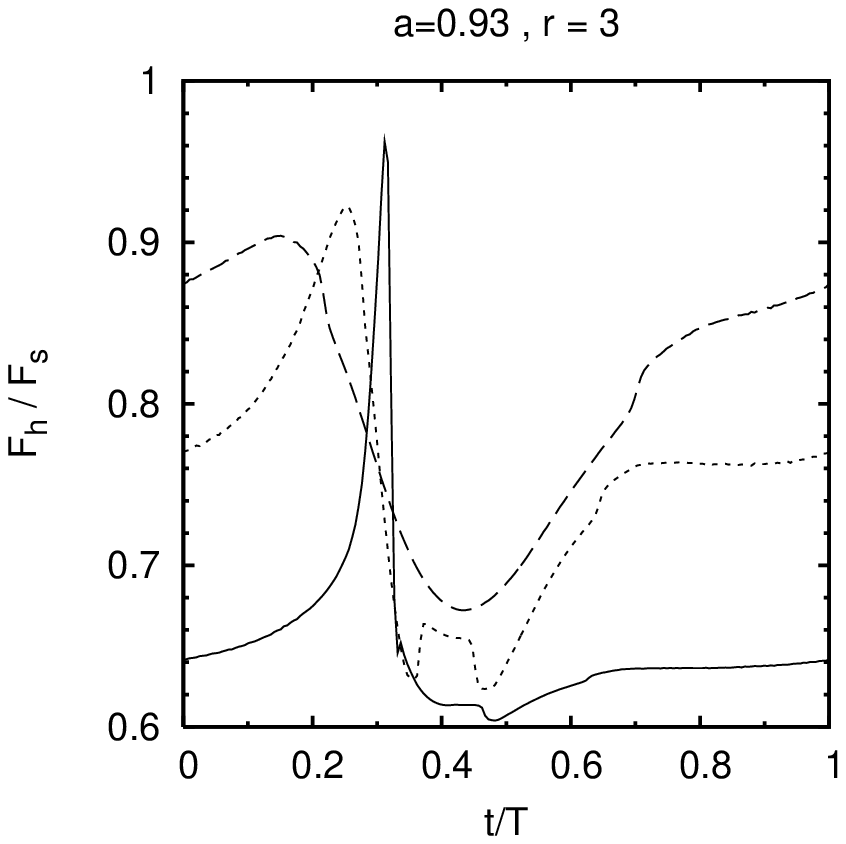}
  \hspace*{0mm}
  \caption{The same as in
   Figs.~\ref{fig-light_curves1}, \ref{fig-spectrum1},
  and \ref{fig-ew1} top rows (i.e.\ the spot
  orbital radius $3\,GM/c^2$) but for the Kerr black hole with
  the spin $0.93\,GM/c^3$.}
  \label{fig-spin}
\end{center}
\end{figure*}

The observed light curves computed for the spot in the vicinity of
the extremally
spinning Kerr black hole ($a=0.998\,GM/c^3$) and the Schwarzschild black hole
($a=0\,GM/c^3$) in the 3--10~keV energy range for different orbital radius can
be seen in Figs.~\ref{fig-light_curves1} and \ref{fig-light_curves2}.
The light curves are influenced mainly by
the overall amplification factor, transfer function and delay amplification
[see paper \citep{dov07} for details], and by the dependence of the local flux
on the emission angle. The primary
emission dominates the observed flux as expected, meanwhile the reflected flux
in the Fe lines from the spot contributes less. There is an exception in
this behaviour, though, for some parts of the orbit close to the black hole
(see top row of Fig.~\ref{fig-light_curves1}). The reflected flux from the spot
exceeds the flux of the primary for the orbital radius $r=3\,GM/c^2$ and for
the inclinations $\theta_{\rm o}=60^\circ$ and $85^\circ$.
The variations of the flux decrease with the orbital radius as expected. Note,
that the amplification of the emission due to the lensing effect is still
relevant as far as $100\,GM/c^2$ for large inclination angles ($85^\circ$).

Figs.~\ref{fig-spectrum1} and \ref{fig-spectrum2} shows the mean spectra taken
over the whole orbit. The line is smeared when taken over the whole orbit.
As it is well known \citep{iwa96} in the Schwarzschild case, if we assume that
the emission comes mainly from above the innermost stable orbit,
the line stays above 3~keV, while in the Kerr case it can be shifted even
below this energy (as is the case for all shown inclinations for the spot
orbit below $4\,GM/c^2$). The iron edge is smeared in all studied cases and the
dominance of the primary emission is evident. As we expect, the line is less
shifted with the increasing orbital radius but it is still substantially
broadened even at the radius $100\,GM/c^2$ due to the large orbital velocity of
the spot.

In order to quantify the properties of the observed spectra let us look at the
equivalent width, ratio of the observed reflected and primary components, and
the hardness ratio (Figs.~\ref{fig-ew1} and \ref{fig-ew2}).

A closer look at the EW, see the left panels in Figs.~\ref{fig-ew1} and
\ref{fig-ew2}, reveals that it does not much differ from its
local value (Fig.~\ref{fig-ew_loc}),
i.e.\ $EW(t)\approx EW_{\rm loc}(\mu_{\rm e}(t))$ (for the
dependence $\mu_{\rm e}(t)$ for some radii see \citet{dov07}).
This is not, however, true for the case of the low orbital radius with an
observer inclination of $85^\circ$ when the EW is magnified due to the
lensing effect.
For the spot close to the black hole ($r=3\,GM/c^2$) the EW is changing
with respect to its mean value by 30\% even for a low inclination angle
$30^\circ$. For an almost edge-on disc it can vary by as much as 200\%.
Similar to the flux variations at the larger radii the variation of the EW
decreases. This is true also for all the other studied spectral
characteristics.

The observed ratio of the reflected flux to the primary flux is amplified when
compared to the local one, see the middle panels in Figs.~\ref{fig-ew1} and
\ref{fig-ew2} and the left panel in Fig. \ref{fig-ratio_loc}. The amplification
is the highest for the lowest
orbital radius and highest inclination angle --- the ratio in this case is
increased by more than one order of magnitude.
Note that in the Kerr case with the orbital radius $3\,GM/c^2$ and for the
inclinations $60^\circ$ and $85^\circ$ the ratio of the observed reflected flux
to the observed primary
flux is larger than unity, meaning the reflected component prevails over the
primary one in a certain part of the orbit.

To evaluate the hardness ratio we compared the fluxes in between 3--6.5 keV
(soft component, $F_{\rm s}$) and 6.5--10 keV (hard component, $F_{\rm h}$).
The hardness ratio is also amplified when we compare it with the local hardness
ratio (the right panels in Figs.~\ref{fig-ew1}, \ref{fig-ew2} and
\ref{fig-ratio_loc}).
As in other spectral characteristics, the amplification and the variation of
the hardness ratio is the largest for the lowest orbital radius
in the extremally rotating Kerr case and they decrease with the increasing
orbital radius.

To see how the different values of the spin parameter influence the studied
properties of the observed signal we can compare the results for the same
radius, see the bottom panels of Figs.~\ref{fig-light_curves1},
\ref{fig-spectrum1} and \ref{fig-ew1}
for the extremally rotating Kerr black hole and compare them with the top
panels of Figs.~\ref{fig-light_curves2}, \ref{fig-spectrum1} and \ref{fig-ew2}
for the Schwarzschild black hole. For the closer radius, $r=3\,GM/c^2$, we have
computed the results for the spin $a=0.93\,GM/c^3$ (the spin cannot be lower
if the spot should be above the innermost stable circular orbit), see
Fig.~\ref{fig-spin} and compare it with the top panels in
Figs.~\ref{fig-light_curves1}, \ref{fig-spectrum1} and \ref{fig-ew1}.
It is clear from all of these comparisons that if we
fix the orbital radius our results do not depend on the spin of the black hole.

\section{Conclusions}

We have studied the light curves, spectra and several spectral characteristics
in the flare--spot model for different orbital radii of the flare.
The primary flux was included and the mutual
normalizations of the primary and reflected emission were treated within the
framework of a simple, yet self-consistent scheme. About half of the isotropic
primary flux hits the disc below the flare and is reprocessed there, creating a
radiating spot. A part of the reprocessed radiation is re-emitted towards the
observer. The radiation is influenced by the relativistic effects before
reaching the observer.

We can sum up our results in several conclusions:
\vspace*{-\smallskipamount}
\begin{enumerate}
 \item The EW, apart for the extreme cases of high inclinations,
   does not differ significantly from the local EW.
   However, close to the black hole it varies even for low inclination of
   30$^\circ$ by up to 30\% when
   compared with its mean value for the whole orbit. The EW could be
   significantly amplified in our model only if the primary emission were
   beamed towards the disc, thus decreasing the observed primary emission.
 \item Both the ratio of the observed reflected to the observed primary flux
   and the hardness ratio are amplified when compared to the values for
   the intrinsic (local) emission.
 \item The variations of all of the studied spectral characteristics are
   the highest for close orbits and higher inclination angles.
 \item The spin of the black hole affects significantly our results only as far
   as it determines the location of the marginal stable orbit.
\end{enumerate}
Here, we would like to remind the reader, that these results apply for a flare
arising very near above the disc and thus they can heavily differ from the
results of the light-bending model by \citet{min04}
where the flare orbits far above the disc and the resulting spot is much
larger.

It follows from our results that the studied flux ratios could be used for
estimating the lower limit of possible values of the spin parameter if the
flare arises in the close vicinity of the black hole.

% Acknowledgments are created using the command \ack:
\ack%%%%%%%%%%%%%%%%%%%%%%%%%%%%%%%%%%%%%%%%%%%%%%%%%%%%%%%%%%%%%%%%%%%%%%%

This research is supported by the ESA PECS project No.~98040. MD and VK
gratefully acknowledge support from the Czech Science Foundation grants
205/05/P525 and 202/06/0041. GM acknowledges financial support
from Agenzia Spaziale Italiana (ASI). RG is grateful for financial support
to the Centre of Theoretical Astrophysics (LC06014).

% Here we specify the basename of the bibliography database file,
\bibliography{spotII}

\begin{thebibliography}{24}
\expandafter\ifx\csname natexlab\endcsname\relax\def\natexlab#1{#1}\fi
\expandafter\ifx\csname url\endcsname\relax
  \def\url#1{\texttt{#1}}\fi
\expandafter\ifx\csname urlprefix\endcsname\relax\def\urlprefix{URL }\fi
\providecommand{\selectlanguage}[1]{\relax}
\providecommand{\eprint}[2][]{\url{#2}}

\bibitem[{Collin et~al.(2003)Collin, Coup\'e, Dumont, Petrucci and
  R\'o\.za\'nska}]{col03}
Collin, S., Coup\'e, S., Dumont, A.-M., Petrucci, P.-O. and R\'o\.za\'nska, A.
  (2003), {Evolution of the X-ray spectrum in the flare model of Active
  Galactic Nuclei}, \emph{Astronomy \& Astrophysics}, \textbf{400}, pp.
  437--449.

\bibitem[{Dabrowski and Lasenby(2001)}]{dab01}
Dabrowski, Y. and Lasenby, A.~N. (2001), {Reflected iron line from a source
  above a Kerr black hole accretion disc}, \emph{Monthly Notices of the Royal
  Astron. Society}, \textbf{321}, pp. 605--614.

\bibitem[{Dov{\v c}iak(2004)}]{dov04a}
Dov{\v c}iak, M. (2004), \emph{{Radiation of Accretion Discs in Strong
  Gravity}}, PhD Thesis, Charles University, Prague, arXiv:astro-ph/0411605.

\bibitem[{Dov{\v c}iak et~al.(2007)Dov{\v c}iak, Karas, Matt and
  Goosmann}]{dov07}
Dov{\v c}iak, M., Karas, V., Matt, G. and Goosmann, R.~W. (2007), {Variation of
  the primary and reprocessed radiation from an orbiting spot around a black
  hole}, \emph{Monthly Notices of the Royal Astron. Society}, accepted.

\bibitem[{Dumont et~al.(2000)Dumont, Abrassart and Collin}]{dum00}
Dumont, A.-M., Abrassart, A. and Collin, S. (2000), {A code for optically thick
  and hot photoionized media}, \emph{Astronomy \& Astrophysics}, \textbf{357},
  pp. 823--838.

\bibitem[{Fabian et~al.(2000)Fabian, Iwasawa, Reynolds and Young}]{fab00}
Fabian, A.~C., Iwasawa, K., Reynolds, C.~S. and Young, A.~J. (2000), {Broad
  Iron Lines in Active Galactic Nuclei}, \emph{Publications of the Astron.
  Society of the Pacific}, \textbf{112}, pp. 1145--1161.

\bibitem[{Galeev et~al.(1979)Galeev, Rosner and Vaiana}]{gal79}
Galeev, A.~A., Rosner, R. and Vaiana, G.~S. (1979), {Structured coronae of
  accretion disks}, \emph{Astroph. Journal}, \textbf{229}, pp. 318--326.

\bibitem[{George and Fabian(1991)}]{geo91}
George, I.~M. and Fabian, A.~C. (1991), {X-ray reflection from cold matter in
  active galactic nuclei and X-ray binaries}, \emph{Monthly Notices of the
  Royal Astron. Society}, \textbf{249}, pp. 352--367.

\bibitem[{Ghisellini et~al.(1991)Ghisellini, George, Fabian and C.}]{ghi91}
Ghisellini, G., George, I.~M., Fabian, A.~C. and C., D. (1991), {Anisotropic
  inverse Compton emission}, \emph{Monthly Notices of the Royal Astron.
  Society}, \textbf{248}, pp. 14--19.

\bibitem[{Goosmann(2006)}]{goo06}
Goosmann, R.~W. (2006), \emph{{Accretion and Emission close to supermassive
  Black Holes in Quasars and AGN: Modeling the UV-X-spectrum}}, PhD Thesis,
  Universit{\"{a}}t Hamburg, Germany.

\bibitem[{Guainazzi(2003)}]{gua03}
Guainazzi, M. (2003), {The history of the iron Kalpha line profile in the
  Piccinotti AGN ESO 198-G24}, \emph{Astronomy \& Astrophysics}, \textbf{401},
  pp. 903--910.

\bibitem[{Iwasawa et~al.(1996)Iwasawa, Fabian, Reynolds, Nandra, Otani, Inoue,
  Hayashida, Brandt, T.Dotani, Kunieda, Matsuoka and Tanaka}]{iwa96}
Iwasawa, K., Fabian, A.~C., Reynolds, C.~S., Nandra, K., Otani, C., Inoue, H.,
  Hayashida, K., Brandt, W.~N., T.Dotani, Kunieda, H., Matsuoka, M. and Tanaka,
  Y. (1996), {The variable iron K emission line in MCG-6-30-15}, \emph{Monthly
  Notices of the Royal Astron. Society}, \textbf{282}, pp. 1038--1048.

\bibitem[{Martocchia et~al.(2000)Martocchia, Karas and Matt}]{mar00}
Martocchia, A., Karas, V. and Matt, G. (2000), {Effects of Kerr space-time on
  spectral features from X-ray illuminated accretion discs}, \emph{Monthly
  Notices of the Royal Astron. Society}, \textbf{312}, pp. 817--826.

\bibitem[{Martocchia and Matt(1996)}]{mar96}
Martocchia, A. and Matt, G. (1996), {Iron Kalpha line intensity from accretion
  discs around rotating black holes}, \emph{Monthly Notices of the Royal
  Astron. Society}, \textbf{282}, pp. L53--L57.

\bibitem[{Matt et~al.(1993)Matt, Fabian and Ross}]{mat93}
Matt, G., Fabian, A.~C. and Ross, R.~R. (1993), {Iron K-alpha lines from X-ray
  photoionized accretion discs}, \emph{Monthly Notices of the Royal Astron.
  Society}, \textbf{262}, pp. 179--186.

\bibitem[{Merloni and Fabian(2001)}]{mer01}
Merloni, A. and Fabian, A.~C. (2001), {Accretion disc coronae as magnetic
  reservoirs}, \emph{Monthly Notices of the Royal Astron. Society},
  \textbf{321}, pp. 549--552.

\bibitem[{Miniutti and Fabian(2004)}]{min04}
Miniutti, G. and Fabian, A.~C. (2004), {A light bending model for the X-ray
  temporal and spectral properties of accreting black holes}, \emph{Monthly
  Notices of the Royal Astron. Society}, \textbf{349}, pp. 1435--1448.

\bibitem[{Miniutti et~al.(2003)Miniutti, Fabian, Goyder and Lasenby}]{min03}
Miniutti, G., Fabian, A.~C., Goyder, R. and Lasenby, A.~N. (2003), {The lack of
  variability of the iron line in MCG-6-30-15: general relativistic effects},
  \emph{Monthly Notices of the Royal Astron. Society}, \textbf{344}, pp.
  L22--L26.

\bibitem[{Misner et~al.(1973)Misner, Thorne and Wheeler}]{mis73}
Misner, C.~W., Thorne, K.~S. and Wheeler, J.~A. (1973), \emph{{Gravitation}},
  Freeman, San Francisco.

\bibitem[{Niedzwiecki and Zycki(2007)}]{nie07}
Niedzwiecki, A. and Zycki, P.~T. (2007), {On variability and spectral
  distortion of the fluorescent iron lines from black-hole accretion discs},
  \emph{Monthly Notices of the Royal Astron. Society}.

\bibitem[{Poutanen and Fabian(1999)}]{pou99}
Poutanen, J. and Fabian, A.~C. (1999), {Spectral evolution of magnetic flares
  and time lags in accreting black hole sources}, \emph{Monthly Notices of the
  Royal Astron. Society}, \textbf{306}, pp. L31--L37.

\bibitem[{Reynolds and Nowak(2003)}]{rey03}
Reynolds, C.~S. and Nowak, M.~A. (2003), {Fluorescent iron lines as a probe of
  astrophysical black hole systems}, \emph{Phys.\ Reports}, \textbf{377}, pp.
  389--466.

\bibitem[{Turner et~al.(2002)Turner, Mushotzky, Yaqoob, George, Snowden,
  Netzer, Kraemer, Nandra and Chelouche}]{tur02}
Turner, T.~J., Mushotzky, R.~F., Yaqoob, T., George, I.~M., Snowden, S.~L.,
  Netzer, H., Kraemer, S.~B., Nandra, K. and Chelouche, D. (2002), {Narrow
  Components within the Fe Kα Profile of NGC 3516: Evidence of the Importance
  of General Relativistic Effects?}, \emph{Astroph. Journal}, \textbf{574}, pp.
  L123--L127.

\bibitem[{Yaqoob et~al.(2003)Yaqoob, George, Kallman, Padmanabhan, Weaver and
  Turner}]{yaq03}
Yaqoob, T., George, I.~M., Kallman, T.~R., Padmanabhan, U., Weaver, K.~A. and
  Turner, T.~J. (2003), {Fe XXV and Fe XXVI Diagnostics of the Black Hole and
  Accretion Disk in Active Galaxies: Chandra Time-resolved Grating Spectroscopy
  of NGC 7314}, \emph{Astroph. Journal}, \textbf{596}, pp. 85--104.

\end{thebibliography}

\end{document}